\documentclass[11pt]{article}

\usepackage{setspace}
\usepackage[margin=1.5in]{geometry}

\usepackage[english]{babel}

\usepackage{amsmath}
\usepackage{amsfonts}
\usepackage{amssymb}
\usepackage{graphicx}
\usepackage{tikz-cd}
\usepackage{mathrsfs}
\usepackage{xfrac}
\usepackage{url}
\usepackage{color}
\usepackage[colorlinks=true, citecolor=blue]{hyperref}
\usepackage{enumitem}
\usepackage{accents}

\numberwithin{equation}{section}
\date{\today}

\def\a{\alpha}   
   
\def\t{\theta}
\def\be{\begin{equation}}
\def\ee{\end{equation}}
\def\ba#1\ea{\begin{align}#1\end{align}}
\def\no{\nonumber\\}
\def\ra{\rangle}
\def\la{\langle}

\def\diff{\text{\sl Diff}^+\!(S^1)}
\def\adiff{\mathfrak{diff}(S^1)}
\def\ddiff{\mathfrak{diff}^*(S^1)}
\def\whddiff{\accentset{\circ}{\mathfrak{diff}}^*(S^1)}
\def\psl{\text{\sl PSL}(2, \bb R)}
\def\apsl{\mathfrak{psl}(2, \bb R)}

\def\ad{\text{ad}}
\def\coad{\text{coad}}
\def\hodge{\,^\star\!}
\def\ads{\text{\sl AdS}_3}
\def\mink{\text{\sl Mink}_3}
\def\lads{\ell_\text{\sl AdS}}

\newcommand{\ca}[1]{\mathcal{#1}}
\newcommand{\bb}[1]{\mathbb{#1}}
\newcommand{\fr}[1]{\mathfrak{#1}}
\newcommand{\scr}[1]{\mathscr{#1}}
\newcommand{\ii}{\imath}
\newcommand{\id}{\mathrm{I}}
\newcommand{\wt}[1]{\widetilde{#1}}
\newcommand{\bs}[1]{\boldsymbol{#1}}
\newcommand{\wh}[1]{\widehat{#1}}
\newcommand{\ol}[1]{\overline{#1}}
\newcommand{\ac}[1]{\accentset{\circ}{#1}}
\newcommand{\tensortype}[2]{\left( \hspace{-0.2cm} \begin{smallmatrix} #1 \vspace{-0.08cm} \\ \hspace{0.3cm} #2 \end{smallmatrix} \hspace{-0.05cm} \right)}

\graphicspath{{Figures/}}

\title{\bf Quantization of causal diamonds\\ in (2+1)-dimensional gravity\\Part I: Classical reduction
\vspace{5mm}
}
\date{\today}
\author{Rodrigo Andrade e Silva\thanks{rasilva@umd.edu}
\vspace{-5mm}
\\
\and
{\it \small Center for Fundamental Physics,
University of Maryland}\\
{\it\small College Park, MD, 20742, USA}\\
{\it \small and}\\
{\it \small Perimeter Institute for Theoretical Physics}\\
{\it\small Waterloo, ON, N2L 2Y5, Canada}}

\begin{document}
\begin{titlepage}
\maketitle
\thispagestyle{empty}

\begin{abstract}
We develop the non-perturbative reduced phase space quantization of {\sl causal diamonds} in $(2+1)$-dimensional gravity with a nonpositive cosmological constant. In this Part I we focus on the classical reduction process, and the description of the reduced phase space, while in Part II we discuss the quantization of the phase space and quantum aspects of the causal diamonds. The system is defined as the domain of dependence of a spacelike topological disk with fixed boundary metric. By solving the constraints in a constant-mean-curvature time gauge and removing all the spatial gauge redundancy, we find that the phase space is the cotangent bundle of $\diff/\psl$, i.e., the group of orientation-preserving diffeomorphisms of the circle modulo the projective special linear subgroup. Classically, the states correspond to causal diamonds embedded in $\ads$ (or $\mink$ if $\Lambda = 0$), with fixed corner length, and whose Cauchy surfaces have the topology of a disc.
\end{abstract}

\end{titlepage}

\tableofcontents

\section{Introduction}
\label{sec:intro}

One of the grand goals of modern physics is to develop a consistent theory of quantum gravity, a problem that has confounded physicists for almost a century and is still unresolved despite intense research. 
Without the help of experimental evidence to steer us in the right direction, possibly the best course of action is to commit to a set of principles and proceed through logic and mathematical rigor to explore the ultimate consequences of these choices. 
Here we adopt the principle that quantum gravity is a quantum mechanical theory of gravity, and moreover that it is obtained from a canonical quantization of general relativity. The first principle derives from the perspective that quantum mechanics is based on a very rigid structure (namely, the complex linear structure of operators on Hilbert spaces), which has been tested in many different scenarios with no hint of violation --- before adventuring into the exploration of more radical fundamental theories,
it is fair to take the conservative stand and simply trust quantum mechanics until further conceptual revision is called for.
The second principle is that of canonical quantization, a prescription proposed by Dirac to infer the quantum theory underlying a given classical theory, which has been remarkably successful in many situations.
Although it is possible (and even likely) that the ``fundamental'' theory of quantum gravity does not correspond to a quantization of general relativity, it is still plausible that such a quantization could provide a partial, approximated picture for quantum gravity and yield sufficient insight to motivate the next leap forward in this endeavor. 

In this and a following paper \cite{rodrigo2023Pt2}, we develop a non-perturbative canonical quantization of causal diamonds in (2+1)-dimensional gravity. By causal diamonds we mean a class of finite-sized, globally-hyperbolic spacetimes whose Cauchy slices have the topology of a ball --- each ``diamond state'' is defined as the maximal development of Einstein's equation from initial data given on a Cauchy slice. The motivations, which will be further explained in this introduction, are two-fold. The first is that we want to better understand quantum gravity in a quasi-local sense. That is, what is the proper notion of ``spacetime subregion'' (or even ``spacetime'' itself) in quantum gravity, given that there are no compactly-supported gauge-invariant observables in gravity and therefore no clear notion of locality? At least from the classical perspective, causal diamonds are the natural object to study as they best represent a finite self-contained subsystem of spacetime. The second motivation is to explore a particular program for canonically quantizing gravity which seems promising in many situations, in arbitrary dimensions, based on the symplectic reduction of the phase space using a convenient gauge-fixing of time by constant-mean-curvature surfaces. If one can prove that this gauge-fixing is well-posed for the class of spacetimes under consideration, the constraints simplify and, and in many cases, can be solved (in principle) yielding a universal characterization of the reduced phase space as the cotangent bundle of the space of conformal geometries on the Cauchy slice. However, for this method to apply to the problem, we need to assume some energy condition (at the classical level) and also some boundary condition. Here we assume pure gravity with a non-positive cosmological constant and ``Dirichlet condition'' for the metric induced on the boundary of the Cauchy slice. Moreover, to have a better handle on the problem, and particularly on its quantization, we assume 2+1 spacetime dimensions. 

This work is divided into two parts: in ``Part I'' we describe the classical system, the structure of constraints and gauge transformations, and carry out the reduction process to find the reduced phase space of the theory;
in ``Part II'' \cite{rodrigo2023Pt2} we describe Isham's group-theoretic method of quantization, find an appropriate group to carry the quantization, and discuss some aspects of the resulting quantum theory. The results are also summarized, in a brief but fairly explicit manner, in \cite{e2023causal}.

\subsection{Motivations}
\label{subsec:motiv}

It has long been known that perturbative canonical quantization of general relativity, expanded around a fixed background geometry, does not lead to a complete theory. In particular, it is not renormalizable since the coupling parameter, $G$, has negative mass dimension in spacetime dimensions equal or greater than three.\footnote{Notwithstanding this power-counting argument, gravity in three spacetime dimensions is actually perturbatively renormalizable, which can be seen when expressed as a topological Chern-Simons theory~\cite{witten19882+}. For a closed Cauchy slice this is expected since the theory has a finite number of degrees of freedom, but in asymptotically AdS or in the presence of spatial boundaries the conclusion may not be as clear---in fact, since the theory has no local degrees of freedom, the question of pertubative renormalizability, in the context of local quantum field theories, might be ill-posed.}  There remains the hope that a careful, non-perturbative quantization of general relativity could still be meaningful. 
Among the other challenges encountered in the quantization of gravity, one is the famous {\sl problem of time} \cite{isham1993canonical,anderson2012problem,kuchavr2011time}. As time is a dynamical aspect of gravity, as opposed to a background entity, its role and place in the quantum theory is enigmatic.
Another issue is the non-linearity of the constraints of general relativity, particularly in the momentum variables, which severely complicates attempts to proceed with the quantization exactly and non-perturbatively.
Lastly, a notable peculiarity of gravity is the absence of local, or even compactly supported, observables. This is because any physical observables must be invariant under gauge transformations, and therefore must be invariant under general spacetime diffeormorphisms (at least those that are supported away from the boundary). Thus, the general notion of locality, and even the meaning of subregions, is particularly fuzzy in quantum gravity.

In view of these general challenges, we attempt to the address the following two main points in this work:
\begin{enumerate}
\item We wish to better understand how to describe quantum gravity in finite regions of space(time). In particular, what can be learned if we take the classical notion of a self-contained subregion of spacetime, i.e. a {\sl causal diamond}, and quantize (Einstein-Hilbert) gravity inside it. More precisely, we wish to quantize the class of spacetimes consisting of causal diamonds in pure general relativity (including a cosmological constant), where a causal diamond is defined as the maximal development of initial data, satisfying the constraints, given on a bounded acausal spatial slice.

\item We wish to continue the exploration of a program for quantizing gravity non-perturbatively by explicitly reducing the phase space via a particular gauge-fixing for time defined by a CMC (constant-mean-curvature) condition \cite{moncrief1990solvable,moncrief1989reduction,fischer1997hamiltonian}. If one is quantizing a class of spacetimes in which each spacetime admits a regular CMC foliation (i.e., where the leaves are defined by having a constant trace of extrinsic curvature, a.k.a, mean-curvature), then the constraints of general relativity can be cast into a more manageable form, in terms of the {\sl Lichnerowicz equation}; if one can prove certain existence and uniqueness properties for the solution of this equation, then quite generally the reduced phase space (of pure gravity) ends up being $T^*[\text{ConGeo}(\Sigma)]$, i.e., the cotangent bundle of the space of conformal geometries\footnote{The space of conformal geometries, $\text{ConGeo}(\Sigma)$, is the space of equivalence classes of metrics $h$ on a manifold $\Sigma$ where two metrics are identified if they can be related by a combined Weyl scaling and diffeomorphism push-forward, $h \sim \Psi_*\Omega h$; the class of metrics $h$, positive functions $\Omega$ and diffeomorphisms $\Psi$ participating in this quotient depends on the details of gravitational system being reduced.} on the Cauchy slice. This yields a non-perturbative characterization of the reduced phase space, which completely resolves the issues related to the constraints and gauge invariance of gravity at the classical level; accordingly, it is in principle easier to take the next step and try to quantize the resulting (reduced) theory non-perturbatively.
\end{enumerate}

Another reason for attempting to combine these two points is that certain spacetimes do not admit CMC foliations (or even a maximal slice), due to global reasons. For example, any non-flat spacetime with topology $T^3 \times \bb R$, satisfying the timelike convergence condition (i.e., $\text{\it{\textbf{Ric}}}(u, u) \ge 0$ for all timelike vectors $u$), does not posses a maximal slice \cite{marsden1980maximal,fischer1975linearization,ruffini1977proceedings,schoen1978incompressible,schoen1979existence},
and therefore cannot be treated with this Lichnerowicz method. Even if there is a foliation by CMCs, the Lichnerowicz equation may not have the desired existence and uniqueness properties (e.g., see \cite{choquet2004einstein}).
If one could properly quantize local regions of spacetime via this Moncrief-Lichnerowicz approach, and glue the resulting ``quantum causal diamonds'' to form a larger spacetime, then one would have a theory of quantum gravity describing a much larger class of spacetimes.\footnote{The program described in point 2 is closely related to {\it shape dynamics}~\cite{mercati2014shape,barbour2012shape,gomes2011dynamics,barbour2014solution}. The difference is that in shape dynamics the space of conformal geometries is fundamental, and therefore it is only equivalent to gravity in the special cases where this phase space reduction can be implemented. In our approach, we are committed to working entirely within the context of general relativity.}

While promising in principle, the actual implementation of this program for general causal diamonds, in arbitrary dimensions, would be a quite ambitious effort. First, it would involve a classification of the space of conformal geometries in $n$-dimensional manifolds with boundaries, which is not a fully solved mathematical problem; second, one would have to develop a non-perturbative quantization of such space, which would presumably be challenging.
In view of the difficulties, it is worthwhile to study the problem in a simplified setting, where we might have a better handle on both the physics and the mathematics.
In this work, the problem of interest is 2+1 dimensional (Einstein-Hilbert) gravity with a nonpositive cosmological constant in the domain of dependence of a topological disc, hereinafter simply referred to as a {\sl causal diamond}. For reasons that will be explained later, the class of spatial metrics are restricted by a ``Dirichlet boundary condition'', that is, the (induced) metric is fixed at the boundary of the disc. 
We reiterate that this is not a theory of a single causal diamond, but rather a dynamical theory of gravity in the class of globally-hyperbolic spacetimes whose Cauchy slices are topological discs (with fixed induced corner metric). The reason for considering a nonpositive cosmological constant is that it ensures that the spacetime can be entirely foliated by CMC surfaces, thus providing a natural notion of time for causal diamonds, and allowing the application of the CMC gauge-fixing of time, and also that the associated Lichnerowicz equation can be proven to have the desired existence and uniqueness properties.

There exists an extensive literature on (2+1)-dimensional gravity systems. A limited sample of references includes work on spacetimes with closed spatial slices (where the reduced phase space is finite-dimensional) \cite{witten19882+,witten1989topology,moncrief1989reduction,moncrief1990solvable,fischer1997hamiltonian,ashtekar19892+,hosoya19902+,Carlip:1998uc, carlip2005quantum,mondal2020thurston}, on spacetimes with finite timelike boundary \cite{kraus20213d,adami2020sliding,ebert2022field}, and on asymptotically $\ads$ spacetimes\cite{brown1986central,freidel20042+,carlip2005conformal,witten2007three,maloney2010quantum,Scarinci:2011np,kim2015canonical, cotler2019theory}. 
The study of causal diamonds can be valuable to improve the understanding of subsystems or ``regions of spacetime' in quantum gravity --- as mentioned above, the dynamical nature of spacetime and the diffeomorphism invariance of the theory, which precludes the existence of local (or quasi-local) observables, makes the notion of ``subregions of spacetime'' particularly fuzzy --- deservedly, they have received a great deal of attention recently from a variety of different approaches attempting to unveil their quantum properties \cite{chandrasekaran2019symmetries, de2016entanglement,banks2021path,jacobson2019gravitational,jacobson2022entropy}. In this work we intend to push the analysis further by developing a fully non-perturbative quantization of (pure) gravity in causal diamonds spacetimes, via the phase space reduction approach paired with Isham's group-theoretic quantization. While simple enough to be exactly solvable classically, due to the absence of local degrees of freedom, the system has nevertheless an infinite-dimensional reduced phase space of ``boundary gravitons''.  (A recent paper by Witten \cite{witten2022note} 
revitalizes this program for canonically quantizing gravity, applying a similar approach to asymptotically Anti-de Sitter spacetimes.)

\subsection{Summary}
\label{subsec:sum}

We begin with a somewhat detailed summary of the contents. The goal is to provide a quick guide to the main ideas and results, in view of the extensive nature of the paper. 

\vskip 0.2cm
\noindent\emph{\underline{Note}:} Appendix \ref{app:GSandC} contains a compilation of the main symbols, definitions and conventions used in the text.
\vskip 0.2cm

\begin{figure}[h!]
\centering
\includegraphics[scale = 0.6]{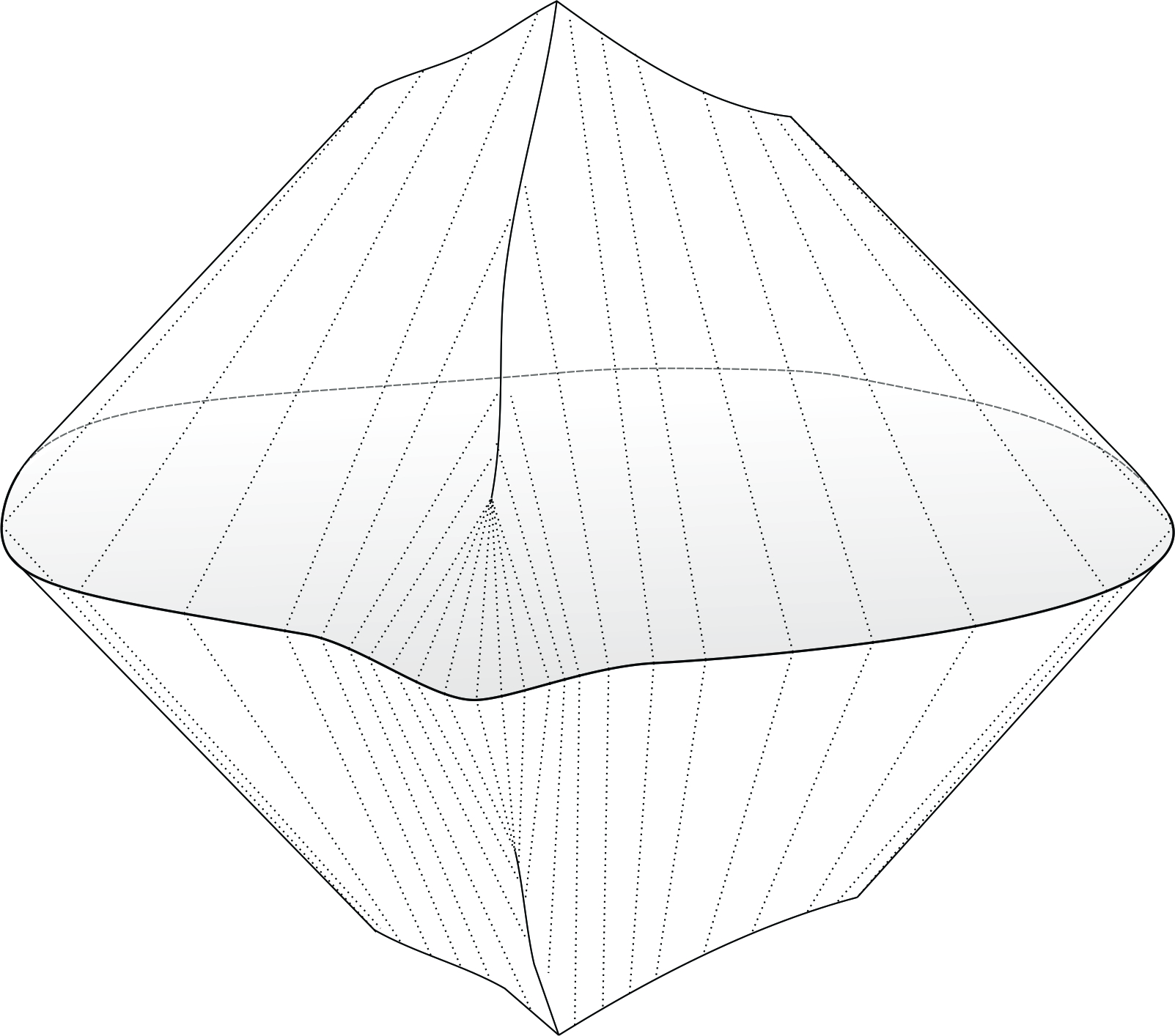}
\caption{A typical causal diamond, obtained by maximally developing geometric data on a Cauchy slice with disc topology.}
\label{fig:typicdia}
\end{figure}

The gravitational system of our interest is defined (Sec. \ref{subsec:system}) as the class of maximal developments, under the vacuum Einstein's equations (with a non-positive cosmological constant), of all initial geometric data (satisfying a certain boundary condition) given on a manifold with the topology of a 2-dimensional disc (with boundary). According to the ADM (Arnowitt-Deser-Misner) formalism, the geometrical data consists of a spatial metric $h_{ab}$ and an extrinsic curvature $K^{ab}$, which must satisfy the momentum constraint $\nabla_a (K^{ab} - K h^{ab}) = 0$ and the Hamiltonian constraint $K^{ab}K_{ab} - K^2 -  R_{(h)} + 2\Lambda = 0$, where $K := K^{ab}h_{ab}$ is the trace of the extrinsic curvature and $R_{(h)}$ is the Ricci scalar for $h_{ab}$. 
Each given geometric data defines a {\sl causal diamond}. We reiterate that we will be quantizing not a single causal diamond, but the class of all such causal diamonds satisfying a certain condition on the corner (i.e., the boundary of any Cauchy slice) length. Note that a typical causal diamond in this class is not the intersection of the past and future of two timelike-separated points, but rather it will have horizons with a ridged, mountain-like appearance, as in Fig.~\ref{fig:typicdia}.

In quantizing a gauge theory one must, sooner or later, deal with the constraints and gauge invariance of the theory. There are two mainstream routes of proceeding. The first route, called the Dirac approach, is to first quantize the theory ignoring the constraints, and then impose the constraints at the quantum level. That is, if the (unconstrained) phase space is covered by conjugate coordinates $q^i$ and $p_i$, and there is a set of (first-class) constraints $C_\a = 0$, the quantization goes by first constructing a Hilbert space carrying a representation of the Heisenberg algebra $[\widehat q^i, \widehat p_j] = i \hbar \delta^i_{\,j}$, and then restricting to the {\sl physical} subspace defined by states satisfying the conditions $\widehat C_\a |\psi\ra = 0$, and to physical, gauge-invariant observables that commute with the constraints, $[\ca O, \widehat C_\a] = 0$.\footnote{It may be necessary to only impose these constraints weekly, i.e., in between physical states.} The second route, called the reduced phase space approach (Sec. \ref{subsec:redphasespacedef}), is to first impose the constraints at the classical level, remove all the associated gauge ambiguities, and then quantize the resulting (gauge-free) theory. 
There is no a priori guarantee that the two approaches to quantization would lead to the same quantum theory.
Here, for the quantization of  causal diamonds, we shall focus on the reduced phase space approach. In App.~\ref{sec:reduced} we present a brief review of the general concept of a reduced phase space.

We implement a program (Sec. \ref{subsec:outline}) based on the gauge-fixing of time by surfaces of constant mean-curvature, i.e., given any foliation of a causal diamond, one uses the gauge flow generated by the Hamiltonian constraint to re-foliate the diamond into slices of constant $K$, which varies monotonically along the foliation and thus define a suitable time coordinate $\tau = -K$. As $K$ is now a time-dependent constant, on each spatial slice, only the trace-free part of $K^{ab}$, $\sigma^{ab} := K^{ab} - \frac{1}{2}K h^{ab}$, remains as a dynamical variable together with $h_{ab}$. In this gauge, if one starts with some ``seed data'', $(h_{ab}, \sigma^{ab})$, which is taken to satisfy (at least) the momentum constraint, which now reads $\nabla_a \sigma^{ab} = 0$, then it may be possible to deform this seed data into ``valid data'', i.e. so that both the momentum and Hamiltonian constraints are satisfied, via an appropriate Weyl transformation, $(h_{ab}, \sigma^{ab}) \mapsto (e^\phi h_{ab}, e^{-2\phi} \sigma^{ab})$; the condition for this deformed data to satisfy the constraints is that $\phi$ must satisfy a non-linear elliptic equation called the {\sl Lichnerowicz equation}.

Three things must be checked in order to establish the non-perturbative validity of this (partial) gauge-fixing prescription, and a fourth one to ensure utility. First, one needs to ensure that the CMC foliation always exists and is unique for every causal diamond within the class of causal diamonds under consideration. We will argue (Sec. \ref{subsec:nicefoliation}) that the foliation exists and covers each diamond entirely as $\tau$ ranges from $-\infty$ to $\infty$, provided that $\Lambda < 0$; the case $\Lambda = 0$ can be included by a continuity argument. The second point is that the CMC foliation can be attained by a gauge transformation, starting from any other foliation (Sec. \ref{subsec:CMCattain}). It is thus important to make sure that generic smearings of the Hamiltonian constraint, $H[\eta] := \int_\Sigma d^2\!x \sqrt{h}\, \eta (K^{ab}K_{ab} - K^2 -  R_{(h)} + 2\Lambda)$, where $\eta$ vanishes at the boundary $\partial\Sigma$ (since all Cauchy slices $\Sigma$ in a causal diamond meet at the corner), must be gauge-generators. As the constraints are first-class, $H[\eta]$ is a generator of gauge provided that it generates a well-defined flow in the (pre) phase space. With respect to the symplectic form $\Omega = \int_\Sigma d^2\!x\, \delta \pi^{ab} \wedge \delta h_{ab}$, where $\pi^{ab} = \sqrt{h} (K^{ab} - K h^{ab})$, the flow of a phase space function is regular if and only if it has well-defined functional derivatives, i.e., if $\delta H[\eta]$ is of the form $\int_\Sigma d^2\!x (A^{ab} \delta h_{ab} + B_{ab} \delta \pi^{ab})$. In the present case, one notices that for arbitrary boundary conditions, only the $H[\eta]$ with vanishing normal derivative of $\eta$ at the boundary are gauge generators; in other words, corner boosts (which tilt the angle at which the Cauchy slice meets the corner) are not gauge transformations, but rather non-trivial transformations between different states. However, if one imposes a ``Dirichlet boundary'' condition, where the induced metric at the boundary is fixed, then all $H[\eta]$, with $\eta|_\partial = 0$, are gauge-generators. Therefore we restrict the class of spatial metrics in this way, so that the CMC gauge is attainable. Third, one needs to ensure that the Hamiltonian constraint can be solved by a Weyl transformation, that is, for any seed data $(h_{ab}, \sigma^{ab})$ there must exist a solution $\phi$ of the associated Lichnerowicz equation. We show that the equation always has solutions, for all $\tau$, as long as $\Lambda \le 0$ (Sec. \ref{subsec:existunique}). Fourth, in determining the reduced phase space, it is important that there are no residual, unfixed gauge directions; in this language, one needs to prove that there is a unique solution $\phi$ for any given seed data. This can be shown to also follow from $\Lambda \le 0$, together with the fact that the Dirichlet boundary condition requires $\phi|_\partial = 0$, as we consider only seed data that satisfies the boundary condition (Sec. \ref{subsec:existunique}). Uniqueness is essential since it implies that when two seed data related by a Weyl transformation, $(h',\sigma') = (e^\lambda h, e^{-2\lambda}\sigma)$, are used as inputs in the Lichnerowicz algorithm, they will each be deformed into the same output valid data. This means that the constraint surface (i.e., the set of all valid data) can be identified with the set of equivalence classes of seed data under Weyl transformations (acting trivially at the boundary). 

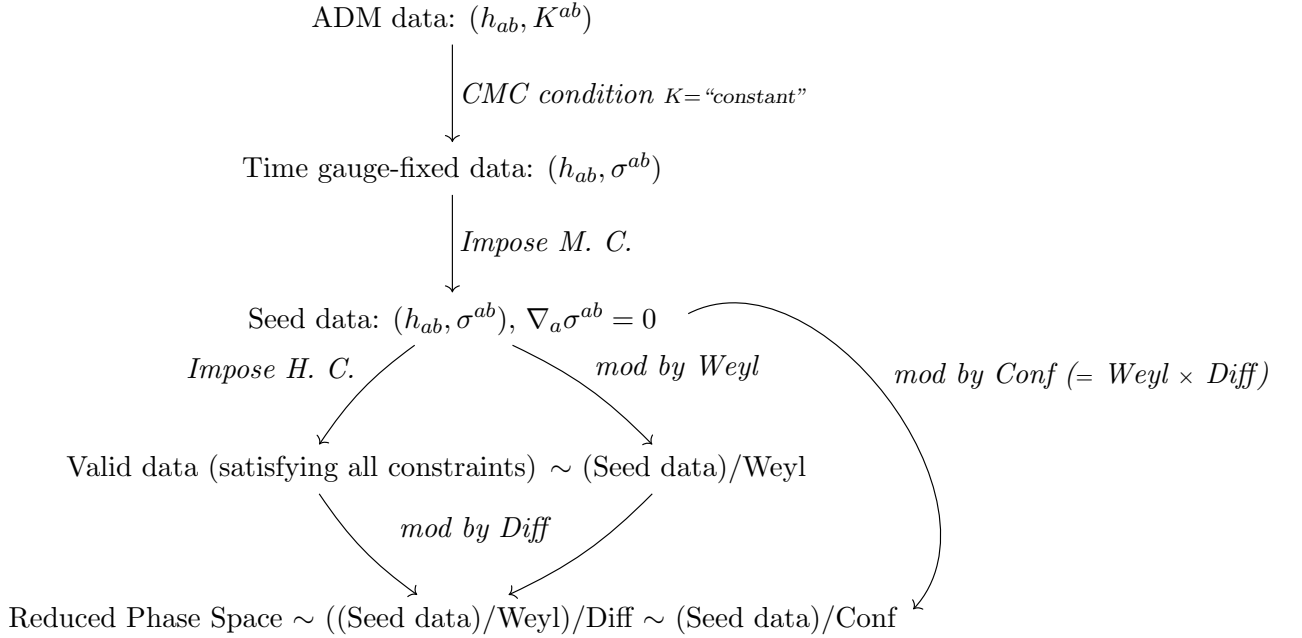
\begin{figure}[h!]
\hspace{-0.5cm}\begin{tikzpicture}[commutative diagrams/every diagram]
\node (ADM) at (0,0) {ADM data: $(h_{ab}, K^{ab})$};
\node (GF) at (0,-2) {Time gauge-fixed data: $(h_{ab}, \sigma^{ab})$};
\node (Seed) at (0,-4) {Seed data: $(h_{ab}, \sigma^{ab})$,  $\nabla_a\sigma^{ab}=0$};
\node (CS1) at (-2,-6) {Valid data (satisfying all constraints)};
\node (CS2) at (3,-6) {$\sim$ (Seed data)/Weyl};
\node (RPS) at (0,-8) {Reduced Phase Space $\sim$ ((Seed data)/Weyl)/Diff $\sim$ (Seed data)/Conf};
\node (RPS2) at (6,-8) {};
\node (Seed2) at (3,-4) {};
\path[commutative diagrams/.cd, every arrow, every label]
(ADM) edge node {\it CMC condition $K = \text{\sl ``constant''}$} (GF)
(GF) edge node {\it Impose M. C.} (Seed)
(Seed) edge[bend right=10] node[swap] {\it Impose H. C.} (CS1)
(Seed) edge[bend left=10] node {\it mod by Weyl} (CS2)
(CS1) edge[bend right=10] node {\it \hspace{7pt} mod by Diff} (RPS)
(CS2) edge[bend left=10] node {} (RPS)
(Seed2) edge[bend left=80] node {\it mod by Conf ($=$ Weyl $\times$ Diff)} (RPS2);
\end{tikzpicture}
\caption{Starting with ADM data, satisfying the Dirichlet condition on the (induced) boundary metric, we gauge-fix ``time'' by imposing the CMC condition. Then we impose the momentum constraint (M. C.) to define ``seed data''. Further imposing the Hamiltonian constraint (H. C.) leads to ``valid data'' (i.e., data satisfying all the constraints); from arguments involving the Lichnerowicz equation, the space of valid data can be identified with the space of seed data modulo (boundary-trivial) Weyl transformations. The reduced phase space is then obtained by further quotienting the space of valid data by (boundary-trivial) diffeomorphisms; this identifies the reduced phase space with the space of seed data modulo (boundary-trivial) conformal transformations.}
\label{red_diag_intro}
\end{figure}

Up to this point we have dealt with the gauge associated with time refoliations, and solved the constraints; we need next to deal with gauge associated with spatial diffeomorphisms. As one could anticipate, it is possible to show that two valid data that can be related by a diffeomorphism that is trivial at the boundary, $(h'_{ab}, \sigma'^{ab}) = (\Psi_* h_{ab}, \Psi_*\sigma^{ab})$, where $\Psi: \Sigma \rightarrow \Sigma$ satisfies $\Psi|_\partial = I$, are gauge related. The reduced phase space is the space of physically distinguishable solutions to the equations of motion, or equivalently the space of valid initial data modulo gauge transformations. Thus, here it can be identified with the set of equivalence classes of valid data under these ``boundary-trivial'' spatial diffeomorphisms, or equivalently the set of equivalence classes of seed data under boundary-trivial conformal transformations. The reduction process can be summarized in a diagram, Fig.~\ref{red_diag_intro}.

The reduced phase space, just identified with the space of seed data modulo conformal transformations, can also be identified with the {\sl cotangent bundle of conformal geometries} on the Cauchy slice (Sec. \ref{subsec:cotanconfgeo}), with its natural symplectic form. To get a ``taste of why'' note that, locally in the space of metrics, the infinitesimal directions that will be quotiented out by conformal transformations are $h_{ab} \mapsto (1 +\lambda)h_{ab} = h_{ab} + \lambda h_{ab}$ and $h_{ab} \mapsto h_{ab} + \pounds_\xi h_{ab} = h_{ab} + 2\nabla_{(a} \xi_{b)}$; then note that $\sigma^{ab}$ in the seed data is characterized by two properties, tracelessness and divergencelessness, and consequently $\sigma^{ab}$ can naturally be thought of as 1-form on the cotangent space of $\text{ConGeo}(\Sigma)$ at $[h] = [\Psi_*\Omega h]$ since the pairing $\la \sigma, \delta h\ra := \int_\Sigma d^2\!x \sqrt{h}\, \sigma^{ab} \delta h_{ab}$ is insensitive to the directions that will be projected out, namely $\la \sigma, \lambda h_{ab} \ra = 0$ and $\la \sigma, \pounds_\xi h_{ab}\ra = 0$.

Particularizing to the case where $\Sigma$ is a disc, $D$, with the Dirichlet condition on the (induced) boundary metric, we can show (Sec. \ref{subsec:confgeodisc}) that the space of conformal geometries is $\diff/\psl$, i.e., the group of (orientation-preserving) diffeomorphisms of the circle modded by the three-dimensional $\psl$ subgroup of projective special linear transformations in two real dimensions. (At the algebra level, $\adiff$ can be identified with the space of vector fields on $S^1$, and  the $\apsl$ subalgebra then corresponds to the three-lowest ``Fourier modes'', $\partial_\theta$, $\sin\theta\, \partial_\theta$ and $\cos\theta\, \partial_\theta$.) The reduced phase space is therefore
\be
\wt P = T^*[\diff/\psl]
\ee
with the natural symplectic form associated with the cotangent bundle structure.

We also present another method for carrying out the reduction of the phase space based on a suitable ``choice of coordinates'' on the (pre) phase space (Sec. \ref{sec:Rcc}). Inspired by the previous argument, we note that by starting with ``conformal coordinates'' one can immediately detect the null directions of the symplectic form, and then identify the reduced phase space as a ``level-set'' in these coordinates. In this part we are more rigorous in discussing gauge transformations in terms of the degenerate directions of the symplectic form. This analysis is particularly relevant when there are boundaries, since transformations that would naively be gauge in the bulk can become physical symmetries (mapping between distinct physical states) when acting non-trivially in a neighborhood of the boundary --- 
in particular, we justify that only diffeomorphisms that act trivially on the boundary are gauge. An advantage of this alternative approach is that it provides a natural ``coordinatization'' of the reduced phase space, along with the explicit quotient map from the redundant but concrete geometrical variables (namely, metrics and extrinsic curvatures) to the physical but abstract variables describing the reduced phase space. This is useful because what characterizes a quantum theory is not the structure of the Hilbert space itself, but the way in which meaningful physical observables are represented in the Hilbert space.\footnote{Recall that any two separable Hilbert spaces, as typically considered in physics, are isomorphic as vector spaces. One can take a countable basis $\{\Psi_n\}$ in one Hilbert space and another countable basis $\{\Phi_n\}$ in the other Hilbert space, and simply define an (invertible) linear map between them by $\Psi_n \mapsto \Phi_n$. Therefore the Hilbert space of a simple Harmonic oscillator is the same as the Hilbert space of the Standard Model of particle physics.}
When we come to the quantization, the variables in the reduced phase space will be those to become (self-adjoint) operators in the Hilbert space; it is therefore essential that we can understand their physical meaning, i.e., what experimental measurement would be described by that particular operator. Having this explicit map from the abstract variables back to concrete, geometrical variables is therefore valuable in this program.

The conformal coordinates are defined (Sec.~\ref{subsec:confcoord}) using the fact that, on a disc, any two metrics are conformally-equivalent. That is, given any reference metric $\ol h_{ab}$, any other metric $h_{ab}$ can be obtained via a conformal transformation, $h_{ab} = \Psi_* \Omega \ol h_{ab}$, for some diffeomorphism $\Psi$ and Weyl factor $\Omega$, which are generally not boundary-trivial. (In App.~\ref{app:unimap} we review the Riemann mapping theorem, a special case of the uniformization theorem, which yields explicit expressions for the $\Psi$ and $\Omega$ that uniformize a given metric on the disc.)
We can then ``pull-back'' the traceless and divergenceless (with respect to $h_{ab}$) extrinsic curvature $\sigma^{ab}$ through this conformal map as $\ol\sigma^{ab} := \Omega^2 \Psi^* \sigma^{ab}$. In this way, $\ol\sigma^{ab}$ is traceless and divergenceless with respect to the reference metric $\ol h_{ab}$, and instead of using geometric variables $(h_{ab}, \sigma^{ab})$ to describe the seed data, we can now use {\sl conformal coordinates} $(\Psi, \Omega, \ol\sigma^{ab})$. The Dirichlet condition on the induced boundary metric fixes the boundary value of $\Omega$ in terms of the boundary action of $\Psi$, $\psi := \Psi|_\partial$, and the Lichnerowicz equation then fixes uniquely the entire value of $\Omega$ in terms of $\psi$ and $\ol\sigma^{ab}$; thus the constraint surface is parametrized (Sec. \ref{subsec:impconst}) only by $(\Psi, \ol\sigma^{ab})$. At this stage, by looking at the symplectic form, it becomes evident that changing $\Psi \mapsto \Psi'$ is a gauge transformation as long as $\Psi|_\partial = \Psi'|_\partial$; which leads (Sec. \ref{subsec:bulkdiff}) to a partial phase space reduction $(\Psi, \ol\sigma^{ab}) \mapsto (\psi, \ol\sigma^{ab})$.  The fact that $\ol\sigma^{ab}$ is traceless and transverse (with respect to $\ol h_{ab}$) implies that it can be described by boundary data. More precisely, the space of such $\ol\sigma^{ab}$ is naturally isomorphic to the subspace of $\ddiff$ (the dual Lie algebra of $\diff$) that annihilates the $\apsl$ subalgebra of $\adiff$; this space will be denoted by $\whddiff$, and its elements by $\ac\sigma$. At this stage, the phase space is thus described by $(\psi, \ac\sigma) \in \diff \times \whddiff$. But by direct inspection of the symplectic form we find that there are still three null directions per phase space point. The component of these directions to the $\diff$ factor are directly related to $\psl$, but they also have a non-trivial component along the $\whddiff$ factor (Sec. \ref{subsec:respsl}). We construct explicitly a map $J$ that quotients out those directions, leading to the fully reduced phase space $\wt P = T^*[\diff/\psl]$.
The classical states can be understood as corresponding to causal diamonds embedded in $\ads$ (or $\mink$ if $\Lambda = 0$), with fixed corner length, whose Cauchy surfaces have the topology of a disc. In App.~\ref{app:embed} we describe this embedding picture in more detail, constructing a map from each point in the reduced phase space to a causal diamond embedded in $\ads$ (or $\mink$ if $\Lambda = 0$). For intuition and artistic purposes, some accurate pictures of causal diamonds are also displayed.

In the last two appendices, App.~\ref{sec:redH} and App.~\ref{sec:approxH}, we study the Hamiltonian generating time-evolution between CMC slices. We first review some basic facts about determining the Hamiltonian in a gauge-fixed, reduced phase space approach (Sec. \ref{subsec:genreview}). In a pure reduction process, the Hamiltonian on the reduced phase space is simply obtained by ``pushing-forward'' the original Hamiltonian. This push-forward is well-defined because, in a consistently formulated dynamical theory, the Hamiltonian is gauge-invariant and therefore constant within the pre-image (under the quotient map) of any point. When there is a gauge-fixing involved, the original and reduced Hamiltonians generating evolution along the ``time parameter'' are not related so simply. The easiest manner to determine the reduced Hamiltonian is by looking at the action (Sec. \ref{subsec:CMCHam}), from which we recover the expected result \cite{york1972role} that the Hamiltonian generating evolution between CMC slices is given, at time $\tau$, by the area of the slice with $K = -\tau$. Despite its simple geometrical interpretation, its expression in terms of reduced phase space variables is highly non-trivial since it is defined implicitly with respect to solutions of the Lichnerowicz equation. In fact, it is only evident from its formula that it is a well-defined function on the partially-reduced phase space, $\diff \times \whddiff$. As a check of consistency (Sec. \ref{subsec:Hwelldef}), we show explicitly that this Hamiltonian is indeed a well-defined function on $T^*[\diff/\psl]$. Because this Hamiltonian is a complicated function on the reduced phase space, it is interesting to explore some regimes where it can be solved analytically, or at least approximated (Sec. \ref{sec:approxH}). This is relevant if one wishes to describe the quantum dynamics of the system. We stress that while the full classical reduction  and the kinematical part of the quantization (i.e., representing a complete algebra of observables on a Hilbert space) are performed non-perturbatively, the dynamics (in CMC time) may well only be only amenable to approximate analysis.

We note again that App.~\ref{app:GSandC} --- {\sl glossary, symbols and conventions} --- may serve as a useful reference throughout the reading of this paper. All aspects concerning the quantization and the quantum theory of causal diamonds will be discussed in Part II \cite{rodrigo2023Pt2}.

\section{The causal diamond}
\label{sec:diamond}

In this section we define the dynamical system of interest, the causal diamond. We also provide a brief outline of the phase space reduction procedure, discussing the details in the subsequent sections.

\subsection{The system}
\label{subsec:system}

We define the causal diamond as the domain of dependence of a spatial slice $\Sigma$ having the topology of a disc $D$ (i.e., a 2-dimensional ball). More precisely, we define the spacetime as the maximal development of the initial-value problem associated with Einstein's equations for pure gravity with a nonpositive cosmological constant $\Lambda$,
\be
\ca G_{ab} + \Lambda g_{ab} = 0
\ee
given a Riemannian metric $h_{ab}$ and extrinsic curvature $K^{ab}$ on $\Sigma \sim D$. Our choice of boundary conditions, to be later justified, is that the induced metric on the spatial boundary is fixed
\be
h \big|_{\partial \Sigma} = \gamma
\ee
where $\gamma$ is a given (fixed) metric on $\partial\Sigma \sim S^1$. Since we can parametrize the points of the boundary by the length with respect to $\gamma$, the only {\sl intrinsic} attribute of this metric is the total length, $\ell$, of the boundary.

According to the ADM (Arnowitt-Deser-Misner) formalism~\cite{arnowitt2008republication}, the (pre)phase space $\ca P$ corresponds to the space of all Riemmanian metrics, satisfying the Dirichlet boundary condition, together with the space of all extrinsic curvatures on the ball $\Sigma \sim D$. For definiteness, let us denote the space of Riemannian metrics on the spatial disc by $\text{Riem}(D; \gamma)$, consisting of all positive-definite symmetric tensors of type $\tensortype{0}{2}$, $h_{ab}$, on $D$ satisfying $h|_{\partial D} = \gamma$; and denote the space of all extrinsic curvatures on $D$ by $\text{Sym}(D, \tensortype{2}{0})$, consisting of all symmetric tensors of type $\tensortype{2}{0}$, $K^{ab}$, on $D$. The (pre)phase space then have the trivial product structure
\be\label{prephasespacetrivial}
\ca P = \text{Riem}(D; \gamma) \times \text{Sym}(D, \tensortype{2}{0})
\ee
One could worry about the degree of smoothness of these function spaces. In the study of partial differential equations, such as the initial value problem of Einstein's equation, it is natural to consider {\sl Sobolev spaces}. The Sobolev space $W^{k,p}(U)$, where $k \in \bb N$, $1 \le p \le \infty$ and $U$ is a open subset of $\bb R^n$, is defined as the space of all functions $f : U \rightarrow \bb R$ such that $f$ and its (weak) derivatives of order equal or less than $k$ are in $L^p(U)$. (The generalization for functions valued in $\bb R^n$ is natural.) These spaces are convenient for they are Banach spaces (complete normed vector spaces), which facilitates certain proofs of existence of solutions by allowing one to construct sequences of approximate solutions that converge to an exact solution. The case $p = 2$ is particularly interesting because it is a Hilbert space, implying that its (topological) dual is isomorphic to itself. This allows us to think of the phase space as a cotangent bundle, in a precise way, as we explain next. Note that $\text{Riem}(D)$ can be seen as the open region of $\text{Sym}(D, \tensortype{0}{2})$ defined by the conditions $\text{det}(h) > 0$ and $\text{tr}(h) > 0$. Since $\text{Sym}(D, \tensortype{0}{2})$ is a linear space, we can assume that it is a Sobolev space $W^{k,2}(D)$ of symmetric-matrix-valued functions on $D$. Then the topology of $\text{Riem}(D)$ is inherited from $\text{Sym}(D, \tensortype{0}{2})$. A tangent vector to $\text{Riem}(D)$ can be seen as a vector in $\text{Sym}(D, \tensortype{0}{2})$, and due to the linearity of this space, the tangent vector can be naturally identified with an element of $\text{Sym}(D, \tensortype{0}{2})$ itself. We can write,
\be
T_h[\text{Riem}(D)] \sim \text{Sym}(D, \tensortype{0}{2})
\ee
where $T_h$ denotes the tangent space at $h \in \text{Riem}(D)$. The space of 1-forms at $h$ can be defined to be the (topological) dual of $\text{Sym}(D, \tensortype{0}{2})$, and since that is a Hilbert space, the dual is isomorphic to itself. Nevertheless, it is most natural to characterize the space of 1-forms as $\text{Sym}^1(D, \tensortype{2}{0})$, the space of symmetric tensor densities of type $\tensortype{2}{0}$ and weight $1$. In this way, the action of a dual vector $\pi$ on a vector $\zeta$, at $h$, is given by contracting them and integrating over $D$,
\be\label{Riemcovectorvectoraction}
\pi(\zeta) := \int_{D} d^n\!x\, \pi^{ab} \zeta_{ab}
\ee
The space of tensor densities is isomorphic to the space of tensors since they can be related by a factor of a power of $\sqrt{\text{det}(h)}$; in particular, $\pi^{ab} = \sqrt{\text{det}(h)} \wt\pi^{ab}$, where $\wt\pi^{ab}$ is a standard tensor.
Therefore,
\be
T^*_h[\text{Riem}(D)] \sim \text{Sym}^1(D, \tensortype{2}{0}) \sim \text{Sym}(D, \tensortype{2}{0})
\ee
where $T^*_h$ denotes the cotangent space at $h \in \text{Riem}(D, \gamma)$. Since $\text{Riem}(D)$ is an open subset in a vector space, its tangent and cotangent bundles are trivial. So,
\be
T^*[\text{Riem}(D)] = \text{Riem}(D) \times \text{Sym}(D, \tensortype{2}{0})
\ee

Our configuration space, however, contains the additional restriction on the induced boundary metric, so it is $\text{Riem}(D, \gamma)$. This condition affects the tangent space, since tangent vectors $\zeta$, tangent to curves $\text{Riem}(D, \gamma)$, are now subjected to the homogeneous boundary condition
\be\label{DBCvec}
\zeta_{ab} t^a t^b \big|_{\partial D} = 0
\ee
where $t^a$ is a vector on $D$ tangent to $\partial D$. Nonetheless, since the condition above refers to a set of measure zero in $D$, it does not affect the space of cotangent vectors, which can still be taken to be elements $\pi$ of $\text{Sym}^1(D, \tensortype{2}{0})$ acting on vectors as in \eqref{Riemcovectorvectoraction}. The cotangent bundle thus continues to have a trivial structure,
\be
T^*[\text{Riem}(D, \gamma)] = \text{Riem}(D,\gamma) \times \text{Sym}(D, \tensortype{2}{0})
\ee
which equals the (pre)phase space $\ca P$ in \eqref{prephasespacetrivial}.
The symplectic structure will be described next.

\subsection{Reduced phase space}
\label{subsec:redphasespacedef}

The {\sl reduced phase space}, $\widetilde{\ca P}$, is the space of physically distinct classical states, defined by the quotient of the constraint surface $\ca S$, within the pre-phase space $\ca P$, under the gauge transformations (or, in other words, the equivalence classes of gauge orbits).
The reduction process will generally go through the following steps:
\begin{enumerate}
\item From the Lagrangian $L$, compute the conjugate momenta, $p = \partial L/\partial \dot q$, and see whether there are constraints among the phase space variables, $C^1(q, p) = 0$. The superscript ``1'' is because these constraints coming directly from the Lagrangian, without imposing the equations of motion, are called {\sl primary} constraints (not to be confused with ``first-class'' --- footnote \ref{firstsecondclass}). 
\item Compute the Hamiltonian $H$ and find all {\sl secondary} constraints $C^2$ coming from the equations of motion, i.e., following from imposing that the primary constraints must be respected by time evolution, $dC^1/dt = \{C^1, H\} \approx 0$, where the approximate sign indicates that the equality holds when all constraints are satisfied (note that the Poisson brackets are evaluate without imposing any constraints). The same process must be repeated for the secondary constraints, and tertiary constraints and so forth until no additional constraints are found. Once this process ends we can forget about this classification into primary, secondary, etc, and simply treat all the constraints on the same footing. 
\item The set of all constraints determines the surface $\ca S$, so we compute the restriction of the original (pre)symplectic form $\Omega$ to $\ca S$, $\omega := \Omega|_{\ca S}$. If $\omega$ is non-degenerate, $\ca S$ is already the reduced phase space (this is the case if all constraints are second-class). If $\omega$ is  degenerate, we must identify its null directions and the corresponding gauge orbits.
\item To find the reduced phase space, we mention two convenient ways to quotient by the gauge orbits that may be useful:
\begin{itemize}
\item The first way is to ``gauge-fix'', which means introducing additional constraints in such a way that all constraints become second-class. In geometrical terms, we would look for a submanifold $\ca S'$ of $\ca S$ with the property that it intersects with each gauge orbit at precisely one point. In this case, the reduced phase space $\widetilde{\ca P}$ can be identified with $\ca S'$, and the symplectic form $\widetilde\omega$ is simply the restriction of the pre-symplectic form $\omega$ on $\ca S$ to $\ca S'$. A gauge-fixing approach is not always possible, something known as the Gribov phenomenon. As an example, consider the case where the gauge orbits are isomorphic to a group, $G$, in such a way that $\ca S$ is a principal $G$-bundle over the reduced phase space $\widetilde{\ca P}$; in this case, the existence of a ``gauge-fixing condition'' corresponds to say that there exists a global cross section $\ca S'$ of $\ca S$, which is only possible if the bundle is trivial, i.e., $\ca S \sim \widetilde{\ca P} \times G$. 
\item The second way is to ``change coordinates'' in a suitable manner. If it is possible to cover $\ca S$ with a coordinate system, there may be classes of coordinates such that the pre-symplectic form becomes simpler, and the null direction become evident. For example, consider $\ca S \sim \bb R^4$, covered with coordinates $\{x_1, x_2, p_1, p_2\}$ and a pre-symplectic form $\omega = \delta(p_1 + p_2) \wedge \delta(x_1 + x_2)$; this symplectic form suggests a change of coordinates to $x_\pm = x_1 \pm x_2$, $p_\pm = p_1 \pm p_2$, so that $\omega = \delta p_+ \wedge \delta x_+$, revealing that the appropriate quotient map is $J(x_1, x_2, p_1, p_2) = (x' := x_+, p' := p_+)$, where $\{x', p'\}$ are coordinates on $\widetilde{\ca P} \sim \bb R^2$. This approach may also not be always possible, as $\ca S$ may not admit a global coordinate system.
\end{itemize}
Note that both ways can be used in conjunction, each one removing part of the gauge ambiguities. Still these two procedures may not be enough, as there may remain some residual gauge ambiguities that have to be addressed in a specialized manner. When treating the causal diamond, we will employ the first way to deal with the ambiguities associated with the time diffeomorphisms, the second way to deal with the spatial diffeomorphisms, and there will remain a small (finite) number of ambiguities that will need to be removed in a third way.
\end{enumerate}

Since the reduced phase space is a ``standard'' phase space, in the sense that it has a (non-degenerate) symplectic structure, it is amenable to the application of canonical quantization. In particular, there are no subtleties associated with gauge ambiguities when it comes to the quantization, for all the gauge has already been eliminated at the classical level. The fact that only the physical, gauge-invariant degrees of freedom are undergoing quantization is a highly appealing feature of this approach, giving some confidence that the class of quantum theories obtained are ``natural''. Nevertheless, this approach does not come without drawbacks. One of them is that the reduced phase space usually has a non-trivial topology, or at least does not have a natural vector space structure, which requires more sophisticated schemes of canonical quantization, such as geometric quantization or group-theoretic quantization. Also, certain observables like the time-evolution Hamiltonian may become complicated when written in terms of the ``canonical observables'' (i.e., the ``$q$'s and $p$'s'') compatible with the phase space topology, and this may lead to severe operator-ordering ambiguities. Since there is no general proof that the Dirac approach is equivalent to the the reduced phase space approach (in fact, there are examples where they are known to disagree --- for a study in the case of general relativity see, e.g., \cite{ashtekar1982canonical}), it is worth to explore both approaches, and in this work we focus on the latter.

\subsection{An outline of the reduction process}
\label{subsec:outline}

Here we present a brief outline of the gauge reduction process for the diamond, based on the gauge-fixing of time by CMC slices,  following a general program of quantization via phase space reduction introduced by Moncrief et al \cite{moncrief1989reduction,fischer1997hamiltonian,fischer2001reduced}. (See also a more recent paper reviving this program~\cite{witten2022note}.)
The details will be explained in the main sections of the paper. 

Let us first discuss the constraints on the phase space. For this class of diamond-shaped spacetimes, it is natural to set up the ADM decomposition with respect to a family of surfaces anchored at the $S^1$ boundary. In fact, these are the only Cauchy surfaces in a diamond. This corresponds to restricting to lapse functions that vanish at the boundary, $N|_{\partial\Sigma} = 0$, and shift vectors that are tangent to the boundary, $N^a|_{\partial\Sigma} \in T(\partial\Sigma) \subset T(\Sigma)$. The action functional can be defined between any two such surfaces as
\be\label{GRaction}
S = \frac{1}{16 \pi G} \int d^3\!x\, \sqrt{-g} (\ca R - 2\Lambda)
\ee
up to boundary terms. (Notice that $\ca R$ stands for the Ricci scalar associated with the spacetime metric g, while the symbol $R$ will be reserved for spatial metrics.) 
We wish to define the system  in such a way that the spacetime solution (within the diamond) can be fully determined from initial data on a slice; we take the (pre)symplectic form $\Omega$ to be the conventional one from the ADM formalism\footnote{In field theories, the bulk part of the symplectic form is uniquely determined from the (bulk part of the) Lagrangian, but there are often ambiguities in choosing its boundary term \cite{speranza2018local,speranza2022ambiguity,jacobson1994black}. In some settings, such as in dynamically-closed theories defined in a spacetime ``cylinder'' with appropriate conditions on the timelike boundary, one can (almost) determine the boundary term for the symplectic form by insisting that the action principle is well-posed (in the sense of admitting stationary points when the configuration variables are fixed at the initial and final Cauchy slices) \cite{harlow2020covariant}. As an open-system, it is unclear how to fix these ambiguities for a causal diamond without further structure; in particular, it seems that one needs to specify how the causal diamonds are to be embedded as subsystems of a larger spacetime, or describe condition for the symplectic flux across the horizons \cite{chandrasekaran2021anomalies}. (See, however, \cite{kirklin2019unambiguous}.) Here we take the simplest choice, where the symplectic form is just the integral over the Cauchy slice of a local symplectic current, which is consistent with a self-contained causal diamond.}
\be\label{GRsym}
\Omega =  \int d^2\!x\, \delta \pi^{ab} \wedge \delta h_{ab} 
\ee
where $\pi^{ab}$ is the momentum conjugate to the spatial metric $h_{ab}$ defined by
\be\label{GRpi}
\pi^{ab} = \sqrt{h} \left(K^{ab} - K h^{ab}\right)
\ee
with $K = K^{ab}h_{ab}$ being the trace of the extrinsic curvature. Note that ``$\delta$'' denotes the exterior derivative in phase space.

The ADM Hamiltonian takes the pure-constraint form
\be\label{GRhamiltonian}
H = \frac{1}{16 \pi G} \int_{\Sigma} d^2\!x  \left[ N \sqrt{h}\left(K^{ab}K_{ab} - K^2 -  R + 2\Lambda\right) - 2 N_b \nabla_a \pi^{ab} \right]
\ee
where $R$ and $\nabla$ are respectively the Ricci scalar and covariant derivative on $\Sigma$ associated with $h_{ab}$. The derivative of a density is defined by converting it to a tensor, applying the derivative and then converting back to a density of the same weight; that is, here we have ${\nabla}_a \pi^{ab} := h^{1/2} {\nabla}_a (h^{-1/2} \pi^{ab})$.
Note that $K^{ab}$ is an implicit function of $h_{ab}$ and $\pi^{ab}$ obtained by inverting \eqref{GRpi}; in 2+1 dimensions it is given by $K^{ab} = h^{-1/2} (\pi^{ab} - \pi h^{ab})$, where $\pi := \pi^{ab} h_{ab}$ is the trace of $\pi^{ab}$. The variation with respect to $N$ yields the {\sl Hamiltonian constraint}
\be
K^{ab}K_{ab} - K^2 -  R + 2\Lambda = 0
\ee
and the variation with respect to $N^a$ yields the {\sl momentum constraint}
\be
{\nabla}_a  (K^{ab} - K h^{ab} ) = 0
\ee
While the momentum constraint is linear in the momentum variables, similarly to electromagnetism, the Hamiltonian constraint is non-linear. This non-linearity is partially responsible for the relative difficulty in dealing with the constraints of gravity.

An interesting method for solving the constraints of general relativity is the {\sl Lichnerowicz method}~\cite{lichnerowicz1944integration,lichnerowicz1952equations,choquet2008general,choquet2004einstein,moncrief1989reduction}. The goal is to convert the Hamiltonian constraint, which is a non-linear differential equation involving tensor fields $K^{ab}$ and $h_{ab}$, into a differential equation for a single scalar field $\phi$, which is accomplished by a suitable Weyl transformation. 
Let us introduce the {\sl traceless part of the extrinsic curvature}, $\sigma^{ab}$, which in 2+1 dimension is given by
\be\label{sigmaKdef}
\sigma^{ab} := K^{ab} - \frac{1}{2} K h^{ab}
\ee
The constraints then become
\begin{align}
&\sigma^{ab}\sigma_{ab}  -  R + 2\Lambda - \frac{1}{2} K^2 = 0 \label{sigmaH}\\
&{\nabla}_a  \left(\sigma^{ab} - \frac{1}{2} K h^{ab} \right) = 0 \label{sigmaM}
\end{align}
Now suppose that we have a set of {\sl input data}, $(h_{ab}, \sigma^{ab}, K)$, that satisfies the momentum constraint but not necessarily the Hamiltonian constraint. The idea is to deform this data by a suitable pointwise conformal transformation (i.e., a Weyl transformation), preserving the momentum constraint, until the Hamiltonian constraint is satisfied. Consider the transformed data $(\widetilde h_{ab}, \widetilde\sigma^{ab}, \widetilde K)$ defined by
\begin{align}
& \widetilde h_{ab} = e^\phi h_{ab} \nonumber\\
& \widetilde \sigma^{ab} = f (\phi)\sigma^{ab} \nonumber\\
& \widetilde K = w (\phi ) K 
\end{align}
where $\phi : \Sigma \rightarrow \bb R$ is a scalar function on $\Sigma$ and $f : \bb R \rightarrow \bb R$ and $w : \bb R \rightarrow \bb R$ are real functions to be specified. (Note that $f(\phi)$ and $w(\phi)$ are to be understood as $f \circ \phi$ and $w \circ \phi$.) If we assume that the metric $h_{ab}$ in the input data satisfies the boundary condition, we can impose the Dirichlet boundary condition on $\phi$,
\be\label{phiDBC}
\phi \big|_{\partial\Sigma} = 0
\ee
so that the deformed data $\wt h_{ab}$ also satisfies the same boundary condition for the induced metric. This will ensure that the physical initial data generated by this method (i.e., the data that satisfies both the momentum and the Hamiltonian constraints) will also satisfy the boundary conditions. The momentum constraint for the deformed data becomes,
\begin{align}
\widetilde{\nabla}_a &  \left(\widetilde\sigma^{ab} - \frac{1}{2}  \widetilde K \widetilde h^{ab} \right) = \nonumber\\
& = \frac{1}{2} \left( f(\phi) + w(\phi) e^{-\phi} \right) {\nabla}^b K + \left( f'(\phi) + 2 f(\phi) \right) \sigma^{ab} {\nabla}_a \phi + \frac{1}{2} w'(\phi) e^{-\phi} K {\nabla}^b \phi = 0 \label{deformedMC}
\end{align}
where it was used that $(h_{ab}, \sigma^{ab}, K)$ satisfies the momentum constraint, i.e., ${\nabla}_a \sigma^{ab} = \frac{1}{2} {\nabla}^b K$. Here $\widetilde{\nabla}$ is the covariant derivative associated with $\widetilde h_{ab}$. The Hamiltonian constraint becomes,
\begin{align}
\widetilde\sigma^{ab}\widetilde\sigma_{ab}  - \widetilde{ R} + 2\Lambda - \frac{1}{2} {\widetilde K}^2 &=  \nonumber\\
= e^{-\phi}  \nabla^2 \phi +  & e^{2\phi}f(\phi)^2 \sigma^{ab}\sigma_{ab}  - e^{-\phi}  R + 2\Lambda - \frac{1}{2}w(\phi)^2 K^2 = 0 \label{deformedHC}
\end{align}
In the left hand side, $\widetilde\sigma^{ab}\widetilde\sigma_{ab} = \widetilde h_{ac}\widetilde h_{bd} \widetilde\sigma^{ab}\widetilde\sigma^{cd}$; in the right hand side, $ \sigma^{ab} \sigma_{ab} =   h_{ac}  h_{bd}  \sigma^{ab} \sigma^{cd}$ and $ \nabla^2 = h^{ab} \nabla_a \nabla_b$.

Given functions $f$ and $w$, the constraints have thus become a pair of coupled differential equations for the scalar $\phi$. For this method to be useful, we would like to be able to choose $f$ and $w$ in such a way that the equations become as simple as possible and, most importantly, that the equations admit solutions for a large class of input data $(h_{ab}, \sigma^{ab}, K)$. The most useful choice comes up in conjunction with a gauge-fixing for the time: we consider the {\sl constant mean curvature} gauge (abbreviated as ``CMC''), in which the spatial slices are taken to have a constant trace of extrinsic curvature,
\be
K \big|_\Sigma = - \tau
\ee
where $\tau$ is a constant parameter on $\Sigma$. In the case of the diamond, we will establish that by varying $\tau$ from $-\infty$ to $\infty$ the spacetime will be foliated with Cauchy slices. Note that the ``initial time'', $\tau = 0$, corresponds to the maximal slice (i.e., the slice with maximal area). In this gauge, the first term of the momentum constraint \eqref{deformedMC} vanishes, suggesting the following convenient choices for the functions $f$ and $w$,
\begin{align}
f'(\phi) + 2f(\phi) = 0 \quad &\rightarrow \quad f(\phi) = e^{-2\phi} \nonumber\\
w'(\phi) = 0 \quad &\rightarrow \quad w(\phi) = 1
\end{align}
With those choices the momentum constraint is automatically satisfied for any $\phi$, provided that the input data satisfies
\be\label{sigmacons}
{\nabla}_a  \sigma^{ab} = 0
\ee
Also, note that $\phi$ is determined as the solution of a single differential equation coming from the Hamiltonian constraint \eqref{deformedHC},
\be\label{LichEq}
\nabla^2 \phi -   R +  e^{-\phi} \sigma^{ab}\sigma_{ab} - e^\phi \chi = 0
\ee
where
\be\label{chidef}
\chi = -2\Lambda + \frac{1}{2}\tau^2
\ee
is a time-dependent parameter. Equation \eqref{LichEq} for $\phi$ is called the {\sl Lichnerowicz equation}. Note that $\Lambda \le 0$ implies that $\chi \ge 0$ for all times, and this will be used for proving existence and uniqueness of solutions $\phi$ to the Lichnerowicz equation (for any Dirichlet boundary condition, such as in \eqref{phiDBC}). This means that for any input data $(h_{ab}, \sigma^{ab}, -\tau)$, in the CMC gauge, we can always generate a unique set of valid initial data given by $(e^\phi h_{ab}, e^{-2\phi} \sigma^{ab}, -\tau)$. 

The Lichnerowicz method thus yields the following characterization for the constrained phase space, $\ca S$. As mentioned above, a set of input data $\scr S = (h_{ab}, \sigma^{ab}, -\tau)$ corresponds to one and only one set of valid initial data, $(e^\phi h_{ab}, e^{-2\phi} \sigma^{ab}, -\tau)$, where $\phi$ is the unique solution of $\eqref{LichEq}$. Now given any scalar field $\lambda : \Sigma \rightarrow \bb R$ vanishing at the boundary, $\lambda|_{\partial\Sigma} = 0$, consider a deformed set of input data $\scr S' = (e^\lambda h_{ab}, e^{-2\lambda} \sigma^{ab}, -\tau)$. The Lichnerowicz problem for $\scr S'$ leads to a unique valid data $(e^{\phi'} e^\lambda h_{ab}, e^{-2\phi'} e^{-2\lambda}\sigma^{ab}, -\tau)$, where $\phi'$ is the unique solution of 
\be\label{Lichtransdata}
{\nabla}^{'2} \phi' -  { R}' +  e^{-\phi'} (e^{-2\lambda}\sigma^{ab})(e^{-2\lambda}\sigma^{cd}) (e^\lambda h_{ac}) (e^\lambda h_{bd}) - e^{\phi'} \chi = 0
\ee
with boundary condition $\phi'|_{\partial\Sigma} = 0$. Here $\nabla'$ and ${ R}'$ are associated with $h'_{ab} := e^\lambda h_{ab}$. This equation can be rewritten in terms of $\nabla^2$ and $ R$ as,
\be
e^{-\lambda} \nabla^2 \phi' - e^{-\lambda} \left(  R - \nabla^2 \lambda \right) + e^{-\phi'} e^{-2\lambda}\sigma^{ab}\sigma_{ab} - e^{\phi'} \chi = 0
\ee
or equivalently,
\be
\nabla^2 (\phi' + \lambda) -  R + e^{-(\phi' + \lambda)} \sigma^{ab}\sigma_{ab} - e^{(\phi' + \lambda)} \chi = 0
\ee
which has the same form as \eqref{LichEq}, with the same boundary condition, but in the variable $\phi' + \lambda$. Therefore $\phi' + \lambda = \phi$ is the unique solution, implying that the initial data obtained from $\scr S'$ is the same one obtained from $\scr S$, i.e., $(e^{\phi'} e^\lambda h_{ab}, e^{-2\phi'} e^{-2\lambda}\sigma^{ab}, -\tau) = (e^\phi h_{ab}, e^{-2\phi} \sigma^{ab}, -\tau)$. Consequently, each Weyl-related equivalence class of input data,
\be
(h_{ab}, \sigma^{ab}, -\tau) \sim (e^\lambda h_{ab}, e^{-2\lambda} \sigma^{ab}, -\tau) \,,\quad \lambda \big|_{\partial\Sigma} = 0
\ee
corresponds to a unique set of valid data $(e^\phi h_{ab}, e^{-2\phi} \sigma^{ab}, -\tau)$. In other words, the constraint surface $\ca S$ can be identified with this space of equivalence classes
\be\label{CSequiv}
\ca S = [(h_{ab}, \sigma^{ab}) \sim (e^\lambda h_{ab}, e^{-2\lambda} \sigma^{ab})]
\ee
where $\tau$ has been omitted since it does not transform under this Weyl transformation. 

The constraint surface $\ca S$ contains null directions, corresponding to the gauge transformations associated with spatial diffeomorphisms on the CMC slices $\Sigma$. Let $\Psi : \Sigma \rightarrow \Sigma$ be a {\sl boundary-trivial} diffeomorphism of $\Sigma$, i.e., a diffeomorphism that acts trivially on the boundary,
\be
\Psi \big|_{\partial\Sigma} = \id
\ee
where $\id$ is the identity map. It is clear that if we apply such a diffeomorphism to a set of initial data we will produce a physically equivalent set of data.\footnote{Note that diffeomorphisms which are not boundary-trivial are not allowed for they generally do not preserve the Dirichlet boundary condition for the induced metric. The only exception is the $SO(2)$ family of isometries of the boundary, but those are true symmetries of the system, not gauge transformations.} In this way, the reduced phase space $\wt {\ca P}$ can be identified with the quotient of $\ca S$ under those boundary-trivial spatial diffeomorphisms. Using the above characterization for the constraint surface, \eqref{CSequiv}, we can identify the reduced phase space as
\be\label{RPSequiv}
\wt{\ca P} = \{(h_{ab}, \sigma^{ab}) \sim (\Psi_* e^\lambda h_{ab}, \Psi_* e^{-2\lambda} \sigma^{ab})\}
\ee
Note that the transformation of the metric, $h_{ab} \mapsto \Psi_* e^\lambda h_{ab}$, is a general boundary-trivial {\sl conformal transformation}, i.e., a combination of a Weyl transformation and a diffeomorphism, $(\Psi, \lambda)$, satisfying the boundary condition $(\Psi, \lambda)|_{\partial\Sigma} = (\id, 0)$.\footnote{The nomenclature here may differ from other references, which sometimes restrict the term ``conformal tranformation'' to those that satisfy $h_{ab} = \Psi_* e^\lambda h_{ab}$; instead we refer to those very special conformal transformations as {\sl conformal isometries}.} We are going to review how the space of equivalence classes in \eqref{RPSequiv} can be identified with the cotangent bundle of the space of conformal geometries on $\Sigma$. The space of conformal geometries on $\Sigma$, denoted by $\text{ConGeo}(\Sigma)$, is the space of equivalence classes  of Riemannian metrics quotiented by boundary-trivial conformal transformations,
\be
\text{ConGeo}(\Sigma) = \{ h_{ab} \sim \Psi_* e^\lambda h_{ab} \}
\ee
Thus the phase space can be identified as
\be
\wt{\ca P} = T^*[\text{ConGeo}(\Sigma)]
\ee
This result quite general and comes up whenever the system admits a CMC gauge and the Lichnerowicz method is ``nicely'' posed (i.e., there are existence and uniqueness theorems for the associated Lichnerowicz equation). In the case of the diamond, we will see that the reduced phase space is given by
\be\label{RPStop}
\wt{\ca P} = T^*[\diff/\psl]
\ee
where $\diff$ is the group of orientation-preserving diffeomorphisms of the boundary $\partial\Sigma \sim S^1$ and $\psl$ is the projective special linear group in 2 real dimensions (which is a 3-dimensional closed subgroup of $\diff$, in a way that will be described later).

We will also discuss another approach for reducing the phase space, in Sec. \ref{sec:Rcc}. The idea is to ``change coordinates'' from $(h_{ab}, \sigma^{ab})$ to a new set of variables $(\Psi, \Phi, \sigma^{ab})$, where $\Psi$ is a diffeomorphism of $\Sigma \sim D$ and $\Phi : \Sigma \rightarrow \bb R$ is a scalar function. This change of coordinates is defined by taking a standard metric $\bar h_{ab}$, such as the Euclidean metric on $\Sigma$ (i.e., the metric of a flat disc, $dr^2 + r^2 d\theta^2$ with unit radius), and considering the conformal transformation $(\Psi, \Phi)$ from $\bar h_{ab}$ into $h_{ab} = \Psi_*\Phi \bar h_{ab}$. This is allowed because of the uniformization theorem, which ensures that any two metrics on a topological disc can be related by a conformal transformation. The contraint surface $\ca S$ can then be covered with ``coordinates'' $(\Psi, \bar\sigma^{ab})$, where $\bar \sigma^{ab}$ satisfies the divergenceless condition $ \nabla_a \bar\sigma^{ab} = 0$ (where here $\nabla$ is the derivative associated with $\bar h_{ab}$). The symplectic form becomes evidently degenerate with respect to ``bulk diffeomorphisms'', suggesting a reduction to a (not fully) reduced phase space that can be covered with ``coordinates'' $(\psi, \bar\sigma^{ab})$, where $\psi := \Psi |_{\partial\Sigma}$ is the boundary action of $\Psi$. This space has the topology $\diff \times \text{\sl ``functions on $S^1$ not containing the Fourier modes $1$, $\sin\theta$ and $\cos\theta$''}$. By a simple analysis, we can determine that there are still three degenerate directions to be removed, corresponding to the $\psl$ subgroup of $\diff$, ultimately leading to the (fully) reduced phase space \eqref{RPStop}. Not only this alternative approach serves as a check of the earlier result, but it is also very useful as it provides an explicit ``change of coordinates'' from the ``natural'' variables describing the reduced phase space to the more easily interpretable geometric quantities like spatial metric and extrinsic curvature.

Before proceeding with more technical developments, it is interesting to understand what the classical states described in \eqref{RPStop} actually represent. First note that in three spacetime dimensions there are no ``local'' gravitational degrees of freedom (i.e., there are no gravitational waves). This is because the Weyl tensor vanishes identically, meaning that the curvature is completely determined by the Ricci tensor,
\be
\ca R_{abcd} = 2 \left( g_{a[c} \ca R_{d]b} - g_{b[c} \ca R_{d]a} \right) - \ca R g_{a[c} g_{d]b}
\ee
and the Ricci tensor is fixed by Einstein's equation. Therefore the metric can be ``locally'' determined using normal coordinates,\footnote{The {\sl Riemann normal coordinates} are defined with respect to the exponential map (associated with the metric) in the following way. Given a manifold $\ca M$ with metric $g_{ab}$, let $U \subset \ca M$ be a neighborhood of $x_0 \in \ca M$ such that the exponential map (based at $x_0$), $\exp_{x_0} : T_{x_0}\ca M \rightarrow \ca M$, is an isomorphism between an open neighborhood $V$ of $T_{x_0}\ca M$ and $U$. This means that for any $x \in U$ there exists a unique $v \in V \subset T_{x_0}\ca M$ such that $x = \exp_{x_0}(v)$. Given a basis $e_\mu$ of $T_{x_0}\ca M$ we can decompose $v = x^\mu e_\mu$. The normal coordinates on $U$, with respect to $x_0$ and $e_\mu$, is defined by assigning coordinates $x^\mu$ to $x$. The metric $g_{ab}$ on $U$ (or a open subset of $U$) can be reconstructed from the value of the curvature (and all its derivatives) at $x_0$, and it is given (to first order) by $g_{\mu\nu}(x) = \eta_{\mu\nu} - \frac{1}{3} R_{\mu\alpha\nu\beta}(0) x^\alpha x^\beta + \cdots$, where $e_\mu$ is taken to be orthogonal.} leaving no physical (gauge-invariant) degree of freedom left. If there is no matter present, Einstein's equation imply that $\ca R_{ab} = 2\Lambda g_{ab}$, so
\be\label{caRAdS3}
\ca R_{abcd} =  2 \Lambda g_{a[c} g_{d]b}
\ee
implying that the metric is maximally symmetric. In the case $\Lambda < 0$ this means that the spacetime is locally Anti-de Sitter ($\ads$), and in the case $\Lambda = 0$ the spacetime is Minkowski ($\mink$). Thus our diamond is ``locally $\ads$'' (resp. ``locally $\mink$), meaning that the neighborhood of any point in the bulk can be isometrically embedded into global $\ads$ (resp. $\mink$) spacetime. In fact, since we are considering trivial topology for the Cauchy slices, the entire diamond can be isometrically embedded as a region of $\ads$ (resp. $\mink$) spacetime. Consequently, the only degrees of freedom are associated with the global shape of the diamond. In other words, the space \eqref{RPStop} must be describing diamond-shaped regions of AdS spacetime, with a fixed boundary length $\ell$. It is reasonable to ask whether the phase space can be identified with a special subset (or classes of equivalence) of embedded diamonds in $\ads$. However, the formal analysis of the reduced phase space paired with the explicit embedding construction (App. \ref{app:embed}) there is no natural one-to-one correspondence between points in the phase space with any special subset of embeddings. We will explain this point in more detail in App. \ref{app:embed}.
\begin{figure}
\centering
\includegraphics[scale = 0.6]{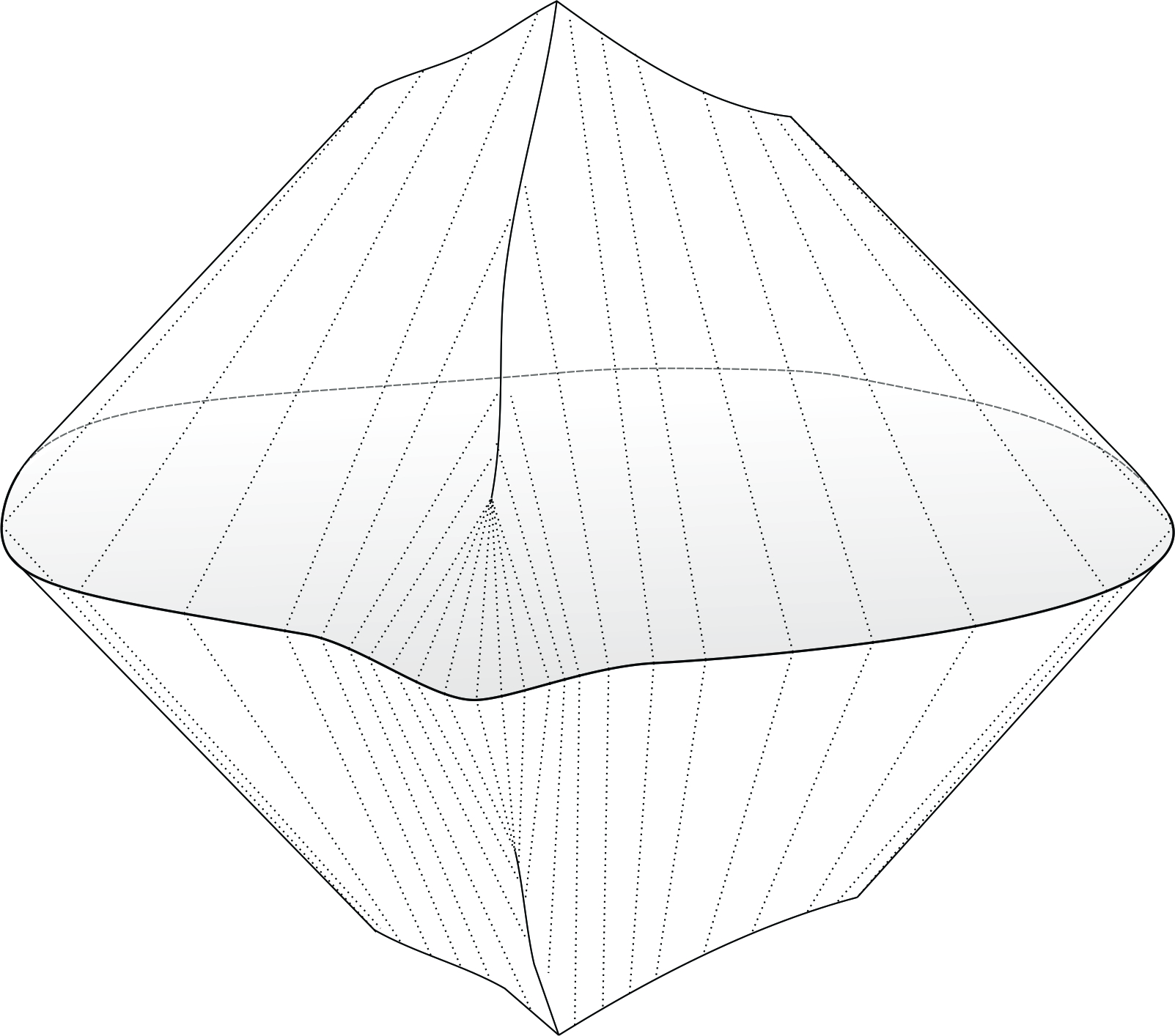}
\caption{The phase space consists of causal diamonds in $\ads$ (or $\mink$ if $\Lambda = 0$) with topologically trivial Cauchy slices (discs) whose corner loops have fixed length $\ell$.}
\label{fig:diashape}
\end{figure}
Finally, note that any such a diamond-shaped region in $\ads$ can be fully specified by giving how the $S^1$ boundary, with length $\ell$, embeds into $\ads$. More precisely, let $C \subset \ads$ be a (spacelike, achronal) loop in $\ads$, with length $\ell$, satisfying the condition that it is the boundary of a spacelike topological disc $D$, then the domain of dependence of $D$ determines a unique diamond; moreover, observe that the diamond so defined depends only on $C$, but not on $D$.

\section{Constant mean curvature foliation}
\label{sec:CMC}

In this section we show that the class of spacetimes consisting of causal diamonds, with nonpositive cosmological constant, admits a nice foliation by surfaces of constant mean curvature. Some pertinent references are \cite{bartnik1988regularity,bartnik1988remarks, brill1976isolated, gerhardt1983h}. Moreover, the Lichnerowicz problem set up with respect to such a foliation is well-posed in the sense that the solutions to the Lichnerowicz equation exist and are unique given the boundary conditions.

\subsection{The foliation is nicely behaved}
\label{subsec:nicefoliation}

One fundamental step in the reduction process is the preliminary gauge-fixing of time by the choice of a constant-mean-curvature (CMC) foliation of the spacetime. It is therefore essential that we can guarantee the a priori existence and regularity of such a foliation for all possible initial data, that is, all causal diamonds with fixed boundary metric. For motivation, we shall begin this section with a very simple argument based on Raychaudhuri's equation establishing some nice properties of CMC slices in our class of spacetimes, such as the fact that there can exist at most one CMC slice with a given $\tau = -\kappa$, and if two CMC slices exist such that $\tau_2 > \tau_1$ then CMC${}_2$ is entirely to the future of CMC${}_1$. In the next subsection, we cite a general theorem ensuring the existence and regularity of a foliation by CMCs. Finally we prove that a slice approaching the future horizon of the diamond has arbitrarily negative supremum of mean curvature ($\sup \kappa \rightarrow -\infty$) while a slice approaching the past horizon has arbitrarily positive infimum of mean curvature ($\inf \kappa \rightarrow \infty$), which implies that the foliation covers the whole causal diamond as $\tau$ ranges from $-\infty$ to $\infty$.

The Raychaudhuri's equation governs how a congruence of geodesic expands, twists and shears. If $\xi$ the unit vector tangent to a congruence of timelike geodesics in a (1+d dimensional) spacetime with metric $g_{ab}$, and $h_{ab} := g_{ab} + \xi_a \xi_b$ is the ``spatial metric'' (i.e., $h^a_{\,\,b}$ is the projector onto the subspace orthogonal to $\xi$), we define the following parameters associated to the congruence: expansion $\theta := \nabla_a \xi^a$, shear $\sigma_{ab} := \nabla_{(a}\xi_{b)} - \frac{1}{d} \theta h_{ab}$ and twist $\omega_{ab} := \nabla_{[a}\xi_{b]}$. The equation describing how the geodesics expand in time is
\be
\frac{d \theta}{ds} = \xi^a \nabla_a \theta = - \frac{1}{d}\theta^2 - {\ca R}_{ab} \xi^a \xi^b - \sigma_{ab} \sigma^{ab} + \omega_{ab} \omega^{ab}
\ee
where $s$ is the proper length along the geodesics and ${\ca R}_{ab}$ is the Ricci curvature associated with $g_{ab}$. The equation for the twist is
\be
\xi^c \nabla_c \omega_{ab} = - \frac{2}{d}\theta \omega_{ab} + 2 \sigma^c_{\,\,[a} \omega_{b] c}
\ee
and we can see that if $\omega_{ab}= 0$ at one point of a geodesic then it will remain zero along that geodesic. Frobenius theorem says that  the congruence is (locally) hypersurface orthogonal if and only if 
 $\omega_{ab} = 0$; hence, if the congruence is defined by shooting geodesics orthogonally from a codimension-1 surface, then $\omega_{ab} = 0$. 

Let $\Sigma_1$ and $\Sigma_2$ be two compact, acausal surfaces, sharing the same boundary, with constant mean curvatures $\kappa_1$ and $\kappa_2$, respectively. 
Take $\Gamma = \Sigma_1 \cap \Sigma_2$ as the set of points where the two surfaces intersect each other; $\Gamma$ will divide $\Sigma_1$ and $\Sigma_2$ into patches, $\Sigma_1^i$ and $\Sigma_2^i$, such that the spacetime region between each  $\Sigma_1^i$ and $\Sigma_2^i$ is ``lens shaped''. More precisely, let us define $\Sigma_1^i$ and $\Sigma_2^i$ to be coverings of $\Sigma_1$ and $\Sigma_2$, respectively, by compact connected regions satisfying the following properties:

$(i)$ Either $\Sigma_1^i \cap \Gamma = \partial\Sigma_1^i = \partial\Sigma_2^i  = \Sigma_2^i \cap \Gamma$ or $\Sigma_1^i = \Sigma_2^i \subset \Gamma$;

$(ii)$ If $\Sigma_1^i \ne \Sigma_2^i$, then $\text{Int}(\Sigma_1^i)$ is either entirely to the future or entirely to the past of $\text{Int}(\Sigma_2^i)$.

Now the argument can be made for each $i$. 
Let us first consider the case where $\kappa_1 \ne \kappa_2$, so $\Sigma_1^i \ne \Sigma_2^i$. Suppose that $\Sigma_2^i$ is to the future of $\Sigma_1^i$ and consider the set of all timelike geodesics from $\Sigma_1^i$ to $\Sigma_2^i$; let $\gamma$ be one geodesic with maximum length. Evidently the length of $\gamma$ is non-zero. Moreover $\gamma$ would be orthogonal to both $\Sigma_1^i$ and $\Sigma_2^i$. Now consider a congruence of geodesics around $\gamma$ starting orthogonal to $\Sigma_1^i$. Since this congruence starts non-twisting ($\omega_{ab} = 0$) it would remain hypersurface orthogonal; to a first-order approximation, $\Sigma_2^i$ would be orthogonal to the congruence at the point where it intersects with $\gamma$. Therefore, at the point where $\Sigma_1^i$ intersects with $\gamma$ we have $\theta = \kappa_1$ and at $\Sigma_2^i$ we have $\theta = \kappa_2$; also, $\sigma_{ab}$ would correspond to the traceless part of the extrinsic curvature of $\Sigma_1^i$ and $\Sigma_2^i$ at the respective points. We can then integrate the Raychaudhuri's equation from $\Sigma_1^i$ to $\Sigma_2^i$ to get
\be
\kappa_2 - \kappa_1 = \int\!ds\,\frac{d \theta}{ds} = - \int\!ds \left( \frac{1}{2}\theta^2 + \sigma_{ab} \sigma^{ab} - 2\Lambda \right)
\ee
where we have particularized to $d=2$, used Einstein's equation ${\ca R}_{ab} = 2\Lambda g_{ab}$ and $\xi_a\xi^a = -1$.
Note that, for a non-positive cosmological constant, the left-hand side is non-positive. Thus $\kappa_2 \le \kappa_1$. That is, if a CMC is entirely to the future of another CMC, then the latter must not have a smaller mean curvature. We can easily strengthen the conclusion if we assume that $(i)$ $\kappa_1$ and $\kappa_2$ are not both zero or $(ii)$ the spacetime is negatively curved, $\Lambda < 0$. In case $(i)$ we see that $\theta^2$ must be non-vanishing on some portion of the maximal curve, and in case $(ii)$ we have $- 2\Lambda > 0$. In both cases the left-hand side of the equation is negative, implying that $\kappa_2 < \kappa_1$. In particular, this means that two CMCs sharing the same boundary, with different $\kappa$'s, satisfying one of these conditions must not intersect at points in their interior; and the CMC with the smallest $\kappa$ will be to the future of the other one. Note also that if the two $\kappa$'s are infinitesimally close to each other then the CMCs must be infinitesimally close to each other, as can be seen from the inequality
\be
\text{length}(\gamma) \le \frac{|\kappa_2 - \kappa_1|}{\text{min}_\gamma \left(\theta^2/2 + \sigma_{ab} \sigma^{ab} \right) - 2 \Lambda}
\ee
We have therefore seen that, if one of these conditions are satisfied, then the CMCs  (if they exist) would never intersect each other (except at their common boundaries), they would be temporally ordered (i.e., as $\kappa$ decreases the CMCs move to the future) and they must foliate a region of the space (i.e., there can be no gap between CMCs with infinitesimally close $\kappa$'s). 

Not covered in the previous argument is the case of maximal slices ($\kappa = 0$) in flat spacetime ($\Lambda = 0$) with zero extrinsic curvature. The previous argument cannot not rule out the possibility that there are more than one maximal slice, and perhaps with a gap between them (i.e., a region between two maximal slices devoid of any CMCs). 
This simplest manner to approach this case is by considering a continuity argument. Namely, 
one can see that the foliation varies smoothly (in a given background manifold) with respect to infinitesimal variation of the $\Lambda$ parameter; as the foliation is well-behaved for all $\Lambda < 0$, with no gap at $\kappa = 0$, which remains true in the limit $\Lambda \rightarrow 0^-$, implies that the foliation is also well-behaved at $\Lambda = 0$.

\subsubsection{Crushing singularity}
\label{subsubsec:Crush}

Consider a globally-hyperbolic connected spacetime, having a compact Cauchy slice $\Sigma$. All Cauchy slices are homeomorphic, so consider an arbitrary homeomorphism between any slice $\Sigma$ to a reference slice $\Sigma_0$, which allows us to compare points $x$ in different slices.
Let $\Sigma_-$ and $\Sigma_+$ be two (sufficiently regular) Cauchy surfaces (closed or with a common boundary) with mean curvatures $K_-(x)$ and $K_+(x)$, respectively, and suppose that $\Sigma^+$ is entirely to the future of $\Sigma^-$. Now, if $K_+(x) < K_-(x)$ then for any (continuous) function $K(x)$ such that $K_+(x) < K(x) < K_-(x)$,  there exists \cite{gerhardt1983h} a slice $\Sigma$, between $\Sigma_-$ and $\Sigma_+$, whose mean curvature is $K(x)$.

Combined with the previous arguments establishing the nice properties of CMC slices, it follows the CMC foliation exists and spans the whole spacetime if one can show that the future and past horizons are {\sl crushing singularities} \cite{gerhardt1983h,eardley1979time}. More precisely, the future horizon $\ca N^+$ (i.e., the boundary of the future domain of dependence of a Cauchy slice) is said to be a crushing singularity if there exists a family of surfaces $\Sigma^+_\lambda$ such that $\lim_{\lambda \rightarrow \infty} \Sigma^+_\lambda = \ca N^+$ and $\lim_{\lambda \rightarrow \infty} \text{sup}_{\Sigma^+_\lambda} K = -\infty$; and the past horizon $\ca N^-$ is similarly said to be a crushing singularity  if there exists a family of surfaces $\Sigma^-_\lambda$ such that $\lim_{\lambda \rightarrow \infty} \Sigma^-_\lambda = \ca N^-$ and $\lim_{\lambda \rightarrow \infty} \text{inf}_{\Sigma^-_\lambda} K = +\infty$. In words, this means that one could find a family of surfaces approaching the future horizon whose mean curvature is arbitrarily negative at every point, and similarly a family a surfaces approaching the past horizon whose mean curvature is arbitrarily positive everywhere. Then for any constant $K \in (-\infty,+\infty)$, there will exist a slice with $K(x) = K$; as shown before, these slices will be unique (for each $K$) and continuously ordered in time, $\tau = -K$, with no gap (i.e., an open region not sliced by any CMC), thereby defining a regular CMC foliation of the whole diamond.

Here we shall argue that for any causal diamond in our phase space the future horizon is a crushing singularity. A completely analogous analysis can be used to show that the past horizon is also a crushing singularity. The proof will not be fully rigorous, but we hope that it will be sufficiently convincing. The idea is to consider a surface $\ca S$ that is very close to a null surface $\ca N$ (whose null generators are geodesic), and in a suitable coordinate system adapted to $\ca N$, argue using a first order approximation (in the parameter describing the nearness of $\ca S$ to $\ca N$) that we can define $S$ with arbitrarily negative $K$. Note that $\ca S$ is not an entire Cauchy slice since $\ca N$ typically corresponds only to a portion of the future horizon, which is not a manifold because of the caustics. In fact, it appears that, in general, the future horizon can always be described by a finite number of null manifolds $\ca N_i$, emanating from the corner $\partial\Sigma$, and meeting at the graph-like caustics---see Fig. \ref{fig:blue1} for a representation of a typical shape of the horizon. We will consider a set of $\ca S_i$, near their respective $\ca N_i$, and join them smoothly in a neighborhood of the caustics, by ``rounding off'' their intersection.

First let us review the coordinate formula for the mean curvature of a surface. Suppose that in some open spacetime region, there are coordinates $t$ and $x^\alpha$ such that the surface $\ca S$ can be described as the zero-set of the function
\be
S := t - f(x)
\ee
where $f$ is some real function of the $x^\alpha$. Let $n$ be the unit vector normal to $S$, which implies that $n^a = c g^{ab} (dS)_b$ for some factor $c$. This factor can be determined by the normality condition; assuming that the surface is spacelike, $-1 = n^a n_a = c^2 g^{ab} (dS)_a (dS)_b$, which gives 
\be
n^a = \pm \frac{g^{ab} dS_b}{\sqrt{-dS_c dS^c}}
\ee
where the $\pm$ sign must be selected based on some choice of orientation. For a Cauchy slice in the diamond, we define $K_{ab} = h^c{}_a\boldsymbol\nabla_c n_b$ with an $n$ pointing to the future. The mean curvature is then given by
\be
K = \boldsymbol\nabla_a n^a = \pm \boldsymbol\nabla_a \left( \frac{g^{ab} dS_b}{\sqrt{-dS_c dS^c}} \right)
\ee
In coordinates, $x^\mu = (t, x^\alpha)$, this reads
\be\label{Kcoordformula}
K =  \pm \frac{1}{\sqrt{-g}} \partial_\mu \left(\sqrt{-g} \frac{g^{\mu\nu} \partial_\nu S}{\sqrt{-g^{\rho\sigma} \partial_\rho S \partial_\sigma S}} \right)
\ee
where $\sqrt{-g} := \sqrt{- \text{det}(g)}$. From the definition of $S$, $\partial_\mu S = (1, -\partial_\alpha f)$.

Now consider a null manifold $\ca N$ (co-dimension 1 in spacetime\footnote{In this section the spacetime is assumed to be of dimension greater than or equal to 3.}) whose null (future-pointing) generators are $u_+^a$. These generators are geodesic, i.e., 
\be
u^a_+ \bs\nabla_a u^b_+ = 0
\ee
We shall recall here the Gaussian null coordinates~\cite{friedrich1999rigidity}, characterizing a neighborhood of $\ca N$.
Let $\lambda^+$ denote the affine parameter along these geodesics, $d\lambda^+(u_+) = 1$.
Let $\ca C$ be a spatial manifold (co-dimension 2 in spacetime) embedded in $\ca N$, orthogonal to $k_+$, and let $x^\alpha$ be coordinates on it. Extend these coordinates $x^i$ to $\ca N$ by taking $x^i$ constant along the null generators,
\be
\pounds_{u_+}(x^i) = 0
\ee
and consider that $\lambda^+ = 0$ at $\ca C$.
These define a coordinate system $(\lambda^+; x^i)$ on $\ca N$, and we denote the vector fields tangent to $x^i$ by $\partial_i$. At every point of $\ca N$ define the null vector $u_-$ orthogonal to $\partial_i$, past-pointing, and satisfying the normalization condition
\be
(u_-)^a (u_+)_a = 1
\ee
Extend $u_-$ away from $\ca N$ by requiring that it is geodesic,
\be
u^a_- \bs\nabla_a u^b_- = 0
\ee
Let $\lambda^-$ be the corresponding affine coordinate, $d\lambda^-(u_-) = 1$, assumed to vanish at $\ca N$. 
Then extend the coordinates $(\lambda^+; x^i)$ away from $\ca N$ by taking them constant along $u_-$,
\be
\pounds_{u_-}(\lambda^+) = \pounds_{u_-}(x^i) = 0
\ee
These define coordinates $(\lambda_-, \lambda^+; x^i)$ in a spacetime neighborhood $\ca V$ of $\ca N$.

Let us investigate some properties of this coordinate system. First note that, at $\ca N$, $u_+$ is everywhere orthogonal to $\partial_i$. This follows from the fact that $u_+\cdot \partial_i := (u_+)_a (\partial_i)^a = 0$ at $\ca C$ and
\be
u^a_+ \bs\nabla_a \left( u_{+b} \partial_i^b \right) = u_{+b} u^a_+ \bs\nabla_a \partial_i^b = u_{+b} \partial_i^a \bs\nabla_a u_+^b = 0
\ee
where it was used that $u_+$ satisfies the geodesic equation,  the coordinate condition $\pounds_{u_+}\partial_i = 0$ and that $u_+$ is everywhere null in $\ca N$.
Second note that since $u_-$ is defined away from $\ca N$ by the geodesic condition, it is null everywhere in $\ca V$. Moreover, the inner product between $u_-$ and any other basis vector is constant within $\ca V$, as follows
\ba
&u^a_- \bs\nabla_a \left( u_{-b} u_+^b \right) = u_{-b} u^a_- \bs\nabla_a u_+^b = u_{-b} u_+^a \bs\nabla_a u_-^b = 0 \nonumber\\
&u^a_- \bs\nabla_a \left( u_{-b} \partial_i^b \right) = u_{-b} u^a_- \bs\nabla_a \partial_i^b = u_{-b} \partial_i^a \bs\nabla_a u_-^b = 0
\ea
where, in each line, we used (in order) the geodesic equation for $u_-$, the coordinate conditions, $\pounds_{u_-}u_+ = 0$ and $\pounds_{u_-}\partial_i = 0$, and the fact that $u_- \cdot u_- = 0$ in $\ca V$. Thus, $u_+ \cdot u_- = 1$ and $\partial_i \cdot u_- = 0$ in $\ca V$. Finally, note that the derivative of $u_+ \cdot u_+$ along $u_-$ vanishes at $\ca N$,
\be
\frac{1}{2}u^a_- \bs\nabla_a (u_{+b} u_+^b) = u_{+b} u^a_- \bs\nabla_a u_+^b = u_{+b} u^a_+ \bs\nabla_a u_-^b =  u^a_+ \bs\nabla_a (u_{+b} u_-^b) - u_{-b} u^a_+ \bs\nabla_a u_+^b = 0
\ee
The reason why this generally only vanishes at $\ca N$ is because $u_+$ is typically only geodesic ($u^a_+ \bs\nabla_a u_+^b = 0$) at $\ca N$.

The metric components in this coordinate system thus satisfy the following properties in $\ca V$,
\ba
&g_{+-} = g(u_+, u_-) = 1 \no
&g_{--} = g(u_-, u_-) = 0 \no
&g_{-i} = g(u_-, \partial_i) = 0
\ea
and the additional properties at $\ca N$,
\ba
&g_{++} = g(u_+, u_+) = 0 \no
&g_{+i} = g(u_+, \partial_i) = 0 \no
&\partial_- g_{++} = u_-^a \bs\nabla_a (u_+ \cdot u_+) = 0 \label{gatN}
\ea
We define $h_{ij} := g(\partial_i, \partial_j)$. Since we wish to study the properties of spacelike surfaces approaching $\ca N$, and the exterior curvature (and the mean curvature) contain one derivative away from the surface, we will consider a first order expansion of the metric in $\lambda^-$. Thus, in matrix form, up to first order in $\lambda^-$,
\be
g \approx 
\left(
\begin{array}{cc|c}
0 & 1 & 0 \\
1 & 0 & g_{+i} \\ \hline
0 & g_{+i} & h_{ij}
\end{array}\right)
\ee
where the components are ordered as $(-+i)$.
Note that $g_{+i}$ is first order in $\lambda^-$, $h_{ij}$ is zeroth order and
$g_{++}$ does not appear since it is second order (due to the last equation in \eqref{gatN}). The inverse metric matrix, to first order in $\lambda^-$, reads
\be
g^{-1} \approx 
\left(
\begin{array}{cc|c}
0 & 1 & - g_{+j}h^{ji}\\
1 & 0 & 0 \\ \hline
- h^{ij}g_{+j} & 0 & h^{ij}
\end{array}\right)
\ee
where $h^{ij}$ denotes the inverse of $h_{ij}$. 
To first order in $\lambda^-$, the determinant is given simply by
\be
\text{det}(g) \approx - \text{det}(h)
\ee
where it was used that $\text{det}(g+\delta g) \approx \text{det}(g) + \text{det}(g) \text{tr}(g^{-1}\delta g)$.

Now let us consider a surface $\ca S$ described by the zero-level of $S$,
\be
S := \lambda^- - \epsilon f(\lambda^+;x^i)
\ee
where $\epsilon$ is a (positive) ``small'' parameter to make it explicit that $\ca S$ is near $\ca N$. Suppose that at $\ca C$, intended to represent a piece of the diamond corner, we have $f(0;x^i) = 0$, indicating that the surface emanates from the corner (as it is intended to represent a portion of a Cauchy slice). In the coordinates constructed above,
\be
\partial_\mu S = (1, -\epsilon\partial_+f, - \epsilon\partial_i f)
\ee
If $f$ is ``order 1'', then we will be interested in a neighborhood with $\lambda^- \lesssim \epsilon$, so we can use the first-order approximations above to write
\be
g^{\mu\nu} \partial_\nu S \approx (-\epsilon\partial_+ f, 1; - h^{ij} g_{+j} - \epsilon h^{ij} \partial_j f)
\ee
where terms such as $\epsilon g_{+i} h^{ij} \partial_jf$, that would appear in the ``$-$'' component, are neglected for being of order $\epsilon^2$. Also, we have
\be
g^{\mu\nu} \partial_\mu S \partial_\nu S \approx - 2 \epsilon \partial_+ f
\ee
The mean curvature of $\ca S$ is therefore
\be\label{KnearN0}
K \approx \frac{1}{\sqrt{h}} \partial_\mu \left( \sqrt{h}\frac{g^{\mu\nu} \partial_\nu S}{\sqrt{2\epsilon \partial_+f}} \right)
\ee
where the $+$ sign was chosen so that $n^\mu = \pm g^{\mu\nu} \partial_\nu S/\sqrt{2\epsilon \partial_+f}$ points to the future --- the reasoning is that, as $\lambda^-$ grows towards the past (i.e., the interior of the diamond), we need
\be
d\lambda^-(n) = n^- = \pm \frac{(-\epsilon\partial_+ f)}{\sqrt{2\epsilon \partial_+f}} < 0 
\ee
which implies that
\be
\pm (-\epsilon\partial_+ f) < 0
\ee
but in order for $\ca S$ to be spacelike, $\partial_\mu S \partial^\mu S \approx - 2 \epsilon \partial_+ f < 0$, which is consistent with the $+$ choice above. Since the argument of the derivative in \eqref{KnearN0} is independent of $\lambda^-$ in this approximation, we have
\be
K \approx \left(\frac{1}{\sqrt{h}} \partial_+ \sqrt{h} \right)  \frac{1}{\sqrt{2\epsilon \partial_+f}} + \partial_+\left( \frac{1}{\sqrt{2\epsilon \partial_+f}} \right) - \frac{1}{\sqrt{h}} \partial_i \left( \sqrt{h} \frac{h^{ij} g_{+j} + \epsilon h^{ij} \partial_j f}{\sqrt{2\epsilon \partial_+f}} \right)
\ee
Note that the first two terms are order $\epsilon^{-1/2}$, while the third term is order $\epsilon^{1/2}$; therefore the first two terms dominate in the limit $\epsilon \rightarrow 0$. In addition, the quantity inside parenthesis in the first term can be identified, in this approximation, with the expansion parameter $\Theta$ of the null generators of $\ca N$, so we have
\be
K \approx  \frac{\Theta}{\sqrt{2\epsilon \partial_+f}} + \partial_+\left( \frac{1}{\sqrt{2\epsilon \partial_+f}} \right)
\ee
Now note that $\Theta$ is bounded from above, as follows. Since the corner is smooth and compact, $\Theta$ at $\ca C$ has compact image. Any causal diamond has a compact null horizon, meaning that $\Theta$ can evolve with respect to the Raychaudhuri equation. In the present case one can show that $\Theta$ must decrease along $u^+$, so it will either run to $-\infty$ (if a conjugate point appears, i.e., nearby null generators of $\ca N$ converge to a point), or it may simply stops at a finite value if $\ca N$ ends before a conjugate point appears (say, when the null generators of $\ca N$ intersect with another $\ca N'$ emanating from another portion of the corner). The conclusion is that there exists a finite $\Theta_0$ such that
\be
\Theta < \Theta_0
\ee
and consequently, within the approximations,
\be
K <  \frac{\Theta_0}{\sqrt{2\epsilon \partial_+f}} + \partial_+\left( \frac{1}{\sqrt{2\epsilon \partial_+f}} \right)
\ee
Now let $f$ be solution of the equation
\be\label{feqkappa0}
\partial_+\left( \frac{1}{\sqrt{ \partial_+f}} \right) + \frac{\Theta_0}{\sqrt{ \partial_+f}}= \kappa_0
\ee
where $\kappa_0$ is some (negative) constant. This will imply that
\be
K < \frac{\kappa_0}{\sqrt{2\epsilon}}
\ee
so by taking $\epsilon \rightarrow 0$ we have that $\ca S$ will have a mean curvature whose supremum is less than an arbitrarily negative number.

We need to make sure that equation \eqref{feqkappa0} has sensible solutions, i.e., representing a spacelike surface within the diamond, for all $\Theta_0 \in \bb R$ and at least one $\kappa_0 < 0$. The equation is linear in $\zeta(\lambda^+;x) := 1/\sqrt{\partial_+f}$, and the general solution is
\be
\zeta(\lambda^+;x) = \zeta(0;x) e^{-\Theta_0 \lambda^+} + \kappa_0 \int_0^{\lambda^+}\!\! d\tau\, e^{-(\lambda^+ -\tau)\Theta_0}
\ee
Then,
\be
f(\lambda^+;x) = \int_0^{\lambda^+}\!\! d\tau\, \frac{1}{\zeta(\tau; x)^2}
\ee
where it was used that $f(0; x) = 0$. Note that $f(\lambda^+;x) > 0$, which is consistent with the surface being inside the diamond. In order for it to be spacelike we need $\partial_+f >0$, which is equivalent to say that $\zeta(\lambda^+,x) > 0$ for $\lambda^+$ within its (finite) range. But since $\zeta(0;x)>0$ can be chosen arbitrarily large (i.e., the surface can be made to start arbitrarily close to being tangent to $\ca N$), then the term involving $\kappa_0$ will not have the opportunity to make $\zeta$ become negative. To make this more precise, say that the maximum value of $\lambda^+$ at given $x\in \ca C$ is $\Lambda^+$. (In what follows we will omit the $x$ argument.) Now consider three cases, where $\Theta_0$ at $x$ is zero, positive or negative. If $\Theta_0 = 0$, then $\zeta(\lambda^+) = \zeta(0) + \kappa_0 \lambda^+$; thus we need $\zeta(0) > -\kappa_0 \Lambda^+$. If $\Theta_0 \ne 0$, then $\zeta(\lambda^+) = (\zeta(0) - \kappa_0/\Theta_0) e^{-\Theta_0\lambda^+} + \kappa_0/\Theta_0$; thus if $\Theta_0 < 0$ we need $\zeta(0) > \kappa_0/\Theta_0$, and if $\Theta_0 >0$, we need $\zeta(0) > - \frac{\kappa_0}{\Theta_0} (e^{\Theta_0\Lambda^+} - 1)$. These impose upper bounds on $\partial_+f$ at $\ca C$; moreover, since $\ca C$ is compact, this bound can be satisfied with $\partial_+f(0;x)$ strictly positive for all $x\in \ca C$.

Lastly we must address the fact that the future horizon of the diamond is not a manifold, since it has singularities at the top. In three spacetime dimensions, the singular subset seems to have a (1-dimensional) graph-like shape.\footnote{The symmetric diamond is exceptional as its future horizon is a cone, with a unique singular point at the top.} Suppose that $\ca C$ and $\ca C'$ are disjoint intervals of the corner, and suppose that the null surfaces emanating from them, $\ca N$ and $\ca N'$, meet at a line segment $\ca J$. The surfaces $\ca S$ and $\ca S'$, respectively approaching $\ca N$ and $\ca N'$ with arbitrarily negative mean curvature, would meet in a singular fashion slightly to the past of $\ca J$. The idea is to ``round off'' this intersection, by interpolating them with a surface $\ca H$ that is approximately a quadratic surface---a piece of an ellipsoid. We wish to show that if at least one of the radii of the ellipsoid tends to zero, the mean curvature diverges; and, in particular, if it is curved so that the ``center'' is to the past of the surface, as $\ca H$ would be, then it diverges negatively. Thus, if we do the rounding off close enough to $\ca J$, then $\ca S \cup \ca H \cup \ca S'$ will have arbitrarily negative mean curvature.

To define $\ca H$, consider a coordinate system $(t;x)$ in a neighborhood of $\ca J$ that is small enough so that the metric can be approximated by the Minkowski metric, $ds^2 = -dt^2 + d(x^1)^2 + d(x^2)^2$.\footnote{The spacetime curvature corrections to this metric will not influence the argument.}
Let $\ca H$ be described as the zero-level of the function
\be
H(t;x) = t - h(x)
\ee
where $h$ is some real function of $x$.
Then formula \eqref{Kcoordformula} applies, yielding
\be
K = \sum_i \partial_i \left( \frac{\partial_i h}{\sqrt{1 - (\partial h)^2}} \right) = \frac{\partial^2 h}{\sqrt{1 - (\partial h)^2}} + \sum_{ij}\frac{\partial_i h \partial_i \partial_j h \partial_j h}{\sqrt{1 - (\partial h)^2}^3}
\ee
where $i,j \in \{1,2\}$, $(\partial h)^2 := \sum_i \partial_i h \partial_i h$, and $\partial^2 h := \sum_i \partial_i \partial_i h$. The plus sign was chosen so that the normal vector $n$ is future-pointing, $dt(n)>0$. Now suppose that $\ca H$ can be approximated by a quadratic surface, meaning
\be
h = - \frac{1}{2}\sum_i a_i (x_i)^2
\ee
for coefficients $a_i$. 
We get,
\be
K = -\frac{\sum_i a_i}{\sqrt{1 - \sum_i a_i^2 (x^i)^2}} -\frac{\sum_i (a_i)^3 (x^i)^2}{\sqrt{1 - \sum_i a_i^2 (x^i)^2}^3}
\ee
Note that for this surface to be spacelike we need $(\partial h)^2 = \sum_i a_i^2 (x^i)^2 < 1$, which implies a upper bound on the range of $x^i$. This could be satisfied if we choose $|x^i| < 1/(\sqrt{2}|a_i|)$. Now let us say that $x^1$ is oriented along $\ca J$ and $x^2$ is orthogonal. If $\ca J$ is a smooth 1-dimensional manifold, and $\ca H$ is going along it, then $a_1$ will be bounded while $a_2$ can be taken to be arbitrarily positive. Therefore the main contribution from the numerators are for $i=2$, which gives,
\be
K \approx -a_2 \frac{1 - a_1^2 (x^1)^2}{\sqrt{1 - \sum_i a_i^2 (x^i)^2}^3} < -a_2
\ee
implying that the mean curvature of $\ca H$ can be made arbitrarily negative, and consequently the mean curvature of $\ca S \cup \ca H \cup \ca S'$ can be made arbitrarily negative by pushing it towards $\ca N \cup \ca J \cup \ca N'$ with a sharp rounding off at the top. 

This argument can be extended to the case where the null surfaces meet at a vertex of the singularity graph, or for the symmetric diamond where the singularity is a point. An analogous argument also applies for the past horizon, showing that a surface with arbitrarily large mean curvature can be constructed.

\subsection{The CMC gauge is attainable}
\label{subsec:CMCattain}

We have seen that for any causal diamond in our class of spacetimes (i.e., a state in our system), the CMC condition defines a unique foliation that refers only to intrinsic geometrical structures of the spacetime. This is, therefore, a good non-perturbative prescription to fix the ``gauge of time''. But why do we have a ``gauge of time'' in the first place? 
If we are inclined to regard the diamond as a ``self-contained'' system, the underlying intuition is that classically the only ingredients we have to construct a diamond spacetime solution is an abstract topological disc and a set of data $(h_{ab}, \pi^{ab})$ on it, and there is nothing physically unique about what time coordinate is used to evolve the equations of motion away from the disc, $(h_{ab}, \pi^{ab}) \mapsto (h_{ab}(t), \pi^{ab}(t))$, so one  could take $(h_{ab}(t), \pi^{ab}(t))$ as the ``initial data'' for any $t$ and the resulting solution should correspond to the same physical spacetime. This is essentially the statement that refoliations of the spacetime are gauge. However, from the perspective of the Hamiltonian formalism, we cannot be this vague about what we choose to regard as gauge or not, as they have a sharp definition in terms of the null directions of the symplectic form.  If one is determined to ascribe gauge status to certain transformations, it is mandatory to make the appropriate modifications to the theory or specific choices of boundary conditions. As we have discussed in the introduction, we wish to understand the implementation of the CMC gauge fixing program,
so we must determine what boundary conditions will imply that refoliations of spacetime are in fact gauge, so that the CMC foliation can be attained via a {\sl gauge transformation}. We will show in this section that this condition is precisely the fixing of the induced boundary metric, $h|_\partial = \gamma$.

As the constraints of General Relativity are of first class, they generate gauge transformations. The Momentum constraint is know to generate spatial diffeomorphisms, while the Hamiltonian constraint generate evolution between Cauchy slices. We are therefore interested in a smearing of the Hamiltonian constraint,
\be
H_0[N] := \int d^2x  N \sqrt{h} (K^{ab} K_{ab} - K^2 - R + 2\Lambda) 
\ee
where the lapse function $N$ vanishes at the boundary, $N|_\partial = 0$, since two Cauchy slices in a causal diamond always meet at the corner. Recall that $\pi^{ab} = \sqrt{h} (K^{ab} - K h^{ab})$. However, this charge will not generate a gauge transformation if it doesn't generate a (regular) symplectic flow in the first place. Recall that a charge $Q$ generates a symplectic flow $X$ according to the equation $\delta Q = - i_X \Omega$. Now consider the (pre)symplectic form $\Omega = \int_\Sigma d^2x \,\delta \pi^{ab} \wedge \delta h_{ab}$, which is purely a bulk integral. For any (smooth) vector field $X$ in phase space, define $(X^h)_{ab} := i_X\delta h_{ab}$ and $(X^\pi)^{ab} := i_X \delta \pi^{ab}$. We have
\be
i_X\Omega = \int_\Sigma d^2x \left[ (X^\pi)^{ab} \delta h_{ab} - (X^h)_{ab} \delta\pi^{ab} \right]
\ee
Thus $\delta Q$ will only be associated with a regular vector field $X$, via $\delta Q = - i_X \Omega$, if
\be
\delta Q = \int_\Sigma d^2x \left[ A^{ab} \delta h_{ab} + B_{ab} \delta\pi^{ab} \right]
\ee
for smooth functions $A^{ab}$ and $B_{ab}$; in particular, $\delta Q$ must contain no boundary terms involving variations of the dynamical fields. If that is the case, we say that $Q$ is {\sl symplectically differentiable}. The conclusion is that $H_0[N]$ is a gauge-generator if and only if it is symplectically differentiable. 

The variation $H_0$ gives, on-shell,
\be
\delta \int d^2x \sqrt{h}\, N (K^{ab} K_{ab} - K^2 - R + 2\Lambda) = \int d^2x \sqrt{h}\, N \delta(K^{ab} K_{ab} - K^2 - R)
\ee
Note that the variation of $N$ itself is irrelevant here since it would end up multiplying a constraint. The variation of $K^{ab} K_{ab} - K^2$ will naturally have the form of a integral of differentials of the dynamical fields over the bulk, so it will not cause issues with differentiability. The only term that could cause issues is
\be
\delta R = - \delta h_{ab} R^{ab} + \nabla^a (\nabla^b\delta h_{ab} - h^{bc} \nabla_a \delta h_{bc})
\ee
since the presence of differentials inside derivatives means that, upon integration by parts and application of Stokes theorem, these differentials may end up in a boundary piece. In fact, the boundary piece one gets is
\begin{eqnarray}
\int d^2x \sqrt{h}\, N (-\delta R) &\overset{\partial}{\sim}& -\int_\partial ds\, n^a N (\nabla^b\delta h_{ab} - h^{bc} \nabla_a \delta h_{bc}) \nonumber\\
&& + \int_\partial ds\, ( n^b \nabla^a N \delta h_{ab} - n_a \nabla^a N h^{bc} \delta h_{bc}) \label{Ndelta}
\end{eqnarray}
where $\overset{\partial}{\sim}$ means that the two sides differ only by bulk integrals not containing spatial derivatives of field differentials, $ds$ is the proper length measure on the boundary and $n^a$ is the unit normal (outward-pointing) vector to the boundary (and tangent to the Cauchy slice).

Imposing that the lapse function vanishes at the corner, the boundary integral in the first line goes away, leaving
\be\label{deltaH0}
\delta H_0 \,\overset{\partial}{\sim}\, \int_\partial ds \left( n^b \nabla^a N \delta h_{ab} - n_a \nabla^a N h^{bc} \delta h_{bc} \right)
\ee
Let $t^a$ be the unit vector tangent to the corner, so $h_{ab} = n_a n_b + t_a t_b$. Since $N=0$ along the corner, $t^a \nabla_a N = 0$ which implies that $\nabla^a N = \lambda n^a$ for some scalar $\lambda$ on the boundary. We therefore get
\be\label{boosts}
\delta H_0 \,\overset{\partial}{\sim}\, - \int_\partial ds\, \lambda ( h^{ab} -n^a n^b )\delta h_{ab} = - \int_\partial ds\, \lambda t^a t^b \delta h_{ab}
\ee
We see that $H_0[N]$ is symplectically differentiable, for arbitrary lapse satisfying $N|_\partial = 0$ (so, in particular, arbitrary $\lambda := n^a \nabla_a N \ne 0$), provided that
\be
t^a t^b \delta h_{ab} = 0
\ee
which is equivalent to say that the induced boundary metric on $\partial\Sigma$ must be fixed.

Note that if we do not fix this boundary condition, one can still evolve from one Cauchy slice to another, but the charge generating that trasformation would be an ``augmented'' version of $H_0$, where we add to it a boundary term, $Q_\partial := 2 \int_\partial ds n^a\nabla_a N$. The boundary term ensures that $H_0 + Q_\partial$ is differentiable (assuming $\delta(n^a\nabla_a N) = 0$) and thus generate a regular symplectic flow. However, on-shell, $H_0 + Q_\partial \approx Q_\partial$, which is not a constant in phase space and thus does not generate a gauge-transformation. In other words, if the Dirichlet boundary condition is not imposed, deformations of the Cauchy slice that tilt the angle that it makes with the corner are non-trivial transformations between distinct physical states. In higher dimensions, the charges generating these corner boosts are  given by analogous expressions, but with $ds$ replaced by the induced volume form on the boundary.

\subsection{Existence and uniqueness for the Lichnerowicz equation}
\label{subsec:existunique}

Here we justify the claim that the Lichnerowicz equation, first appearing in \eqref{LichEq}, always has one and only one solution for any given boundary condition. The proof is a straightforward modification of O'Murchadha and York paper ~\cite{o1973existence} --- they consider 3+1 spacetime dimensions, in which case the Lichnerowicz equation is polynomial in the Weyl factor (in fact, this polynomial form is true for all dimensions greater than 2+1). We wish to adapt their argument to 2+1 spacetime dimensions, in which case the Lichnerowicz equation has an exponential form in terms of the Weyl factor. See also \cite{moncrief1989reduction}.

On the spatial disc $D$, the Lichnerowicz equation associated with pre-initial data $\scr S = (h_{ab},\sigma^{ab},\tau)$ is
\be\label{Licheu}
\nabla^2_{(h)} \phi =  R_{(h)} -   \sigma^{ab}\sigma_{ab} e^{-\phi} + \chi e^\phi
\ee
where $R_{(h)}$ and $\nabla_{(h)}$ are respectively the Ricci scalar and the covariant derivative for the metric $h_{ab}$. Before proceeding, it is worth noticing that if the Lichnerowicz equation associated with the pre-inital data $\scr S$ and boundary condition $\phi|_{\partial D} = \varphi$ has a unique solution $\phi_s$, then the Lichnerowicz equation for a Weyl-tranformed data $\scr S' = (e^\lambda h_{ab},e^{-2\lambda} \sigma^{ab},\tau)$ and boundary condition $\phi'|_{\partial D} = \varphi - \lambda$ also has a unique solution given by $\phi'_s = \phi_s - \lambda$. This can be easily seen from the transformation properties of the Laplacian and the Ricci scalar, as discussed around equation \eqref{Lichtransdata}, since the Lichnerowicz equation for $\scr S'$ can be re-expressed as
\be\label{Licheu2}
\nabla^2_{(h)} \phi'  = \left(R_{(h)} - \nabla^2_{(h)}\lambda\right)  -  e^{-\lambda} \sigma^{ab}\sigma_{ab} e^{-\phi'} + e^\lambda \chi e^{\phi'}
\ee
revealing that $\phi' + \lambda$ satisfies the same equation as $\phi$, with the same boundary condition. 

To prove uniqueness of solution, suppose that \eqref{Licheu} has a solution $\phi_s$ and consider the transformation in which $\lambda = \phi_s$, so that the Lichnewowicz equation for $\phi'$, as in \eqref{Licheu2}, becomes
\be
\nabla^2_{(h)} \phi'  =  e^{-\lambda} \sigma^{ab}\sigma_{ab} \left( 1 - e^{-\phi'}\right) + e^\lambda \chi \left( e^{\phi'} - 1 \right)
\ee
with vanishing boundary condition for $\phi'$. As expected, $\phi' = 0$ is a solution. Now multiply this equation by $\phi'$ on both sides and integrate over $\Sigma$, with respect to the volume form associated with $h_{ab}$,
\be
\int \phi' \nabla^2_{(h)} \phi'  =  \int \left[ e^{-\lambda} \sigma^{ab}\sigma_{ab} \phi' \left( 1 - e^{-\phi'}\right) + e^\lambda \chi \phi' \left( e^{\phi'} - 1 \right) \right] \ge 0
\ee
observing that the right-hand side is nonnegative since $\phi' ( 1 - e^{-\phi'})$ and $\phi' ( e^{\phi'} - 1 )$ are nonnegative for any $\phi'$, and $\sigma^{ab}\sigma_{ab}$ and $\chi$ are also nonnegative. But integrating the left-hand side by parts gives
\be
\int \phi' \nabla^2_{(h)} \phi' = \int \nabla_{(h)} \cdot \left( \phi' \nabla_{(h)} \phi' \right) - \int  \left( \nabla_{(h)} \phi' \right)^2 \le 0
\ee
where the first term vanishes by using Stokes' theorem and imposing $\phi' = 0$ at the boundary. We therefore conclude that $\nabla_{(h)} \phi' = 0$ and, given the boundary condition, also that $\phi' = 0$. That is, this equation has no solutions other than $\phi' = 0$ and thus  \eqref{Licheu} has no solutions other than $\phi_s$.

In O'Murchadha and York paper, two proofs of existence are offered. Here we shall only revisit one of the proofs, based on the construction of a sequence of functions that converge to the solution. To unclutter the notation, let us write \eqref{Licheu} in its general form
\be\label{Licheu3}
\nabla^2_{(h)} \phi  = F(\phi; x) \,,\quad \phi|_{\partial D} = \varphi
\ee
where
\be
F(\phi;x)  = c(x)  - a(x) e^{-\phi} + b(x) e^{\phi}
\ee
in which $a(x)$ and $b(x)$ are nonnegative functions on $\Sigma$, and $c(x)$ is a function on $\Sigma$. 
As explained, for the purposes of proving existence and uniqueness of solutions, we can always perform a transformation $c \rightarrow c - \nabla^2\lambda$, $a \rightarrow a e^{-\lambda}$, $b \rightarrow b e^{\lambda}$ and $\varphi \rightarrow \varphi - \lambda|_{\partial D}$ (which is the form of the problem expressed in \eqref{Licheu2}). With this freedom, $\varphi$ and $c$ can be chosen to be any given functions, according to convenience, and $a$ and $b$ will fall into one of the three cases below:

$(i)\quad$ $a \ge 0$ and $b > 0$. This is the case of main interest, when $\sigma^{ab} \sigma_{ab} \ge 0$ and $\chi > 0$. For reasons that will become clear, it is convenient to choose $c$ to be any negative constant, $c_0 < 0$.

$(ii)\quad$ $a \ge 0$ and $b = 0$. This is the case for the maximal slice ($\tau = 0$) with zero cosmological constant. Here we can choose $c$ as any function such that $c(x)>0$ if $a(x)>0$ and $c(x) = 0$ if $a(x) = 0$.

$(iii)\quad$ $a = 0$ and $b = 0$. The existence of solution for this case is trivial since we can choose $c = 0$ so the equation reduces to $\nabla^2 \phi = 0$. 

In all cases we could choose, e.g., $\varphi = 0$, but this would not amount to any real simplification. 

The core of the proof lies in the existence of functions $\phi_+$ and $\phi_-$ on $\Sigma$, with $\phi_+ \ge \phi_-$, such that
\begin{eqnarray*}
&F(\phi_+) \ge 0 \,,\,\, \phi_+|_{\partial D} \ge \varphi \\
&F(\phi_-) \le 0 \,,\,\, \phi_-|_{\partial D} \le \varphi
\end{eqnarray*}
and we can see that this is true for cases $(i)$ and $(ii)$ above. For case $(i)$, the $b$ term dominates as long as $\phi_+$ is sufficiently large, making $F > 0$; and for $\phi_-$ very negative the $b$ term is suppressed and the $a$ term dominates wherever $a > 0$, while the $c := c_0 < 0$ term ensures that $F < 0$ even in regions where $a = 0$. For case $(ii)$,  the $c(x)$ term dominates when $\phi_+$ is very large, making $F \ge 0$; while the $a$ term dominates when $\phi_-$ is very negative, making $F \le 0$. It will be convenient (and it is possible) to take $\phi_-$ and $\phi_+$ to be constants satisfying $\phi_-|_{\partial D} \le \varphi \le \phi_+|_{\partial D}$. 

Note that if we had a ``case $(iv)$'' in which $a \ge 0$ and $b < 0$ then it would not be generally possible to find $\phi_+$ and $\phi_-$ satisfying the desired conditions. This is would be the case for a positive cosmological constant, as $\chi = -2\Lambda + \tau^2/2 < 0$ for $\tau$ in some interval around $0$, i.e., around the maximal slice. The problem occurs at any points where $\sigma_{ab}$ vanishes, so that $a = 0$. At these points, $F = c + b e^\phi$ is a monotonically decreasing function of $\phi$, so it is not possible to have both $\phi_+ \ge \phi_-$ and $F(\phi_-) \le 0 \le F(\phi_+)$.

The proof begins by constructing a sequence of functions $\phi_n$ recurssively defined by
\be\label{Lichrec}
\nabla^2 \phi_n - \kappa \phi_n = F(\phi_{n-1}) - \kappa \phi_{n-1}
\ee
with boundary condition $\phi_n|_{\partial D} = \varphi$, where 
\be
\kappa := \max_{x \in D} \max_{\phi_- \le \phi \le \phi_+} \left| \frac{\partial F}{\partial \phi} (\phi; x) \right|
\ee
and the starting function is $\phi_0 := \phi_+$. Note that $\kappa$ is well defined because, for each $x$, $\phi(x)$ is in the compact interval $[\phi_-(x), \phi_+(x)]$, and then $x$ is maximized over the compact set $D$. We will omit the reference to the metric in $\nabla = \nabla_{(h)}$ because it only matters to us that the metric is Riemannian, as this ensures that the differential operator in \eqref{Lichrec}, $L := \nabla^2 - \kappa$, is strictly elliptic\footnote{A linear second order differential operator $L = A_{ij}(x) \partial_i \partial_j + B_i(x) \partial_i + C(x)$ is called {\sl strictly elliptic} on a domain $\Omega$ if there is $\lambda > 0$ such that $A_{ij}(x) \xi_i \xi_j \ge \lambda \xi_i \xi_i$, $x \in \Omega$, for all vectors $\xi$. (Sum over repeated indices implied.)\label{FN:ellidiff}}. The goal is to prove that this sequence is monotonically decreasing, in the sense that $\phi_n < \phi_{n-1}$, and also bounded from below by $\phi_-$. This ensures that the sequence converges and, by a simple argument, that the limit is the solution of \eqref{Licheu3}. 

First let us see that, if the limit exists, it should be the solution of \eqref{Licheu3}. As a note of consistency, observe that the solution is a fixed point of the iteration. That is, if $\phi_{n-1}$ is a solution for some $n > 1$, $\nabla^2 \phi_{n-1} = F(\phi_{n-1})$, then $\nabla^2 (\phi_n - \phi_{n-1}) - \kappa (\phi_n - \phi_{n-1}) = 0$, with boundary condition $(\phi_n - \phi_{n-1})|_{\partial D} = \varphi - \varphi = 0$, and it is clear that $\phi_n = \phi_{n-1}$. Now we proceed with the proper proof. Suppose that $\phi$ is the limit of the sequence $\{\phi_n\}$, meaning that for every $\epsilon > 0$ there exists $N$ such that $|\phi - \phi_n|<\epsilon$ for all $n > N$. This assumes that the convergence is uniform, which is guaranteed by Dini's theorem as long as the limit $\phi$ is continuous\footnote{{\sl Dini's theorem} states that if a monotonically increasing (or decreasing) sequence of continuous real functions on a compact topological space converges pointwise to a continuous function, then the convergence is uniform.}. For $n > N+1$ we have $|\phi_{n-1} - \phi_n| < 2\epsilon$, so subtracting $F(\phi_n)$ on both sides of \eqref{Lichrec} we get
\begin{eqnarray}
|\nabla^2 \phi_n - F(\phi_n)| &=& |\kappa (\phi_n - \phi_{n-1}) +  F(\phi_{n-1}) - F(\phi_n)| \nonumber\\
&<& \kappa| \phi_n - \phi_{n-1}| +  |F(\phi_{n-1}) - F(\phi_n)| \nonumber\\
&<& 2\kappa| \phi_n - \phi_{n-1}| \nonumber\\
&<& 4\kappa \epsilon
\end{eqnarray}
showing that $\phi_n$ converges (weakly\footnote{In the theory of differential equations, a function $u$ is said to be a weak solution of the differential operator $E$ on a domain $D$ if $\int_D E[u] \, \xi = 0$ for all smooth functions $\xi$ supported on arbitrary compact subsets of $\text{Int}(D)$; it is assumed that integrations by parts have been formally applied to the integral so as to move all the derivatives from $u$ to  $\xi$ (thus $u$ may be a weak solution for $E$ even if it does not have well-defined derivatives). For example, a weak solution $\phi$ for \eqref{Licheu3} is required to satisfy $\int (\nabla^2\phi - F(\phi))\xi = \int (\phi \nabla^2\xi - F(\phi)\xi) = 0$ for all smooth $\xi$ that vanishes in a neighborhood of the boundary of the disc. The result above shows that $\nabla^2\phi_n - F(\phi_n)$ converges to zero, which implies that $\phi_n$ converges to a weak solution of \eqref{Licheu3}.}) to a solution of $\nabla^2\phi - F(\phi) = 0$. 

Now let us show that the sequence of solutions $\phi_n$ is monotonically decreasing and bounded from below. We start by reviewing a important result from the theory of linear elliptic differential equations~\cite{evans2022partial}. Given a strictly elliptic differential operator $L$, as defined in footnote \ref{FN:ellidiff}, with $C(x) \le 0$, and given a function $f$, let $u$ be the solution of $-L[u] = f$ with boundary condition $u|_{\partial D} = \varphi$. A function $u^+$ is called a {\sl supersolution } of this differential equation if $-L[u^+] \ge f$ and $u^+|_{\partial D} \ge \varphi$; similarly, a function $u^-$ is called a {\sl subsolution } of this differential equation if $-L[u^-] \le f$ and $u^-|_{\partial D} \le \varphi$. If $u^+$ is a supersolution then $u \le u^+$, and if $u^-$ is a subsolution then $u^- \le u$. The equation \eqref{Lichrec} defining $\phi_n$ satisfies the conditions of the theorem since $L = \nabla^2 - \kappa$ is strictly elliptic with $C = -\kappa < 0$, and here $f = f_{n-1} := - F(\phi_{n-1}) + \kappa \phi_{n-1}$. The simplest way to construct a supersolution is to consider a constant function $\phi_n^+$, in which case $-L[\phi_n^+] = \kappa \phi_n^+$, so we can take
\be
\phi_n^+ =  \text{max}(\kappa^{-1}f_{n-1}, \varphi)
\ee
where the maximization also runs over $D$. (Note that $\kappa \ne 0$, except in case $(iii)$ above, which is trivial.) Similarly, we can construct a subsolution by taking
\be
\phi_n^- =  \text{min}(\kappa^{-1}f_{n-1}, \varphi)
\ee
Therefore, the (sub)supersolution theorem implies that
\be\label{subsupertheo}
\text{min}(\kappa^{-1}f_{n-1}, \varphi) \le \phi_n \le \text{max}(\kappa^{-1}f_{n-1}, \varphi)
\ee
As mentioned before, the starting point is $\phi_0 := \phi_+$. In the equation defining $\phi_1$ we have $\kappa^{-1}f_0 = -\kappa^{-1}F(\phi_+) + \phi_+ \le \phi_+$, where we have used that $F(\phi_+) \ge 0$. Thus, since $\varphi \le \phi_+$, it follows from \eqref{subsupertheo} that $\phi_1 \le \phi_+ =: \phi_0$. Also, as $\phi_- \le \varphi \le \phi_+$ and $F(\phi_-) \le 0 \le F(\phi_+)$, we have $F(\phi_+) - F(\phi_-) \le \kappa (\phi_+ - \phi_-)$, so that $\kappa^{-1}f_0 = -\kappa^{-1}F(\phi_+) + \phi_+  \ge -\kappa^{-1}F(\phi_-) + \phi_- \ge \phi_-$ and it follows from \eqref{subsupertheo} that $\phi_1 \ge \phi_-$.

Having established that $\phi_- \le \phi_1 \le \phi_0 = \phi_+$, we now proceed by induction: suppose that $\phi_- \le \phi_{n-1} \le \phi_{n-2} \le \phi_+$ for $n > 1$ and show that $\phi_- \le \phi_n \le \phi_{n-1}  \le \phi_+$. Replicating the argument above, let us first prove that $\phi_n \ge \phi_-$. 
In the equation defining $\phi_n$ we have $\kappa^{-1}f_{n-1} = -\kappa^{-1}F(\phi_{n-1}) + \phi_{n-1}$, so using $|F( \phi_{n-1}) - F( \phi_-)| \le \kappa ( \phi_{n-1} -  \phi_-)$ we see that $\kappa^{-1}f_{n-1} \ge -\kappa^{-1}F(\phi_-) + \phi_- \ge \phi_-$. Hence $\phi_n \ge \text{min}(\kappa^{-1}f_{n-1}, \varphi) \ge \text{min}(\phi_-, \varphi) = \phi_-$.
Next we prove that $\phi_n \le \phi_{n-1}$.
Subtracting equation \eqref{Lichrec} for $\phi_n$ from that for $\phi_{n-1}$ we get
\be
\nabla^2(\phi_n - \phi_{n-1}) - \kappa (\phi_n -  \phi_{n-1}) = F( \phi_{n-1}) - F( \phi_{n-2}) + \kappa ( \phi_{n-2} -  \phi_{n-1})
\ee
This equation satisfies the super/subsolution theorem for the same operator $L = \nabla^2 - \kappa$, where here we have $u = \phi_n - \phi_{n-1}$ and  $f = F( \phi_{n-2}) - F( \phi_{n-1}) - \kappa ( \phi_{n-2} -  \phi_{n-1})$. Since $|F( \phi_{n-2}) - F( \phi_{n-1})| \le \kappa ( \phi_{n-2} -  \phi_{n-1})$ we see that $f \le 0$. Moreover, for $n > 1$, $(\phi_n - \phi_{n-1})|_{\partial D} = \varphi - \varphi = 0$, implying that $\text{max}(\kappa^{-1}f, u|_{\partial D}) = 0$ and thus $\phi_n - \phi_{n-1} \le 0$. This concludes the proof that $\phi_- \le \cdots \le \phi_n \le \phi_{n-1} \le \cdots \le \phi_+$.

\section{Quotienting by conformal transformations}
\label{sec:Rquo}

In this section we will fill in some of the details left out in Sec. \ref{subsec:outline}. In particular we show that the reduced phase space is isomorphic to the cotangent bundle of the space of conformal isometries of the disk (subjected to the appropriate boundary conditions) and that this space of conformal geometries is isomorphic to $\diff/\psl$. This identification of the phase space with the cotangent bundle of space of conformal geometries of the Cauchy slice is known in the literature and quite generic in the context of gravity~\cite{moncrief1989reduction,fischer1997hamiltonian}, but our presentation is slightly different and specialized to the case of our interest (which contains a boundary).

\subsection{The cotangent bundle of the space of conformal geometries}
\label{subsec:cotanconfgeo}

We have explained in Sec. \ref{subsec:outline} that the reduced phase space for the causal diamonds is, as given in \eqref{RPSequiv}, isomorphic to the space of equivalence classes 
\be
\wt{\ca P} = \{(h_{ab}, \sigma^{ab}) \sim (\Psi_* e^\lambda h_{ab}, \Psi_* e^{-2\lambda} \sigma^{ab})\}
\ee
where $\Psi : \Sigma \rightarrow \Sigma$ is a boundary-trivial diffeomorphism of the disc (i.e., $\Psi$ acts as the identity on the boundary of the disc) and $\lambda: \Sigma \rightarrow \bb R$ vanishes at the boundary of the disc. 
Here we shall describe the cotangent bundle of the space of conformal geometries of the disk in a way that makes explicit that it is isomorphic to $\wt{\ca P}$. 

First let us describe the cotangent bundle of a quotient space in general terms. That is, given a manifold $\ca M$ and a Lie group $G$ that acts on $\ca M$, we wish to characterize the cotangent bundle of the space of orbits, $T^*(\ca M/G)$, in terms of the cotangent bundle of the manifold, $T^*\ca M$. The goal is to later particularize to the case where $\ca M$ is the space of Riemmanian metrics on a disc and $G$ is the group of boundary-trivial conformal transformations.
We shall assume that $\ca M/G$ has a manifold structure, which is guaranteed \cite{isham1999modern,helgason2001differential} to happen (at least in finite dimensions) when the stabilizer group of this action is a closed subgroup of $G$.
Let us think of $\ca M$ as a principal $G$-bundle over $\ca M/G$, denoting by $p: \ca M \rightarrow \ca M/G$ the quotient map that takes a point $x \in \ca M$ to the equivalence class $[x] := \{x \sim \Lambda_g x;  g \in G\}$, where $\Lambda_g$ is the diffeomorphism on $\ca M$ produced by $g \in G$. Since $p$ is a projection map, the push-forward $p_*$ is a surjection from $T_x\ca M$ onto $T_{[x]}(\ca M/G)$. Also, if two vectors $\xi$ and $\xi'$ at $x$ project to the same vector under $p_*$, then they must differ by an element of the kernel of $p_*$, i.e., $\xi' = \xi + \eta$ for some $\eta$ satisfying $p_*\eta = 0$. This means that $T_{[x]}(\ca M/G)$ is isomorphic to the quotient of $T_x\ca M$ by $\text{ker} (p_*)$, where two vectors at $x \in \ca M$ are identified if they differ by an element of $\text{ker}( p_*)$. In display,
\be\label{T[x]=Tx/kerp}
T_{[x]}(\ca M/G) = T_{x}\ca M/\text{ker} (p_*)
\ee

There is also another way to characterize $T_{[x]}(\ca M/G)$ by noticing that vectors at any points along the fiber over $[x]$ all project to vectors at the same base point $[x]$. We have $p \circ \Lambda_g = p$, for all $g \in G$, as $p$ projects the whole fiber to the same base point. This implies that $p_* \Lambda_{g*} = p_*$ and therefore the vector $\xi$ at $x$ has the same image under $p_*$ as the vector $\Lambda_{g*}\xi$ at $\Lambda_{g}x$. Moreover, since $\Lambda_g$ is a diffeomorphism of $\ca M$, its derivative $\Lambda_{g*}$ induces an isomorphism between the tangent spaces $T_x\ca M$ and $T_{\Lambda_g x}\ca M$, and this isomorphism preserves the kernel of $p_*$. Thus, if $\xi$ at $x$ projects to a given vector at $[x]$, then the only other vectors on $T\ca M$ that project to the same vector are given by $\Lambda_{g*}(\xi + \eta)$ for all $g \in G$ and all $\eta \in \text{ker}( p_*)$. In display, the tangent bundle of $\ca M/G$ is identified with the following quotient of the tangent bundle of $\ca M$
\be
T(\ca M/G) = \{ \xi \sim \Lambda_{g*}(\xi + \eta);\,  g\in G,\, \xi \in T\ca M \text{ and } \eta \in \text{ker}( p_*) \}
\ee
where the projection on the tangent bundle of $\ca M/G$, $\wt\pi: T(\ca M/G) \rightarrow \ca M/G$, is simply given by $\wt\pi([\xi]) = p(\pi(\xi))$, where $\pi: T\ca M \rightarrow \ca M$ is the projection on the tangent bundle of $\ca M$ and $\xi \in T\ca M$ is any representative of the class $[\xi] \in T(\ca M/G)$.

Now, most importantly, let us describe the cotangent bundle of $\ca M/G$. If $\wt\alpha$ is a 1-form at $[x] \in \ca M/G$, then its pull-back to $x \in \ca M$ via $p$ has the property of annihilating the whole kernel of $p_*$,
\be
p^*\wt\alpha(\eta) = \wt\alpha(p_*\eta) = \wt\alpha(0) = 0
\ee
where $\eta \in \text{ker}( p_*)$. On the other direction, every 1-form $\alpha$ at $x \in \ca M$ that annihilates $\text{ker}( p_*)$ defines a 1-form at $[x] \in \ca M/G$. This can be seen from the characterization of vectors at $[x]$ given in \eqref{T[x]=Tx/kerp}, since such an $\alpha$ defines a linear map to $\bb R$ which satisfies $\alpha(\xi) = \alpha(\xi + \eta)$ for every vector $\xi$ at $x$ and $\eta \in \text{ker}( p_*)$, and therefore is a well-defined linear map from $T_{[x]}(\ca M/G)$ to $\bb R$. Moreover, if we denote by $\wt\alpha$ the 1-form at $[x]$ defined in this way from $\alpha$, it is clear that $\alpha = p^*\wt\alpha$, showing that there is an isomorphism
\be\label{T*[x](M/G)}
T^*_{[x]}(\ca M/G) = \{ \alpha \in T^*_x\ca M; \text{ where } \alpha(\eta) = 0 \text{ for all } \eta \in \text{ker} (p_*) \}
\ee
Similarly to the other characterization of the tangent bundle, 1-forms on $\ca M/G$ can also be described in terms of a quotient over 1-forms at different points along the fiber $p^{-1}([x])$. From the identity $p \circ \Lambda_{g^{-1}} = p$ we have that $\Lambda_{g^{-1}}^* p^* = p^*$, and it follows that the $\text{ker}( p_*)$-annihilating 1-forms at $x$ are related via $\Lambda_{g^{-1}}^*$ to the $\text{ker}( p_*)$-annihilating 1-forms at $\Lambda_g x$. In fact, if $\eta$ is any vector in $\text{ker}( p_*)$ at $\Lambda_g x$ then $\Lambda_{{g^{-1}}*} \eta$ is in $\text{ker}( p_*)$ at $x$; so if $\alpha$ annihilates $\text{ker}( p_*)$ at $x$ then $0 = \alpha(\Lambda_{{g^{-1}}*} \eta) = \Lambda_{g^{-1}}^*\alpha( \eta)$, revealing that $\Lambda_{g^{-1}}^*\alpha$ annihilates $\text{ker}( p_*)$ at $\Lambda_g x$. In addition, if $\alpha$ at $x$ is related to $\wt\alpha$ at $[x]$, then $\Lambda_{g^{-1}}^*\alpha$  are also related to the same $\wt\alpha$, for all $g \in G$. This gives the characterization of the cotangent bundle of $\ca M/G$ as the following set of equivalence classes
\be\label{T*(M/G)}
T^*(\ca M/G) = \{ \alpha \sim \Lambda_{g^{-1}}^*\alpha; g\in G, \text{ where } \alpha \in T^*\ca M \text{ satisfies } \alpha(\text{ker}( p_*)) = 0 \}
\ee

Now we can particularize the results above to the case of interest, where $\ca M$ is taken to be space $\text{Riem}(D; \gamma)$ of Riemmanian metrics $h$ on the disc $D$ satisfying the desired boundary condition for the induced boundary metric, $h|_{\partial} = \gamma$, and $G$ is the group of boundary-trivial conformal transformations on the metric. The topology of $\text{Riem}(D; \gamma)$ can be defined by seeing it as an (open) subset of the vector space $\text{Sym}^0_2(D)$ of symmetric type-$\tensortype{0}{2}$ tensors on $D$ (which can be assumed to be some Sobolev space, although we shall not worry about these details). Naturally a tangent vector $\xi$ at any point $h \in \text{Riem}(D; \gamma)$ can be identified with $\xi_{ab} \in \text{Sym}^0_2(D^*)$, where $\text{Sym}^0_2(D^*)$ is the subset of $\text{Sym}^0_2(D)$ for which $\xi_{ab}t^at^b = 0$ for any vector $t^a$ tangent to the boundary $S^1$. The tangent bundle of $\text{Riem}(D; \gamma)$ is trivial and equal to
\be
T[\text{Riem}(D; \gamma)] = \text{Riem}(D; \gamma) \times \text{Sym}^0_2(D^*)
\ee
Dual vectors $\alpha$ at any point $h \in \text{Riem}(D; \gamma)$, taken to be continuous linear functions from vectors to $\bb R$, can be naturally identified with symmetric type-$\tensortype{2}{0}$ tensors $\alpha_{ab}$ on $D$. The pairing will be defined as
\be
\alpha(\xi) := \int \vartheta_h \alpha^{ab} \xi_{ab}
\ee
where $\vartheta_h$ is the volume form associated with $h$.\footnote{As the boundary $S^1$ has measure zero in this integration, the boundary condition on the vectors implies no constraints on the dual vectors, which is why they are identified with matrices in $\text{Sym}^2_0(D)$ instead of $\text{Sym}^2_0(D^*)$.}
The cotangent bundle has the trivial structure 
\be
T^*[\text{Met}(\ca M)] = \text{Riem}(D; \gamma) \times \text{Sym}^2_0(D)
\ee
The group of boundary-trivial conformal transformations, $G = \text{Con}(D^*)$, act on the metric as $\Lambda_{(\Psi, \Omega)}h := \Psi_*\Omega h$, where $\Psi$ is a diffeomorphism of the disc acting as the identity at the boundary and $\Omega$ is a positive real function that equals $1$ at the boundary. The kernel of $p_*$ is composed of vectors tangent to the orbits of this action, i.e., the vectors in $\text{Riem}(D; \gamma)$ induced from the algebra of $G$. That is, if $X$ is an element of the algebra of $G$, the induced vector $\eta^X$ is
\be
\eta^X_{ab} := \left.\frac{\partial}{\partial t} \Lambda_{\exp(tX)}h_{ab}\right|_{t=0}
\ee
and this is in $\text{ker}( p_*)$. We can separately consider vectors induced from infinitesimal diffeomorphisms and from infinitesimal Weyl scalings. Let us begin with (boundary-trivial) Weyl scalings, where a typical algebra element is denoted by $W$ and the exponentiation defines a curve on the group $\exp(tW) = (I, e^{tw})$, in which $w$ is a real function on the disc vanishing at the boundary. The induced vector at a point $h$ in $\text{Riem}(D; \gamma)$ is
\be
\eta^W_{ab} = \left.\frac{\partial}{\partial t} e^{tw} h_{ab}\right|_{t=0} = w h_{ab}
\ee
Next, an infinitesimal (boundary-trivial) diffeomorphism is labeled by a vector field $V$ on the disc, vanishing at the boundary, and the exponential defines a curve on the group $\exp(tV) = (\Psi_{tV}, 1)$, in which $\Psi_{tV}$ is the diffeomorphism defined by running along the integral curves of $V$ for a parameter length $t$. The induced vector at a point $h$ in $\text{Riem}(D; \gamma)$ is
\be
\eta^V_{ab} = \left.\frac{\partial}{\partial t} \Psi_{tV*} h_{ab}\right|_{t=0} = - \pounds_V h_{ab}= - 2 \nabla_{(a}V_{b)}
\ee
where $\nabla$ is the derivative associated with $h$. 
The set of vectors $\eta^W$ and $\eta^V$, for all $W$ and $V$, spans the kernel of $p_*$. With this we can determine the space of $\text{ker}( p_*)$-annihilating 1-forms at $h$, thus providing a characterization for 1-forms at $[h]$ according to \eqref{T*[x](M/G)}. So
\be
0 = \alpha(\eta^W) = \int \vartheta_h \alpha^{ab} w h_{ab}
\ee
for all $w$ implies that $\alpha$ must be traceless with respect to $h$, i.e., $\alpha^{ab}h_{ab} = 0$; and
\be\label{intbypartsetaV}
0 = \alpha(\eta^V) = - 2\int \vartheta_h \alpha^{ab} \nabla_{(a}V_{b)} = 2\int  \vartheta_h \nabla_a\alpha^{ab} V_b
\ee
for all $V$ implies that $\alpha$ must be divergenceless with respect to $h$, i.e., $\nabla_a\alpha^{ab}= 0$. Note that there is no boundary term in the integration by parts performed above because $V$ vanishes at the boundary. Those conditions on $\alpha$ are clearly familiar to us: if $\alpha^{ab}$ is interpreted as an extrinsic curvature on the disc, the traceless condition would correspond to the gauge-fixing of time by CMC surfaces (which leaves only the traceless part of the extrinsic curvature as dynamical), and the divergenceless condition would correspond to the momentum constraint.

Finally, we are interested in describing the cotangent bundle of $\text{Riem}(D; \gamma)/\text{Con}(D^*)$ globally according to \eqref{T*(M/G)}. To this end, it only remains to compute the pullback of 1-forms by $\Lambda_{(\Psi, \Omega)}$. But first we should compute the push-forward of a vector $\xi_{ab}$ by $\Lambda_{(\Psi, \Omega)}$. Naturally $\xi_{ab}$ is tangent at $h_{ab}$ to the curve $h_{ab} + t \xi_{ab}$, so its push-forward should be tangent to the curve $\Psi_*\Omega(h_{ab} + t \xi_{ab})$, that is,
\be
\Lambda_{(\Psi, \Omega)*}\xi_{ab} = \Psi_* \Omega \xi_{ab}
\ee
If $\alpha$ is a 1-form at $h$, its pull back $\Lambda_{(\Psi, \Omega)}^*\alpha$ is based at $\Lambda_{(\Psi, \Omega)}^{-1}h = \Omega^{-1}\Psi_*^{-1}h$ and satisfies
\be
\Lambda_{(\Psi, \Omega)}^*\alpha(\xi) = \alpha(\Lambda_{(\Psi, \Omega)*}\xi) = \int \vartheta_h \alpha^{ab} \Psi_* \Omega \xi_{ab}
\ee
where here $\xi$ is a vector at $\Lambda_{(\Psi, \Omega)}^{-1}h$. Using that $\vartheta_{\Omega^{-1} \Psi_*^{-1} h} =  \Omega^{-1} \Psi_*^{-1} \vartheta_h$, we can rewrite the last integral as
\be
\int \vartheta_h \alpha^{ab} \Psi_* \Omega \xi_{ab} = \int \Psi_* \left[ \vartheta_{\Omega^{-1} \Psi_*^{-1} h} ( \Omega^2 \Psi_*^{-1} \alpha^{ab}) \xi_{ab} \right]
\ee
and from the diffeomorphism invariance of integrals,
\be
\Lambda_{(\Psi, \Omega)}^*\alpha(\xi) = \alpha(\Lambda_{(\Psi, \Omega)*}\xi) = \int \vartheta_{\Omega^{-1} \Psi_*^{-1} h} ( \Omega^2 \Psi_*^{-1} \alpha^{ab}) \xi_{ab}
\ee
which allows us to read off
\be
(\Lambda_{(\Psi, \Omega)}^*\alpha)^{ab} = \Omega^2 \Psi_*^{-1} \alpha^{ab}
\ee
or, taking the inverse transformation,
\be
(\Lambda_{(\Psi, \Omega)^{-1}}^*\alpha)^{ab} =  \Psi_* \Omega^{-2} \alpha^{ab}
\ee
Therefore, as a point in $T^*\text{Riem}(D; \gamma)$ is labeled by the pair $(h_{ab}, \alpha^{ab})$, the characterization \eqref{T*(M/G)} for the cotangent bundle of  $\text{Riem}(D; \gamma)/\text{Con}(D^*)$ gives
\begin{eqnarray}
T^*(\text{Riem}(D; \gamma)/\text{Con}(D^*)) &=& \{ (h_{ab}, \alpha^{ab}) \sim (\Psi_*\Omega h_{ab}, \Psi_* \Omega^{-2} \alpha^{ab}) ; \nonumber\\
&&(\Psi, \Omega) \in \text{Con}(D^*), \text{ where } (h_{ab}, \alpha^{ab}) \in T^*\text{Riem}(D; \gamma) \nonumber\\
&& \text{ satisfies } h_{ab} \alpha^{ab} = 0 \text{ and } \nabla_a^{(h)} \alpha^{ab} = 0 \}
\end{eqnarray}
This is precisely the characterization of the reduced phase space for the causal diamonds displayed in \eqref{RPSequiv}, which proves the claim that
\be
\wt{\ca P} =  T^*(\text{Riem}(D; \gamma)/\text{Con}(D^*))
\ee
We stress that this result relies on that fact that the group of conformal transformations acts trivially at the boundary, as it was important that there was no boundary term coming from Stokes' theorem in \eqref{intbypartsetaV}.

\subsection{The space of conformal geometries on a disc}
\label{subsec:confgeodisc}

Now we wish to provide a more direct characterization for the configuration space, i.e., the space of conformal geometries on a disc $\ca Q := \text{Riem}(D; \gamma)/\text{Con}(D^*)$. 
According to the {\sl Riemann mapping theorem}, in complex analysis, given any non-empty simply-connected open subset of the complex plane $\mathbb C$, which is not all of $\mathbb C$, there exists a biholomorphic map onto the open unit disc $\bb D := \{z \in \mathbb C , |z| < 1 \}$. The {\sl Carath\'eodory theorem} extends this theorem to closed sets, stating that if $f$ maps the open unit disc $\bb D$ conformally onto the (bounded) open set $U \subset \mathbb C$, then $f$ has a continuous one-to-one extension to a function from the closure of $\bb D$, $\overline{\bb D}$, onto the closure of $U$, $\overline U$, if (and only if) $\partial U$ is a Jordan curve\footnote{A {\sl  Jordan curve} is a simple closed curve, i.e., a non-self-intersecting continuous loop on the plane.}. This can be applied to our problem, yielding that for any two matrics $h$ and $h'$ there exists an orientation-preserving diffeomorphism $\Psi$ and a Weyl factor $\Omega$ such that $h' = \Psi_*\Omega h$. As before, we shall denote a general conformal transformation by the pair $(\Psi, \Omega)$. See App.~\ref{app:unimap} for a review of this theorem, including an algorithm for explicitly constructing the pair $(\Psi, \Omega)$.

This means that if we had no constraint on the form of $(\Psi, \Omega)$ at the boundary, all metrics on the disc would be conformally equivalent. In other words, $\text{Riem}(D; \gamma)/\text{Con}(D)$ consists of a single point. However, because of the constraints, some states will no longer be conformally connected, so $\text{Riem}(D; \gamma)/\text{Con}(D^*)$ will be non-trivial. 
In order to determine it, note that $\ca Q =\text{Riem}(D; \gamma)/\text{Con}(D^*)$ is a homogeneous space for the group $\diff$. Namely, given a diffeomorphism acting on the boundary of the disc $\psi \in \diff$, define its left action on a conformal geometry $[h] \in \text{Riem}(D; \gamma)/\text{Con}(D^*)$ by
\be
\psi[h] := [\Psi_* \Omega h]
\ee
where $\Psi \in \text{Diff}^+(D)$ is any extension of $\psi$ to the interior of the disc, so that $\Psi|_\partial = \psi$ and $\Omega$ is any positive function on the disc such that its boundary value $\Omega|_\partial$ satisfies
\be
\Omega|_\partial  \gamma = \psi^{-1}_* \gamma
\ee
that is, so that the boundary condition on the metric is preserved. This action is well-defined, for suppose one chooses different extensions $\Psi'$ and $\Omega'$, associated with the same $\psi$, then
\ba
\psi[h] &= [\Psi'_* \Omega' h] \no
&= [(\Psi' \circ \Psi^{-1})_* \Psi_* (\Omega' \Omega^{-1}) \Omega h] \no
&= [(\Psi' \circ \Psi^{-1})_*  ((\Omega' \Omega^{-1}) \circ \Psi^{-1}) \Psi_*\Omega h] \no
&= [\Psi_*\Omega h]
\ea
where from the second to the third line we used the relation $\Psi_* \Omega T = (\Psi_*\Omega) \Psi_* T = (\Omega \circ \Psi^{-1}) \Psi_* T$, for any diffeomorphism $\Psi$, multiplicative scalar $\Omega$ and tensor $T$; and from the third to the fourth line we used that $\Psi' \circ \Psi^{-1}$ and $(\Omega' \Omega^{-1}) \circ \Psi^{-1}$ are trivial at the boundary, due to the boundary conditions, so their action is within the $\text{Con}(D^*)$ classes. Thus the action depends only on $\psi$, not on the extensions. It also does not depend on the metric representative, i.e., $\psi[h] = \psi[\Phi_* \Lambda h]$ if $(\Phi, \Lambda) \in \text{Con}(D^*)$, stabilising that the action is indeed well-defined. 
Finally, note that this action is transitive on $\ca Q$, due to Riemann mapping theorem, so that $\ca Q$ is a homogeneous space for $\diff$. Consequently, $\ca Q = \diff/H$ for some little group $H \in \diff$. The little group is the subgroup of $\diff$ that leaves any particular metric $h$ on the disc invariant.\footnote{The group of conformal isometries for any two metrics on the disc are homomorphic: if $(\Upsilon, \Theta)$ is a conformal isometry for the metric $h$, then $(\Psi, \Omega) (\Upsilon, \Theta) (\Psi, \Omega)^{-1}$ is a conformal isometry for $h' := \Psi_*\Omega h$.} For convenience, and without loss of generality, choose the metric $\bar h$ corresponding to the Euclidean round disc,
described in polar coordinates as
\be
\bar h = dr^2 + r^2d\theta^2
\ee
where $r \in [0,\ell/2\pi]$ and $\theta = [0, 2\pi)$.  If $\psi \in H$, then it must extend to a {\sl conformal isometry} of the disc, 
\be
\text{ConIso}(D) := \text{ConIso}(\bar h) := \{ (\Upsilon, \Theta) \in \text{Con}(D),\,\, \Upsilon_* \Theta \bar h = \bar h\}
\ee
and any conformal isometry of $\bar h$ must correspond to a $\upsilon = \Upsilon|_\partial\in H$. Therefore, we conclude that $H$ is isomorphic to $\text{ConIso}(D)$, so that
\be\label{CGJ}
\ca Q = \text{Riem}(D; \gamma)/\text{Con}(D^*) = \diff/\text{ConIso}(D)
\ee
where $\text{ConIso}(D)$ is seen as a subgroup of $\diff$, defined by restricting $(\Upsilon, \Theta) \mapsto \Upsilon|_\partial$, and the quotient is from the right, i.e., $\psi \sim \psi \circ \upsilon$ iff $\upsilon \in \text{ConIso}(D)$.

The group of conformal isometries of the round Euclidean disc is known to be $\psl$ \cite{bak2010complex}.
Here we present a quick review of the proof. We assume for simplicity that $\ell = 2\pi$, so the disc has unit radius --- the group of conformal isometries is clearly insensitive to the boundary length. 
The idea is to rephrase the problem in the language of complex analysis: the unit Euclidean disc can be naturally identified with the unit complex disc
\be
\bb D = \{z \in \bb C,\,\, |z| \le 1\}
\ee
with the standard flat metric on $\bb C$. An (orientation-preserving) conformal isometry of the disc then translates into a biholomorphic map $f: \bb D \rightarrow \bb D$, as can be seen in App. \ref{app:unimap}, particularly equation \eqref{conisoanalytic}.
Any such $f$, which we call a {\sl conformal automorphism} of $\bb D$, must map the interior of $\bb D$ onto itself, $|f(z)| < 1 \iff |z|<1$, and its boundary $\partial \bb D$ onto itself, $|f(z)| = 1 \iff |z|=1$. Consider the Mobius transformation given by
\begin{equation}
M(z) = \frac{z-b}{1 - \bar b z}
\end{equation}
where $b := f(0) \in \bb D$, with $|b|<1$. Notice that this is a conformal automorphism of $\bb D$, as $|M(z)| = 1 \iff |z| = 1$ and $M(0) = -b \in \text{\sl int}(\bb D)$. Since the space of conformal transformations form a group (with the multiplication being the map composition), we have that the map $F = M \circ f$ is a conformal automorphism of $D$. Moreover, it satisfies $F(0) = 0$. By Schwarz lemma, applied to both $F$ and $F^{-1}$, we must have
\begin{equation}
F(z) = e^{\imath \beta} z
\end{equation}
for some $\beta \in \mathbb R$. Hence,
\begin{equation}
f(z) = M^{-1} \circ F (z) = M^{-1}(e^{\imath \beta} z) =  e^{\imath \beta}\, \frac{ z+e^{-\imath \beta}b}{1+e^{\imath \beta}\bar b  z} =  e^{\imath \beta}\, \frac{ z - a}{1 - \bar a  z}
\end{equation}
where $a := - e^{-\imath \beta}b$ is just another number in $\bb D$ satisfying $|a|<1$. This is, therefore, the most general form of a conformal automorphism of the unit disc, characterizing $\text{ConIso}(D)$.

Note that $\text{ConIso}(D)$ forms a 3-dimensional group with topology $S^1 \times \mathbb R^2$. In fact, this group is precisely $\psl = \text{SL}(2,\bb R)/Z_2$, the projective special linear group in two real dimensions, i.e., consisting of $2\!\times\!2$ real matrices with unit determinant modded by the center $\{I, -I\}$.  The way in which $\psl$ appears most explicitly is by studying the group of conformal automorphisms of the complex upper plane
\be
\bb H = \{z \in \bb C,\,\, \text{\sl Im}(z) \ge 0\}
\ee
The group of automorphisms consists of transformations
\be
z \mapsto \frac{az + b}{cz + d}
\ee
where $a,b,c,d \in \bb R$ and $ad - bc > 0$. But notice that an overall scaling of $(a,b,c,d) \mapsto \lambda (a,b,c,d)$ does not affect a transformation, so we can restrict to $ad - bc = 1$; however, this leaves a residual $(a,b,c,d) \mapsto -(a,b,c,d)$ that needs to be modded out. This set of transformations thus corresponds to $\psl$. 
Since $\bb H$ is conformally equivalent to $\bb D$ (i.e., they  are related by a biholomorphic map),  $\text{ConIso}(D)$ is homomorphic to $\psl$. 

The group of conformal automorphisms of the disc can be seen as a subgroup of $\diff$ by restricting its action to the boundary. 
If a point at coordinate $\theta$ in $\partial\Sigma$ is represented by $z = e^{\imath\theta}$ in $D$, then
\begin{equation}
f(e^{\imath\theta}) = e^{\imath \beta}\, \frac{ e^{\imath\theta} - \rho e^{\imath\alpha}}{\rho e^{\imath (\theta - \alpha)} - 1} = e^{\imath(\theta + \beta + 2\gamma)}
\end{equation}
where $a = \rho e^{\imath\alpha}$, with $\rho \in [0,1)$ and $\alpha \in [0, 2\pi)$, and
\begin{equation}
\gamma = \mathrm{arcsin}\! \left[ \frac{\rho \sin (\theta - \alpha)}{\sqrt{1 + \rho^2 - 2\rho \cos (\theta - \alpha)}} \right]
\end{equation}
with $\gamma \in (-\pi/2,\pi/2]$.
Hence, the associated map $\upsilon = \Upsilon|_\partial$ on $\partial\Sigma$, defined from $e^{i\upsilon(\theta)} = f(e^{i\theta})$, is
\begin{equation}
\upsilon(\theta) = \theta + \beta + 2\, \mathrm{arcsin}\! \left[ \frac{\rho \sin (\theta - \alpha)}{\sqrt{1 + \rho^2 - 2\rho \cos (\theta - \alpha)}} \right]
\end{equation}
which sits inside $\diff$.

\section{Reduction via conformal coordinates}
\label{sec:Rcc}

In this section we consider another approach to the phase space reduction. This is based on a suitable ``coordinate change'' motivated by the previous result: instead of parametrizing the configuration space by spatial metrics, we parametrize it by conformal maps, as we will explain. This approach serves as a confirmation for the previous results and it also has the advantage of providing an explicit map between the physical, gauge-invariant degrees of freedom and the more concrete (but redundant) geometrical variables such as the spatial metric and extrinsic curvature.

\subsection{Conformal coordinates}
\label{subsec:confcoord}

By virtue of the uniformization theorem, any Riemannian metric $h_{ab}$ on $\Sigma \sim D$ can be obtained from a reference metric $\bar h_{ab}$ via some conformal transformation. That is, there exists a (orientation-preserving) diffeomorphism $\Psi : D \rightarrow D$ and a positive scalar $\Phi : D \rightarrow \bb R^+$ such that
\be\label{confmap}
h_{ab} = \Psi_* \Phi \bar h_{ab}
\ee
Because of the boundary condition on $h$,  $h|_{\partial D} = \gamma$, the boundary value of $\Phi$ is determined from the boundary action of $\Psi$, $\psi := \Psi|_{\partial D}$,
\be\label{PhiBC}
\left. \Phi \bar h \right|_{\partial D} = \psi^{-1}_* \gamma
\ee
In this way, we can use conformal maps $(\Psi, \Phi) \in \text{Diff}^+(D) \times_\gamma C^\infty(D, \bb R^+)$, where $\times_\gamma$ indicates that the boundary condition \eqref{PhiBC} on $\Phi$ is satisfied,\footnote{More precisely, $\text{Diff}^+(D) \times_\gamma C^\infty(D, \bb R^+) := \{ (\Psi, \Phi) \in \text{Diff}^+(D) \times C^\infty(D, \bb R^+); \left. \Phi \bar h \right|_{\partial D} = \psi^{-1}_* \gamma\}$.} as ``coordinates'' for the configuration space $\ca Q \sim \text{Riem}(D, \gamma)$. Note that the map $(\Psi, \Phi)$ transforming $\bar h$ into $h$ is not unique, since we can always compose it (on the right) with a conformal isometry $(\Psi_0, \Phi_0)$ of $\bar h$, that is, if $\bar h_{ab} = \Psi_{0*} \Phi_0 \bar h_{ab}$, then $(\Psi', \Phi') := (\Psi, \Phi) \circ (\Psi_0, \Phi_0)$ also maps $\bar h$ into $h$. But this is not a problem since these ``conformal coordinates'' still cover the whole configuration space, and this non-uniqueness only amounts to additional gauge ambiguities being introduced in the description, which will all be removed in the end.

To cover the phase space we need also the momentum ``coordinates'' associated with the extrinsic curvature. As our choice of time gauge has eliminated the trace-part of $K_{ab}$, we must consider only its traceless part, $\sigma^{ab}$. It is convenient to use the conformal map $(\Psi, \Phi)$ to transform $\sigma^{ab}$ to the reference disc $D$. So we define the transformed traceless extrinsic curvature, $\bar\sigma$, as
\be
\bar\sigma^{ab} := \Phi^{-s} \Psi_*^{-1} \sigma^{ab}
\ee
where $s$ is a power which will be chosen so as to simplify the momentum constraint. This can be equivalently written as
\be
\sigma^{ab} =  \Psi_* \Phi^s \bar\sigma^{ab}
\ee
evidencing the isomorphism between $\sigma$'s and $\bar\sigma$'s defined by the conformal map $(\Psi, \Phi)$. In this way, we see that the triplet $(\Psi, \Phi, \bar\sigma)$ can be used as ``coordinates'' for the (unconstrained) phase space. This can be seen as an {\sl enlargement} of the phase space from $\ca P = \text{Riem}(D, \gamma) \times \text{Sym}(D, (2,0))$ to 
\be
\bar{\ca P} := \text{Diff}^+(D) \times_\gamma C^\infty(D, \bb R^+) \times \text{Sym}(D, (2,0))
\ee
where $\times_\gamma$ refers to condition \eqref{PhiBC}. As mentioned before, this enlargement consists of extra gauge directions being introduced, which will be eventually removed.

\subsection{Imposing the constraints}
\label{subsec:impconst}

Now we impose the constraints. First, consider the momentum constraint, \eqref{sigmaM}, which in the CMC gauge becomes
\be\label{sigmatransv}
\nabla_a \sigma^{ab} = 0
\ee
that is, $\sigma$ must be {\sl transverse} with respect to $h$. Now we want to investigate how this condition translates when transformed to $D$, i.e., when expressed in terms of $\bar h$ and $\bar \sigma$. Since $\nabla$ is covariantly constructed from $h$, equation \eqref{sigmatransv} transforms nicely under diffeomorphisms,
\be
\Psi^{-1}_* \left( \nabla^{(h)}_a \sigma^{ab} \right) = \nabla^{(\Psi^{-1}_*h)}_a\! \left(\Psi^{-1}_* \sigma^{ab}\right) = 0
\ee
so we have
\be
\nabla^{(\Phi \bar h)}_a\! \left(\Phi^s \bar\sigma^{ab}\right) = 0
\ee
The left-hand side corresponds to a Weyl transformation, which yields
\be
\nabla^{(\Phi \bar h)}_a\! \left(\Phi^s \bar\sigma^{ab}\right) = \Phi^s \bar\nabla_a \bar\sigma^{ab} + (s + 2) \Phi^{s-1} \bar\sigma^{ab} \bar\nabla_a \Phi - \frac{1}{2} \Phi^{s-1} \bar h_{cd} \bar\sigma^{cd} {\bar h}^{ab} \bar\nabla_a \Phi = 0
\ee
where $\bar\nabla$ is the covariant derivative associated with $\bar h$. The last term vanishes because the tracelessness of $\sigma$ (with respect to $h$) implies that $\bar\sigma$ is traceless (with respect to $\bar h$). Thus we see that the choice $s = -2$ is particularly convenient as it makes condition \eqref{sigmatransv} equivalent to
\be\label{barsigmaM}
\bar\nabla_a \bar\sigma^{ab} = 0
\ee
that is, $\bar\sigma$ must be transverse with respect to $\bar h$. Second, consider the Hamiltonian constraint \eqref{sigmaH}, which in the CMC slice with $K = -\tau$ becomes
\be\label{sigmaH2}
\sigma^{ab}\sigma_{ab}  - R^{(h)} - \chi= 0
\ee
where $\chi = -2\Lambda + \frac{1}{2} \tau^2$ and $R^{(h)}$ is the Ricci scalar associated with $h$. As before we apply $\Psi^{-1}_*$ to this expression to get
\be
(\Psi^{-1}_* h_{ac})( \Psi^{-1}_* h_{bd})( \Psi^{-1}_*\sigma^{ab})( \Psi^{-1}_*\sigma^{cd})  - R^{(\Psi^{-1}_* h)} - \chi= 0
\ee
that is,
\be
(\Phi \bar h_{ac})( \Phi \bar h_{bd})( \Phi^{-2} \bar \sigma^{ab})( \Phi^{-2}\bar \sigma^{cd})  - R^{(\Phi \bar h)} - \chi= 0
\ee
which yields
\be
\bar\nabla^2 \lambda - \bar R + e^{-\lambda} \bar\sigma^{ab} \bar\sigma_{ab} - e^\lambda \chi = 0
\ee
where $\lambda = \log \Phi$, $\bar R$ is the Ricci scalar associated with $\bar h$ and $\bar\sigma^{ab} \bar\sigma_{ab}$ is contracted using $\bar h$. For convenience, we can choose $\bar h$ to be the metric on a flat unit disc
\be
\bar h = dr^2 + r^2 d\theta^2
\ee
where $(r, \theta) \in [0, 1] \times [0, 2\pi)$ are the usual polar coordinates on the disc. In this way, $\bar R = 0$. 

To sum up, we are considering a change of coordinates on the unconstrained phase space from $(h_{ab}, \sigma^{ab})$ to $(\Psi, \lambda, \bar\sigma^{ab})$ defined by
\begin{align}
h_{ab} &= \Psi_* e^\lambda \bar h_{ab} \label{hconftrans}\\
\sigma^{ab} &=  \Psi_* e^{-2\lambda} \bar\sigma^{ab}\label{sigmaconftrans}
\end{align}
where $\bar h_{ab}$ is the Euclidean metric on the reference disc $D$ and $\bar\sigma^{ab}$ is a traceless (with respect to $\bar h$) symmetric tensor on $D$. Because of the boundary condition on $h$,  $h|_{\partial \Sigma} = \gamma$, the boundary value of $\lambda$ is determined from the boundary action of $\Psi$, $\psi := \Psi|_{\partial D}$,
\be\label{lambdabdycond}
\left. e^\lambda \bar h \right|_{\partial D} = \psi^{-1}_* \gamma
\ee
The constraint surface $\ca S$ is determined by imposing condition \eqref{barsigmaM} on $\bar\sigma$ and taking $\lambda$ to satisfy
\be\label{lambdaLich}
\bar\nabla^2 \lambda + e^{-\lambda} \bar\sigma^{ab} \bar\sigma_{ab} - e^\lambda \chi = 0
\ee
The arguments of the previous sections imply that \eqref{lambdaLich} has unique solution depending on $\bar\sigma$ (via the $\bar\sigma^{ab} \bar\sigma_{ab}$ term) and on $\Psi$ (via the boundary conditions). Therefore, $\ca S$ can be ``covered'' with coordinates $(\Psi, \bar\sigma^{ab})$, where $\bar\sigma$ is traceless and transverse with respect to $\bar h$. More precisely, the map $(\Psi, \bar\sigma^{ab}) \mapsto (h_{ab}, \sigma^{ab})$ is a projection map from $\text{Diff}^+(D) \times \text{Sym}(D, (2,0); TT[\bar h])$, where $TT[\bar h]$ means ``traceless and transverse with respect to $\bar h$'', onto $\ca S$. 
Note that we can think of 
\be
\bar{\ca S} := \text{Diff}^+(D) \times \text{Sym}(D, (2,0); TT[\bar h])
\ee
as an {\sl enlargement} of $\ca S$, where additional gauge ambiguities have been introduced. In fact, it is equivalent to think that we are first enlarging $\ca P$ into $\bar{\ca P}$ and then applying the constraints to get $\bar{\ca S}$, or to think that we are first applying the constraints on $\ca P$ to get $\ca S$ and then enlarging it into $\bar{\ca S}$. For concreteness, let us call the projection from $\bar{\ca S}$ to $\ca S$ by $T : \bar{\ca S} \rightarrow \ca S$, so that
\be\label{barStoS}
T(\Psi, \bar\sigma^{ab}) = (h_{ab}, \sigma^{ab}) = (\Psi_* e^\lambda \bar h_{ab}, \Psi_* e^{-2\lambda} \bar\sigma^{ab})
\ee
where $\lambda$ satisfies \eqref{lambdaLich}. 

Now we want to pull-back the (pre)symplectic form $\omega$ on $\ca S$ to $\bar{\ca S}$,
\be
\bar\omega = T^*\omega
\ee
which corresponds to ``writing $\omega$ in $(\Psi, \bar\sigma)$ coordinates''. 
The tangent vector $\bar \eta$ at a point $(\Psi, \bar\sigma)$ of $\bar{\ca S}$ can be expressed as $(X, \alpha) \in T_{\Psi} \text{Diff}^+(D) \oplus T_{\bar\sigma} [\text{Sym}(D, (2,0); TT[\bar h])]$. Since $\text{Sym}(D, (2,0); TT[\bar h])$ is a vector space, it can be naturally identified with its tangent space, so that $T_{\bar\sigma}[\text{Sym}(D, (2,0); TT[\bar h])] \sim \text{Sym}(D, (2,0); TT[\bar h])$. Informally speaking, $X$ can be seen as the vector tangent to a one-parameter family of diffeomorphisms, $t \mapsto \Psi_t$. To be more precise, we can use the group structure of $\text{Diff}^+(D)$ to left-translate $X$ to the identity $I$, and note that the tangent space to $I$ can be naturally identified with vector fields on $D$ (satisfying the condition that the vectors at the boundary are tangent to it, since automorphisms of $D$ must map $\partial D$ into itself). Introducing some notation, let us call
\be
\bar X := l_{\Psi^{-1}*} X
\ee
where $l_{\Psi}(\Psi') = \Psi \circ \Psi'$ is the left multiplication on $\text{Diff}^+(D)$. As $\bar X \in T_{I} \text{Diff}^+(D)$, let us denote the associated vector field on $D$ by $\xi \in \text{Vect}_0(D)$, where $\text{Vect}_0(D)$ is the space of vector fields on $D$ satisfying the parallel boundary condition $\xi |_{\partial D} \in T(\partial D)$. In this way, we can define $X$ as being the vector tangent to the curve $t \mapsto \Psi \circ \Gamma_t$ at $t = 0$, where $\Gamma_t = \text{Exp}(t\xi)$ is the diffeomorphism corresponding to flowing along the integral curves of $\xi$ for a parameter $t$. Thus, we have the identification
\be
T_{(\Psi, \bar\sigma)} \ca S \sim \text{Vect}_0(D) \oplus \text{Sym}(D, (2,0); TT[\bar h])
\ee
which allows us to express $\bar\eta$ in the form 
\be
\bar\eta = (\xi^a, \alpha^{ab})
\ee
where $\xi \in \text{Vect}_0(D)$ and $\alpha \in \text{Sym}(D, (2,0); TT[\bar h])$. 
If $\bar\eta$ and $\bar\eta'$ are vectors tangent to $\bar{\ca S}$ at some point $(\Psi, \bar\sigma)$, and $\eta := T_* \bar\eta$ and $\eta' := T_* \bar\eta'$ are the corresponding pushed vector to $\ca S$, then
\be
\bar\omega (\bar\eta, \bar\eta') = \omega(\eta, \eta')
\ee
Since $\omega$ is the restriction to $\ca S$ of $\Omega$, given in \eqref{GRsym}, and $\eta$ and $\eta'$ are tangent to $\ca S$, we also have
\be
\bar\omega (\bar\eta, \bar\eta') = \Omega(\eta, \eta')
\ee
Note that $\Omega$ involves the conjugate momentum $\pi^{ab}$, which from definition \eqref{GRpi} and relation \eqref{sigmaKdef} can be written as
\be\label{pidef2}
\pi^{ab} = \sqrt{\text{det}(h)} \left( \sigma^{ab} + \frac{1}{2} \tau h^{ab} \right)
\ee
Taking the exterior derivative (on the unconstrained phase space $\ca P$) gives
\be
\delta\pi^{ab} = \sqrt{\text{det}(h)} \left[ \delta\sigma^{ab} + \frac{1}{2} \left( \sigma^{ab} h^{cd} - \tau h^{ac} h^{bd} + \frac{1}{2} \tau h^{ab} h^{cd}   \right) \delta h_{cd} \right]
\ee
and replacing this on $\Omega$ yields
\be\label{Omegarep}
\Omega = \int_D d^2\!x\, \delta \pi^{ab} \wedge \delta h_{ab} = \int_D \vartheta_h \left( \delta \sigma^{ab} + \frac{1}{2} \sigma^{ab} h^{cd} \delta h_{cd} \right) \wedge \delta h_{ab}
\ee
where
\be
\vartheta_h = d^2\!x \sqrt{\text{det}(h)}
\ee
is the natural volume form associated with $h$.\footnote{In a covariant language, given an orientable manifold $\ca M$ with orientation $n$-form $\vartheta$, the natural volume form $\vartheta_h$ associated with a metric $h$ is defined as $\vartheta_h = w \vartheta$, where the scalar $w > 0$ is such that $\vartheta_h(e_1, e_2, \ldots e_n) = 1$ for any (oriented) orthonormal basis $\{e_1, e_2, \ldots e_n \}$.}

In order to evaluate $\Omega(\eta, \eta')$, we compute the ``variations'' $\delta h_{ab}(\eta)$ and $\delta \sigma^{ab}(\eta)$ for a pushed vector $\eta = T_* \bar\eta$. First note that since $\text{Riem}(D; \gamma)$ is a subspace of $\text{Sym}(D, (0,2))$, we can think of a ``variation of $h$'' as a difference between nearby metrics, so that  $\delta h_{ab}(\eta) \in \text{Sym}(D, (0,2))$.\footnote{To say this more precisely, we are regarding $h_{ab}(x)$ as a function from $\ca S$ into $T_x^{(2,0), \text{sym}}D$, the space of symmetric $(2, 0)$ tensors at $x \in D$, where $h_{ab}(x)$ takes a point $p \in \ca S$ and returns the corresponding value of the metric at $x \in D$. Since the target space is a vector space, $\delta h_{ab}(x)$ can be defined in the usual manner, i.e., given a vector $\eta \in T_p\ca S$ we define $\delta h_{ab}(x)(\eta) := \eta(h_{ab}(x)) = \frac{d}{dt}h_{ab}(x)[p_t]$, where $p_t$ is a curve on $\ca S$ tangent to $\eta$.} Similarly, $\text{Sym}(D, (2,0))$ is a vector space, so that a ``variation of $\sigma$'' can also be thought of as a difference between nearby tensors and $\delta \sigma^{ab}(\eta) \in \text{Sym}(D, (2,0))$. Thus, for $\bar\eta = (\xi, \alpha) \in T_{(\Psi, \bar\sigma)} \bar{\ca S}$,  $\delta h_{ab}(\eta)$ and $\delta \sigma^{ab}(\eta)$ are given by
\begin{align}
\delta h_{ab}(\eta) &=  \frac{d}{dt} \left[ (\Psi \circ \Gamma_t)_* e^{\lambda_t} \bar h_{ab} \right] \\
\delta \sigma^{ab}(\eta) &= \frac{d}{dt} \left[ (\Psi \circ \Gamma_t)_* e^{-2\lambda_t} \bar \sigma_t^{ab} \right]
\end{align}
where the derivative is evaluated at $t = 0$, $\Gamma_t = \text{Exp}(t\xi)$, $\sigma_t^{ab} = \alpha^{ab} t$ and $\lambda_t$ is the solution of \eqref{lambdaLich} associated with $(\Psi \circ \Gamma_t, \bar\sigma_t)$. 
By distributing the derivative we get
\begin{align}
\delta h_{ab}(\eta) &= \Psi_* \left[ \frac{d}{dt} \Gamma_{t*} (e^{\lambda} \bar h_{ab}) + e^\lambda \bar h_{ab} \frac{d\lambda_t}{dt} \right] \\
\delta \sigma^{ab}(\eta) &= \Psi_* \left[ \frac{d}{dt} \Gamma_{t*} ( e^{-2\lambda} \bar \sigma^{ab}) - 2 e^{-2\lambda} \bar \sigma_t^{ab} \frac{d\lambda_t}{dt} + e^{-2\lambda}  \frac{d\bar \sigma_t^{ab}}{dt} \right]
\end{align}
which yields
\begin{align}
\delta h_{ab}(\eta) &= \Psi_* \left[ - \pounds_\xi (e^{\lambda} \bar h_{ab}) + \kappa e^\lambda \bar h_{ab} \right] \label{deltah} \\
\delta \sigma^{ab}(\eta) &= \Psi_* \left[ - \pounds_\xi ( e^{-2\lambda} \bar \sigma^{ab}) - 2 \kappa e^{-2\lambda} \bar \sigma^{ab} + e^{-2\lambda}  \alpha^{ab} \label{deltasigma} \right]
\end{align}
where $\kappa$ is defined as
\be\label{kappadef}
\kappa(x; \Psi, \bar\sigma, \eta) := \left. \frac{d\lambda_t}{dt} \right|_{t=0}
\ee
which is a function of $x \in D$ depending implicitly on $(\Psi, \bar\sigma)$ and $\eta$. Thus for the first term in \eqref{Omegarep} we obtain
\begin{align}
\delta\sigma^{ab} \wedge \delta h_{ab}(\eta, \eta') &= \Psi_* e^{-\lambda} \left[  \pounds_\xi \bar \sigma^{ab} \pounds_{\xi'} \bar h_{ab} - 2 ( \pounds_\xi \lambda - \kappa)  \bar\sigma^{ab} \pounds_{\xi'} \bar h_{ab} \,+ \right. \nonumber\\
&\left. +\, ( \pounds_{\xi'} \lambda - \kappa')  \pounds_{\xi} \bar\sigma^{ab} \bar h_{ab} -  \pounds_{\xi'} \bar h_{ab} \alpha^{ab} - \{ (\xi, \alpha) \leftrightarrow (\xi', \alpha') \} \right] 
\end{align}
where $\kappa' = \kappa(x; \Psi, \bar\sigma, \eta')$ and $\{ (\xi, \alpha) \leftrightarrow (\xi', \alpha') \}$ consists of the previous terms but with $(\xi, \alpha)$ exchanged with $(\xi', \alpha')$. Also, for the second term in \eqref{Omegarep}, we have
\begin{align}
\frac{1}{2}\sigma^{ab} h^{cd} \delta h_{cd} \wedge \delta h_{ab}(\eta, \eta') &= \Psi_* e^{-\lambda} \bigg[ \frac{1}{2} \bar\sigma^{ab} \pounds_{\xi'} \bar h_{ab} \bar h^{cd} \pounds_{\xi} \bar h_{cd} \,+ \nonumber\\
& +\, ( \pounds_{\xi} \lambda - \kappa)   \bar\sigma^{ab} \pounds_{\xi'}\bar h_{ab}  - \{ (\xi, \alpha) \leftrightarrow (\xi', \alpha') \} \bigg] 
\end{align}
Therefore,
\be
\Omega(\eta, \eta') = \int_D \vartheta_{e^\lambda \bar h} e^{-\lambda} \left[ \pounds_\xi \bar \sigma^{ab} \pounds_{\xi'} \bar h_{ab} - \pounds_{\xi'} \bar h_{ab} \alpha^{ab} + \frac{1}{2} \bar\sigma^{ab} \pounds_{\xi'} \bar h_{ab} \bar h^{cd} \pounds_{\xi} \bar h_{cd}  - \{ (\xi, \alpha) \leftrightarrow (\xi', \alpha') \} \right]
\ee 
where we have used that $\vartheta_h = \vartheta_{\Psi_* e^\lambda \bar h} = \Psi_* \vartheta_{e^\lambda \bar h}$ and that integrals are invariant under automorphisms, i.e., $\int A = \int \Psi_* A$. Note that the factor $( \pounds_\xi \lambda - \kappa)$ does not appear because it multiplies $ \pounds_{\xi'}(\bar\sigma^{ab}  \bar h_{ab})$, which vanishes due to the tracelessness of $\bar\sigma$. Now using that
\be\label{varthetarel}
\vartheta_{e^\lambda \bar h} = e^\lambda  \vartheta_{\bar h}
\ee
we get
\be
\Omega(\eta, \eta') = \int_D \vartheta_{\bar h} \left[ \pounds_\xi \bar \sigma^{ab} \pounds_{\xi'} \bar h_{ab} - \pounds_{\xi'} \bar h_{ab} \alpha^{ab} + \frac{1}{2} \bar\sigma^{ab} \pounds_{\xi'} \bar h_{ab} \bar h^{cd} \pounds_{\xi} \bar h_{cd}  - \{ (\xi, \alpha) \leftrightarrow (\xi', \alpha') \} \right]
\ee 
which is an integral defined entirely on the reference disc $(D; \bar h)$. Writing the Lie derivatives in terms of covariant derivatives (with respect to $\bar h$), i.e., $\pounds_\xi \bar h_{ab} = 2\bar\nabla_{(a} \xi_{b)}$ and $\pounds_\xi \bar\sigma^{ab} = \xi^c \bar\nabla_c \bar\sigma^{ab} - 2\bar\nabla_c \xi^{(a} \bar\sigma^{b)c}$, we obtain
\begin{align}\label{baromega}
\bar\omega (\bar\eta, \bar\eta') = \Omega(\eta, \eta') &= 2\int_D \vartheta_{\bar h} \bar\nabla_a \left[ \left( \xi_b \alpha'^{ab} - \xi'_b \alpha^{ab} \right) - \left( \xi^c \bar\nabla_c \xi'_b - \xi'_c \bar\nabla_c \xi_b \right) \bar\sigma^{ab} \right] \nonumber\\
&= 2\int_{\partial D} d\theta\, n_a \left[ \left( \xi_b \alpha'^{ab} - \xi'_b \alpha^{ab} \right) - [\xi, \xi']_b \bar\sigma^{ab} \right] 
\end{align}
In the second line we have used Gauss's law, so $n^a$ is a unit (outward-pointing) normal vector at $\partial D \sim S^1$ and $d\theta$ is the measure induced on $\partial D$. Since $\bar h$ was chosen to be the unit-radius metric on the disc, $d\theta$ is just the differential of the angle coordinate, $\theta \in [0, 2\pi)$.

Before fully appreciating the previous result, it is useful to understand better a quantity like $\int\! d\theta\, n_a \bar\sigma^{ab} \xi_b$, where $n$ is normal and $\xi$ is tangent to the boundary. In Cartesian coordinates $\{x, y\}$ on $(D, \bar h)$, the symmetry and traceless of $\bar\sigma$ implies that its components have generic form
\be
\bar\sigma^{\mu\nu} =  \left( \begin{array}{cc}
v & u \\
u & -v \\
\end{array} \right)
\ee
for $u, \, v \in C^\infty(D, \bb R)$. It is convenient to think of these components as parts of a complex function $f : D \rightarrow \bb C$ defined by
\be
f(z) := u(z) + i v(z)
\ee
where $z = x + iy$, $u(z) := u(x, y)$ and similarly for $v$. In this notation, we immediately see that the condition that $\bar\sigma$ is transverse to $\bar h$, $\partial_\mu \bar\sigma^{\mu\nu} = 0$, translates into the condition that $f$ is analytic,
\begin{align}
\frac{\partial u}{\partial x} &= \frac{\partial v}{\partial y} \nonumber\\
\frac{\partial u}{\partial y} &= - \frac{\partial v}{\partial x}
\end{align}
In particular this implies that $f$ is completely determined on $D$ from its value on the boundary $\partial D$. To see this note that $u$ and $v$ are harmonic functions, that is, they satisfy
\be
\frac{\partial^2 u}{\partial x^2} + \frac{\partial^2 u}{\partial y^2} = \frac{\partial^2 v}{\partial x^2} + \frac{\partial^2 v}{\partial y^2} = 0
\ee
Therefore, the general solution for $u$ on $D$ is
\be
u = \sum_{n \ge 0} r^n \left( a_n \cos n\theta + b_n \sin n\theta \right)
\ee
where $\{r, \theta\}$ are the usual polar coordinates (defined by $x = r\cos\theta$ and $y = r\sin\theta$). It is clear that $u$ is uniquely determined from its boundary value $\hat u(\theta) := u(r=1, \theta)$. Now note that 
\be
dv = \frac{\partial v}{\partial x} dx + \frac{\partial v}{\partial y} dy = - \frac{\partial u}{\partial y} dx + \frac{\partial u}{\partial x} dy
\ee
so that
\be
v(x, y) = v(0,0) + \int_\gamma \left( - \frac{\partial u}{\partial y} dx + \frac{\partial u}{\partial x} dy \right)
\ee
where $\gamma$ is any curve joining $(0,0)$ to $(x, y) \in D$. We see that $v$ is completely determined from $u$, modulo the choice of its value $v_0$ at the origin, 
\be
v = v_0 + \sum_{n \ge 0} r^n \left( b_n \cos n\theta - a_n \sin n\theta \right)
\ee
Note that $v_0$ can be absorbed in $b_0$, so that the general solution for $\bar\sigma$ can be written as
\be\label{sigmaab}
\bar\sigma^{\mu\nu} =  \sum_{n\ge 0} r^n \left[
a_n 
\left( \begin{array}{cc}
\sin n\theta & \cos n\theta \\
\cos n\theta & -\sin n\theta \\
\end{array} \right)
+ b_n 
\left( \begin{array}{cc}
-\cos n\theta & \sin n\theta \\
\sin n\theta & \cos n\theta \\
\end{array} \right)
\right]
\ee
for coefficients $a_n$ and $b_n$. This expansion \eqref{sigmaab} will be taken as defining the coefficients $a_n$ and $b_n$, which will be called the Fourier coefficients of $\bar\sigma$. Let $n$ be the unit normal vector field and $t = \partial_\theta$ be the unit tangent vector field on the boundary, $\partial D \sim S^1$. That is, at the point $(x, y) = (\cos\theta, \sin\theta)$, we have $n = \cos \theta \partial_x + \sin\theta \partial_y$ and $t = \cos\theta \partial_x - \sin\theta \partial_y$. Then,
\be
\bar\sigma^{ab} n_a t_b =  u \cos 2\theta - v\sin 2\theta  = \sum_{n\ge 0} \big[ a_n \cos(n+2)\theta + b_n \sin(n+2)\theta \big] 
\ee
If $\xi = f \partial_\theta$ is a tangent vector field on the boundary, where $f\in C^\infty(S^1, \bb R)$, we have
\be\label{sigmanxi}
\int\! d\theta\, \bar\sigma^{ab} n_a \xi_b =  \int\! d\theta\, \sum_{n\ge 0} \big[ a_n \cos(n+2)\theta + b_n \sin(n+2)\theta \big] f(\theta)
\ee
From the Fourier coefficients of $\bar\sigma$, we define the quadratic form $\ac\sigma$ on the boundary by
\be\label{hatsigma}
\ac\sigma(\theta) := - 2 \sum_{n\ge 0} \big[ a_n \cos(n+2)\theta + b_n \sin(n+2)\theta \big] d\theta^2
\ee
This provides a convenient identification between $\bar\sigma^{ab}$ (a symmetric, traceless and transverse tensor on $D$) and $\ac\sigma$ (a quadratic form on $\partial D \sim S^1$ ``missing'' the Fourier modes $1$, $\sin\theta$ and $\cos\theta$). It is also convenient to introduce some notation for this space of $\ac\sigma$'s. The Lie algebra of $\diff$, denoted $\adiff$, is naturally represented by vector fields $\hat\xi$ on $S^1$. We can write it as $\hat\xi = f \partial_\theta$, where $f \in C^\infty(S^1, \bb R)$. The dual Lie algebra, denoted $\ddiff$, is naturally represented by quadratic forms $\alpha$ on $S^1$. We can write it as $\alpha = a(\theta) d\theta^2$, where $a \in C^\infty(S^1, \bb R)$. The reason for representing dual vectors in this way is because it leads to the natural pairing
\be\label{pairing}
\alpha(\hat\xi) := \int a d\theta^2(f\partial_\theta) = \int d\theta \, a(\theta) f(\theta)
\ee
and this will simplify certain formulas (especially in the part about coadjoint orbits). Now observe that the space of $\ac\sigma$'s, as defined in \eqref{hatsigma}, is the subspace $\whddiff$ of $\ddiff$ ``missing'' the Fourier modes $1$, $\sin\theta$ and $\cos\theta$. Another way to characterize $\whddiff$ is to notice that its elements annihilate the generators $\hat\xi = \partial_\theta,\, \sin\theta\,\partial_\theta, \, \cos\theta\,\partial_\theta$ of $\apsl \subset \adiff$, that is,
\be\label{tildediff*}
\whddiff := \{ \ac\sigma \in \ddiff;\, \text{where } \ac\sigma(\hat\xi) = 0 \text{ for all } \hat\xi \in \apsl \subset \adiff \}
\ee
To sum up, we have identified
\be
\text{Sym}(D, (2,0); TT[\bar h]) = \whddiff
\ee
in a natural way. In this language, the quantity in \eqref{sigmanxi} can be expressed as
\be\label{hatsigmadef}
\int\! d\theta\, \bar\sigma^{ab} n_a \xi_b =  - \frac{1}{2} \ac\sigma(\hat\xi)
\ee
where $\hat\xi \in \adiff$ is simply the restriction of $\xi$ to $\partial D \sim S^1$, which is well-defined since $\xi$ is tangent to $\partial D$. Correspondingly, the symplectic form in \eqref{baromega} takes the form
\be
\bar\omega (\bar\eta, \bar\eta') = \ac\alpha(\hat\xi') - \ac\alpha'(\hat\xi) + \ac\sigma([\hat\xi, \hat\xi'])
\ee
where $\ac\alpha \in \whddiff$ is related to $\alpha^{ab}$ in the same way as in \eqref{hatsigma}, and analogously for $\hat\alpha'$ and $\alpha'^{ab}$. Note that the Lie brackets is independent on taking the restriction to $\partial D$ before or after computing the bracket, i.e., $[\hat\xi, \hat\xi'] = \widehat{[\xi, \xi']}$.

\subsection{Removing the bulk diffeomorphisms}
\label{subsec:bulkdiff}

The expression above for the symplectic form on $\bar{\ca S} = \text{Diff}^+(D) \times \text{Sym}(D, (2,0); TT[\bar h])$ clearly reveals a ``huge'' part of the gauge ambiguities. That is, bulk diffeomorphisms acting trivially on the boundary are pure gauge transformations. In particular, this means that any $\bar \eta = (\xi, \alpha)$ where $\xi$ {\sl vanishes at the boundary} corresponds to a degenerate direction of $\bar\omega$. Therefore, any two points $(\Psi, \bar\sigma)$ and $(\Psi', \bar\sigma)$ in $\bar{\ca S}$ such that the diffeomorphisms $\Psi$ and $\Psi'$ have the same boundary action, $\Psi |_{\partial D} = \Psi' |_{\partial D}$, correspond to the same physical state. Thus, we have the following reduction
\begin{align}
\text{Diff}^+(D) \times \text{Sym}(D, (2,0); TT[\bar h]) &\rightarrow \diff \times \whddiff \nonumber\\
(\Psi, \bar\sigma^{ab}) &\mapsto (\psi, \ac\sigma)
\end{align}
where $\psi \in \diff$,
\be
\psi := \Psi \bigg|_{\partial D}
\ee
is the boundary action of $\Psi$. Note that the quotient is really acting only on this diffeomorphism factor, $\Psi \mapsto \psi$, while the association $\bar\sigma^{ab} \mapsto \ac\sigma$ is just the isomorphism discussed above. For concreteness, let us refer to this phase space as 
\be
\hat{\ca S} := \diff \times \whddiff
\ee
and to the corresponding quotient map as $R : \bar{\ca S} \rightarrow \hat{\ca S}$. The symplectic form on $\hat{\ca S}$, denoted by $\hat\omega$, must satisfy $\bar\omega = R^* \hat\omega$. First note that the push-forward of a vector $\bar\eta = (\xi, \alpha^{ab})$ in $\bar{\ca S}$ is given simply by\footnote{The tangent space to a point $(\psi, \ac\sigma) \in \hat{\ca S}$ is being characterized in a way analogous to how we characterized the tangent space for $\bar{\ca S}$, that is, $T_{(\psi, \ac\sigma)} \hat{\ca S} \sim \adiff \oplus \whddiff$, so a generic vector is represented by a pair $\hat\eta = (\hat\xi, \hat\alpha)$, where $\hat\xi$ is a vector field on $S^1$ and $\hat\alpha$ is an element of $\whddiff$.\label{fn:TbarS}}
\be
R_*(\xi, \alpha^{ab}) = (\hat\xi, \ac\alpha)
\ee
where $\hat\xi$ is, as before, the restriction of $\xi$ to the boundary. Thus, one can easily verify that
\be\label{hatomega}
\hat\omega(\hat\eta, \hat\eta') = \ac\alpha(\hat\xi') - \ac\alpha'(\hat\xi) + \ac\sigma([\hat\xi, \hat\xi'])
\ee
defines the desired symplectic form $\hat\omega$ on $\hat{\ca S}$, where $\hat\eta = (\hat\xi, \ac\alpha)$ and $\hat\eta' = (\hat\xi', \ac\alpha')$ are tangent vectors at $(\psi, \ac\sigma)$. 

We must investigate if there are remaining gauge directions to be eliminated, that is, if $\hat\omega$ has is degenerate (and thus possess null directions). Suppose that $\hat\eta = (\hat\xi, \hat\alpha)$ is a null direction of $\hat\omega$, $\ii_{\hat\eta} \hat\omega = 0$, that is, $\hat\omega(\hat\eta, \hat\eta') = 0$ for all vectors $\hat\eta'$. First let $\hat\eta' = (0, \ac\alpha')$, for a generic $\ac\alpha' \in \whddiff$. This implies that
\be
\hat\omega (\bar\eta, \bar\eta') = - \ac\alpha'(\hat\xi) = 0
\ee
But from the definition of $\whddiff$, \eqref{tildediff*}, we have that only the $\apsl$ subset of $\adiff$ is annihilated by all $\ac\alpha'$. Therefore, if $\hat\eta$ is a null direction, we must have $\hat\xi \in \partial_\theta \oplus \sin\theta\,\partial_\theta \oplus \cos\theta\,\partial_\theta$. Now consider $\hat\eta' = (\hat\xi', 0)$, for a generic $\hat\xi' \in \adiff$. This implies that
\be
\hat\omega (\bar\eta, \bar\eta') = \ac\alpha(\hat\xi') + \ac\sigma([\hat\xi, \hat\xi']) = 0
\ee
Let $\ac\alpha = a d\theta^2$, $\hat\xi = f\partial_\theta$, $\hat\xi' = f' \partial_\theta$ and, with a slight abuse of notation, $\ac\sigma = \sigma d\theta^2$. Then, according to the pairing defined in \eqref{pairing}, we have
\be
\hat\omega (\bar\eta, \bar\eta') = \int\!d\theta \left( a f' + \sigma (f \partial_\theta f' - f' \partial_\theta f ) \right) = \int\!d\theta \, f' \left( a  - f  \partial_\theta \sigma -2 \sigma \partial_\theta f  \right) 
\ee
where we used $[f \partial_\theta, f' \partial_\theta] = (f  \partial_\theta f' - f' \partial_\theta f) \partial_\theta$ and, in the last equality, we integrated by parts so as to factorize $f'$. Note that since $f'$ is arbitrary, we conclude that $\hat\eta$ is null only if
\be
a  = f  \partial_\theta \sigma + 2 \sigma \partial_\theta f 
\ee
Therefore, we have showed that $\hat\omega$ has exactly three degenerate directions at a generic point $(\hat\psi, \sigma d\theta^2)$,
\begin{align}
\hat\eta_1 &= (\partial_\theta, \partial_\theta \sigma d\theta^2) \nonumber\\
\hat\eta_2 &= (\sin\theta\,\partial_\theta, (\sin\theta\,\partial_\theta \sigma + 2 \cos\theta\, \sigma ) d\theta^2) \nonumber\\
\hat\eta_3 &= (\cos\theta\,\partial_\theta, (\cos\theta\,\partial_\theta \sigma - 2 \sin\theta\, \sigma) d\theta^2) \label{etanull}
\end{align}
Note that they must be somehow associated with the $\psl$ subgroup of $\diff$, since the $\hat\xi$-component of the null directions coincide with the generators of $\apsl$. If the $\ac\alpha$-component of these null directions were equal to zero, then the reduced phase space would be simply $[\diff/\psl] \times \whddiff$. Note that this is a trivial vector bundle which is locally isomorphic to the cotangent bundle of $\diff/\psl$, since the cotangent space at the projection of the identity, $[I] \in \diff/\psl$, is isomorphic to $\whddiff$. Based on the first approach for reducing the phase space, we actually know that the correct topology for the reduced phase space is actually the cotangent bundle of $\diff/\psl$, that is, $T^*[\diff/\psl]$. Intuitively, we can think that the fact that the null directions are ``tilted'' instead of purely ``horizontal'' (i.e., their $\hat\alpha$-component is non-zero) leads to a quotient that is not just the trivial bundle $[\diff/\psl] \times \whddiff$, but rather the ``twisted'' bundle $T^*[\diff/\psl]$.\footnote{It is interesting to note there is a surprising symplectomorphism between $T^*[\diff/\psl]$ and $[\diff/\psl] \times [\diff/\psl]$, for the natural symplectic structures on each space, provided by the Moss map. Despite this classical equivalence, the quantization may depend on the particular ``presentation'' of the phase space. In our approach $T^*[\diff/\psl]$ is certainly the more natural presentation.}

\subsection{Removing the residual $\psl$}
\label{subsec:respsl}

With the insight above, we will prove that the (fully) reduced phase space is 
\be\label{redphase}
\widetilde{\ca P} = T^*[\diff/\psl]
\ee
by showing that $(i)$ there is a natural projection map $J : \diff \times \whddiff \rightarrow T^*[\diff/\psl]$ and that $(ii)$ the canonical symplectic form $\wt\omega$ on $T^*[\diff/\psl]$, associated with its cotangent bundle structure, which is closed and non-degenerate, pulls-back to $\diff \times \whddiff$, i.e., $\hat\omega = J^* \wt\omega$.

\subsubsection{Proof of $(i)$: Existence of the projection map $J$}
\label{subsubsec:Jexist}

We begin by proving that this projection map exists. Consider the following theorem, which we will prove below.

\emph{Theorem:} Let $G$ be a Lie group and let $H$ be a (closed) subgroup of $G$. Denote the Lie algebra of $G$ by $\fr g$ and the Lie algebra of $H$ by $\fr h$ (naturally, $\fr h$ is a subalgebra of $\fr g$). Define the {\sl annihilator} of $\fr h$, denoted by $\ac{\fr g}^*$, as the subspace (not necessarily a subalgebra) of the dual Lie algebra of $G$, $\ac{\fr g}^* \subset {\fr g}^*$, such that $\fr h$ (and only $\fr h$) is in the kernel of all elements of $\ac{\fr g}^*$. That is,
\be
\ac{\fr g}^* := \{ \ac\sigma \in {\fr g}^*,\, \ac\sigma(\xi) = 0 \text{ if } \xi \in \fr h \subset \fr g \}
\ee
Let $G/H$ be (left) coset space of $H$ in $G$ and denote by $q: G \rightarrow G/H$ the corresponding quotient map, $g \mapsto [g] = gH$. Then there is a natural, well-defined quotient map $J : G \times \ac{\fr g}^* \rightarrow T^*(G/H)$, 
\be
J(g, \ac\sigma) = \wt\sigma
\ee
where $g \in G$, $\ac\sigma \in \ac{\fr g}^*$ and $\wt\sigma \in T^*_{[g]}(G/H)$ is defined by 
\be\label{Jdef}
\wt\sigma(\wt X) := \ac\sigma(\Xi(X))
\ee
where $\wt X \in T_{[g]} (G/H)$ is a tangent vector at $[g]$, $ X \in T_g G$ is any vector at $g$ in the pre-image of $\wt X$ under $q_*$ (i.e., $q_*  X = \wt X$), and $\Xi$ is the Maurer-Cartan form\footnote{The Maurer-Cartan form, $\Xi$, is a $\fr g$-valued 1-form on $G$ defined as follows. Let $l_g : G \rightarrow G$ be the left translation by $g$, i.e., $l_g(g') := gg'$. Given $ X \in T_g G$, we have $\Xi( X) := l_{g^{-1}*} X$, where $l_{g^{-1}*} X \in T_e G \sim \fr g$.}. It is interesting to visualize this map as part of the following commutative diagram
\begin{center}
\begin{tikzcd}
G \times \ac{\fr g}^* \arrow{d}[swap]{p_1} \arrow{r}{J} & T^*(G/H) \arrow{d}{\pi} \\
G  \arrow{r}{q} & G/H
\end{tikzcd}
\end{center}
where $p_1: G \times \ac{\fr g}^* \rightarrow G$ is the projector on the first Cartesian factor, $p_1(g, \sigma) = g$, and $\pi : T^*(G/H) \rightarrow G/H$ is the projection map of the cotangent bundle (which maps a dual vector to the point it is based at). 

We must show that the map $J$ is well-defined and surjective.\footnote{Rigorously, a map $f$ is a {\sl quotient map} if it is surjective and the topology of the codomain is induced from $f$ (i.e., $U$ is open iff $f^{-1}(U)$ is open). We are only proving surjection here, but subsequently we will show that $\text{ker}(J_*)$ has constant dimension which can be used to prove this topological condition.} The only arbitrariness in the definition of $J$ is the choice of $ X$ in the pre-image of $\wt X$ under $q_*$, so we must show that if we choose another $ X'$ satisfying $q_* X' = \wt X$ then the right-hand side of \eqref{Jdef} is unchanged. First let us characterize the kernel of $q_*$, i.e., given a point $g \in G$, what are the tangent vectors $\gamma \in T_gG$ such that $q_*\gamma = 0$? If $t \mapsto \Gamma_t$ is a curve starting at $g$ and tangent to $\gamma$, the condition $q_*\gamma = 0$ implies that, to first order in $t$, $\Gamma_t$ must be projected to a fixed point, i.e., $q(\Gamma_t) \approx [g]$. This implies that there must exist a curve $t \mapsto h_t$ in $H$, starting at $e$, such that $\Gamma_t \approx gh_t$, where $\approx$ means up to first order in $t$ (i.e., the two curves define the same tangent vector at $t=0$). But this implies that $\gamma = l_{g_*} \zeta$ for some $\zeta \in T_e H \sim \fr h$. The conclusion is therefore
\be\label{q*xi0}
q_* X = 0 \quad \text{iff}\quad \Xi( X) \in \fr h \subset \fr g
\ee
Given a vector $\wt X$ at $[g]$, let $ X$ and $ X'$ be two vectors at $g$ in the pre-image of $\wt X$ under $q_*$, i.e., $q_* X = q_* X' = \wt X$. Thus $q_*( X' -  X) = 0$ and, from the result above, $\Xi( X' -  X) \in \fr h$. That is, $\Xi( X') = \Xi( X) + \zeta$ for some $\zeta \in \fr h$. If we use $ X'$ in the right-hand side of \eqref{Jdef}, we get $\sigma(\Xi( X')) = \sigma(\Xi( X)) + \sigma(\zeta) = \sigma(\Xi( X))$, since $\sigma \in \ac{\fr g}^*$ annihilates the $\fr h$ subspace, which yields the same result as if we had used $ X$. We conclude that $J$ is well-defined. To prove surjectiveness, we must show that for any $\wt\sigma \in T^*(G/H)$, there exists a point $(g, \sigma) \in G \times \ac{\fr g}^*$ such that $J(g, \sigma) = \wt\sigma$. Since $q: G \rightarrow G/H$ is surjective, there is always a $g \in G$ such that $\pi(\wt\sigma) = [g]$. Now let $\wt\sigma$ be a generic dual vector at $[g]$ on $G/H$ and pull it back to $g$, $q^*\wt\sigma$, and further pull it back to $e$, $l_g^*q^*\wt\sigma$. This defines an element of $\ac{\fr g}^*$ because, given any $\zeta \in \fr h$ we have that $q_* l_{g*} \zeta = 0$, so $l_g^*q^*\wt\sigma (\zeta) = \wt\sigma(q_* l_{g*} \zeta) = 0$. Thus, for any $\wt X \in T_{[g]}(G/H)$,
\be
J(g, l_g^*q^*\wt\sigma)(\wt X) = l_g^*q^*\wt\sigma (\Xi( X)) = \wt\sigma(q_* l_{g*} l_{g^{-1}*}  X) = \wt\sigma(q_* X) = \wt\sigma(\wt X) 
\ee
which means that $(g, l_g^*q^*\wt\sigma)$ is in the pre-image of $\wt\sigma$ under $J$, so we conclude that $J$ is surjective. This finishes the proof of the theorem.

An interesting property of the map $J$ is that it has constant kernel dimension equal to $\text{dim}(\fr h)$, or more precisely, there is an isomorphism between $\fr h$ and $\text{ker}(J_*)$ everywhere in $G \times \ac{\fr g}^*$. This follows from the fact that, in the course of proving the theorem above, we have constructed an isomorphism between $T^*_{[g]}(G/H)$ and $\ac{\fr g}^*$. In particular, given any $g \in G$, we can see that the map $\sigma \in \ac{\fr g}^* \mapsto \wt\sigma \in T^*_{[g]}(G/H)$, where $\wt\sigma(\wt X) := \sigma( X)$, for any $ X$ satisfying $q_* X = \wt X$, and the map $\wt\sigma \in T^*_{[g]}(G/H) \mapsto l^*_g q^* \wt\sigma \in \ac{\fr g}^*$ are actually inverses of each other. Therefore, since the fibers of these two bundles are isomorphic and the base space $G$ is collapsed to $G/H$, we see that a number $\text{dim}(\fr h)$ of directions are mapped to zero at each point. In fact, we can derive an explicit formula for $\text{ker}(J_*)$. Let $\eta = ( X, \alpha) \in T_gG \oplus \ac{\fr g}^*$ be a vector in $\text{ker}(J_*)$ at $(g, \sigma) \in G \times \ac{\fr g}^*$, 
\be
J_*\eta = J_*( X, \alpha) = 0
\ee
From the commutative diagram, $\pi \circ J = q \circ p_1$, we get $\pi_*J_*\eta = 0 = q_* p_{1*} \eta = q_* X$, so from \eqref{q*xi0} we have $\Xi( X) \in \fr h$. Let $X = l_{g*}\zeta$, where $\zeta \in \fr h$, and let $t \mapsto h_t$ be a curve in $H \subset G$ starting at $e$ and tangent to $\zeta$. Thus the curve $t \mapsto gh_t$ in $G$ starts at $g$ and is tangent to $X$. Also, the curve $t \mapsto \sigma + t\alpha$ in $\ac{\fr g}^*$ starts at $\sigma$ and is tangent to $\alpha$. Thus the curve $t \mapsto (gh_t, \sigma + t\alpha)$ in $G \times \ac{\fr g}^*$ starts at $(g, \sigma)$ and is tangent to $\eta = (X, \alpha)$. Since this curve projects entirely to the fiber over $[g]$ under $J$, we can define the pushed vector by the derivative
\be
J_*(X, \alpha) = \left. \frac{d}{dt}J(gh_t, \sigma + t\alpha) \right|_{t=0}
\ee
where we are making use of the natural identification between {\sl vertical} vectors in $T_{\wt \sigma}[T^*(G/H)]$ and the fiber itself $T^*_{\pi(\wt \sigma)}(G/H)$. 
Let $\wt Y$ be a vector at $[g]$ and let $Y$ is a vector at $g$ satisfying $q_*Y = \wt Y$. We note that the vector $Y_t := r_{h_t*}Y$, where $r_g$ is the right-translation in $G$ by $g$, is a vector at $gh_t$ satisfying $q_*Y_t = \wt Y$. This follows from the fact that, as $[g] = [gh]$ for $h \in H$, then $q = q \circ r_h$, which implies that $q_*Y_t = q_*r_{h_t*}Y = q_*Y = \wt Y$. Consequently,
\be
J(gh_t, \sigma + t\alpha)(\wt Y) = (\sigma + t\alpha)\left(\Xi(Y_t)\right) = (\sigma + t\alpha)\left(l_{(gh_t)^{-1}*}r_{h_t*}Y\right)
\ee
Since the right and left translations commute, the argument of the dual vector above can be written as $l_{(gh_t)^{-1}*}r_{h_t*}Y = l_{h_t^{-1}*}l_{g^{-1}*}r_{h_t*}Y = l_{h_t^{-1}*}r_{h_t*}l_{g^{-1}*}Y$. 

Before further manipulations of the expression above, let us introduce the {\sl adjoint map}, a recurrent object for the remainder of this paper.
Given a group $G$ the {\sl adjoint action} of $g \in G$ on $g' \in G$ is defined as
\be\label{Addef}
\text{Ad}_g(g') = gg'g^{-1}
\ee
Noting that $\text{Ad}_g(e) = e$, the push-forward operation maps $T_e G$ into itself. This gives rise to a natural action of $G$ on its Lie algebra $\fr g$, which we also refer to as the adjoint action of $G$ on $\fr g$ and denote as
\be\label{addef}
\ad_g \xi := \text{Ad}_{g*} \xi
\ee
where $\xi \in \fr g \sim T_e G$. For a fixed element $\xi$, we can think of $\ad_g \xi$ as a map from $G$ into $\fr g$, that is, we can define $\ad\, \xi(g) := \ad_g\xi$. The push-forwards of this map sends a vector $\eta \in T_e G \sim \fr g$ to a vector in $T_\xi \fr g$; but since $\fr g$ is a vector space, any of its tangent spaces can be naturally identified with $\fr g$ itself; therefore $(\ad\,\xi)_*$ can be seen as a map from $\fr g$ to itself. We then define the adjoint action of $\fr g$ on $\fr g$ as
\be\label{aaddef}
\ad_\eta \xi := (\ad\,\xi)_* \eta
\ee
This adjoint action is directly related to the Lie algebra product, 
\be
\ad_\eta \xi = [\eta, \xi]
\ee 
While it is straightforward to show this using the formal definition of the Lie product in terms of the Lie brackets of left-invariant vector fields, we can trivially see it from a matrix realization of $G$: in this case the group exponential coincides with the matrix exponential, $\exp(\eta) = e^\eta$, and a vector tangent to a matrix-valued curve is simply given by the standard parameter-derivative of this curve; then $\ad_\eta \xi = (\ad\,\xi)_* \eta = \frac{d}{dt} \ad\,\xi(e^{t\eta}) = \frac{d}{dt} \ad_{e^{t\eta}} \xi = \frac{d}{dt} e^{t\eta} \xi e^{-t\eta} = \eta\xi - \xi\eta = [\eta, \xi]$, where the $t$-derivatives are all evaluated at $t=0$.

Let us return to the expression $l_{h_t^{-1}*}r_{h_t*}l_{g^{-1}*}Y$. Note that $l_g \circ r_{g^{-1}} = r_{g^{-1}} \circ l_g = \text{Ad}_g$. Therefore $l_{h_t^{-1}*}r_{h_t*}l_{g^{-1}*}Y = \text{Ad}_{h_t^{-1}*}\Xi(Y) = \ad_{h_t^{-1}}\Xi(Y)$, which gives
\be
J(gh_t, \sigma + t\alpha)(\wt Y) = (\sigma + t\alpha)\left(\text{Ad}_{h_t^{-1}*}\Xi(Y)\right)
\ee
so
\be\label{J*adformula}
\left( J_*(X, \alpha) \right) (\wt Y) = \alpha(\Xi(Y)) + \left. \frac{d}{dt}\sigma \left(\text{Ad}_{h_t^{-1}*}\Xi(Y)\right) \right|_{t=0} =  \alpha(\Xi(Y)) - \sigma \left(\ad_\zeta \Xi(Y)\right) 
\ee
where we have used that $h_t^{-1}$ is tangent to $-\zeta$ at $t=0$, so $\frac{d}{dt} \text{Ad}_{h_t^{-1}*}\Xi(Y) = \frac{d}{dt} \ad_{h_t^{-1}}\Xi(Y) = \left(\ad\,\Xi(Y)\right)_*(-\zeta) = - \ad_\zeta \Xi(Y)$, where $\frac{d}{dt}$ is evaluated at $t=0$.

We will also define the {\sl coadjoint action} of $g \in G$ on $\sigma \in {\fr g}^*$ by
\be\label{coaddef}
\coad_g \sigma := \text{Ad}^*_{g^{-1}}\sigma
\ee
where on the right-hand side $\sigma$ is seen as a 1-form at $e$.
The reason for taking the inverse of $g$ in $\text{Ad}^*$ is so that the coadjoint action composes nicely, i.e., $\coad_g \circ \coad_{g'} = \coad_{gg'}$.
Also, analogous to the definition of the adjoint action of $\fr g$ on $\fr g$, we define the coadjoint action of $\fr g$ on ${\fr g}^*$ by
\be\label{coaddef2}
\coad_\zeta \sigma := \left. \frac{d}{dt} \coad_{g_t}\sigma \right|_{t=0}
\ee
where $t \mapsto g_t$ is a curve in $G$ starting at $e$ and tangent to $\zeta \in \fr g$. It is worth noting, for later reference, that $\coad_g \sigma(\xi) =  \text{Ad}^*_{g^{-1}}\sigma(\xi) = \sigma( \text{Ad}_{g^{-1}*}\xi ) = \sigma (\ad_{g^{-1}}\xi)$ and, by taking $g$ as a curve $g_t$ tangent to a vector $\eta$ at $e$ and evaluating the $t$-derivative we get that 
\be\label{coadvsad}
\coad_\eta \sigma(\xi) = - \sigma (\ad_\eta \xi)
\ee
revealing that $\coad_\eta$ is {\sl minus} the algebraic dual of $\ad_\eta$.

Returning again to the evaluation of $J_*$, we can equivalently write \eqref{J*adformula} as 
\be
\left( J_*(X, \alpha) \right) (\wt Y) = \left( \alpha + \coad_{\Xi(X)} \sigma \right) (\Xi(Y))
\ee
recalling that $\Xi(X) = \zeta$ is the vector tangent to $t \mapsto h_t$. So, given the identification $\text{Ver}(T_{\wt \sigma}[T^*(G/H)]) \sim T^*_{\pi(\wt \sigma)}(G/H)$, we have
\be
J_*(l_{g*}\zeta, \alpha) \cong J(g, \alpha + \coad_\zeta \sigma)
\ee
where $\zeta \in \fr h$. Therefore, $\eta = (X, \alpha) \in \text{ker}(J_*)$ if and only if $\Xi(X) \in \fr h$ and
\be
\alpha = - \coad_{\Xi(X)} \sigma
\ee
This confirms that, at every point $(g, \sigma) \in G \times \ac{\fr g}^*$, the kernel of $J_*$ is isomorphic to $\fr h$ via the map
\be\label{kerJ*general}
\fr h \rightarrow \text{ker}(J_*)\,,\quad \zeta \mapsto \eta = (l_{g*}\zeta, - \coad_\zeta \sigma)
\ee
Note that these collapsing directions are ``tilted'' in the bundle $G \times \ac{\fr g}^*$, in the sense that $p_{2*}\eta \ne 0$, which is reminiscent of the the null directions in \eqref{etanull}. 

We are interested in the particularization of the theorem for $G = \diff$ and $H = \psl$. Let us express it in the notation we have been using before. The theorem states that there exists a projection map $J : \diff \times \whddiff \rightarrow T^*[\diff/\psl]$ defined by
\be
J(\psi, \ac\sigma) = \wt\sigma \,,\quad \wt\sigma(\wt\eta) = \ac\sigma(l_{\psi^{-1}*}\hat\eta)
\ee
where $\wt\eta \in T_{[\psi]} (\diff/\psl)$ and $\hat\eta \in T_\psi \diff$ is any vector satisfying $q_*\hat\eta = \wt\eta$ (with $q : \diff \rightarrow \diff/\psl$ being the quotient map). The kernel of $J_*$ is given by
\be\label{kerJ*}
\hat\eta \in \text{ker}(J_*) \quad\text{iff}\quad \hat\eta = (\zeta, - \coad_\zeta \ac\sigma)\,,\,\, \zeta \in \apsl
\ee
Note a slight difference in notation: as explained in footnote \ref{fn:TbarS}, a tangent vector at a point $(\psi, \ac\sigma) \in \diff \times \whddiff$ is being characterized as $(\xi, \hat\alpha) \in \adiff \oplus \whddiff$; but during the discussion of the above theorem a tangent vector at $(g, \sigma) \in G \times \ac{\fr g}^*$ was characterized as $(X, \alpha) \in T_g G \oplus \ac{\fr g}^*$. Of course, these two characterizations are naturally equivalent since $\fr g$ can be identified with $T_gG$ via the map $l_{g*}$; moreover, despite the use of similar notation, we expect that the context should prevent any confusion as it should be clear whether the first component of the pair belongs to $\fr g$ or $T_gG$. We can explicitly compute the coadjoint action using the relation \eqref{coadvsad},
\be\label{coadsigma0}
\coad_\zeta \ac\sigma(\xi) = - \ac\sigma(\ad_\zeta \xi) = - \ac\sigma([\zeta, \xi])
\ee
where $\zeta \in \apsl$ and $\xi \in \adiff$. 
As discussed previously, we can characterize $\adiff$ and $\ddiff$ using vector fields and quadratic form fields on $S^1$, so we write $\ac\sigma = \sigma d\theta^2$, $\zeta = z \partial_\theta$ and $\xi = f \partial_\theta$. Before we proceed, we must explain an unfortunate (and potentially confusing) aspect of this notation: the Lie brackets on $\adiff$, $[\xi, \xi']_\fr{diff}$, differs by a sign from the vector field brackets on $S^1$, $[\xi, \xi']_{S^1}$. That is,
\be\label{LiediffLievec}
[\xi, \xi']_\fr{diff} = - [\xi, \xi']_{S^1}
\ee
In order to clarify this point, let us use a more explicit notation. Let $G = \text{Diff}(\ca M)$ be the group of diffeomorphisms on a manifold $\ca M$. Let $f_x : G \rightarrow \ca M$ be the action of $\psi \in G$ on the point $x \in \ca M$, i.e., $f_x(\psi) := \psi(x)$. As with any group action, to any element of the algebra $\xi \in \fr g$ we can associate a vector field on $\ca M$ defined by $V_\xi |_x := f_{x*}\xi$. The group of diffeomorphisms is special in the sense that this map is an isomorphism, allowing us to identify $\fr g \sim \text{Vect}(\ca M)$.\footnote{The are some subtleties associated with the infinite dimensionality of the group of diffeomorphisms, such as the fact that the exponential map is not surjective in any neighborhood of the identity (although it has dense image). So we can think of this identification of $\fr{diff}(\ca M)$ with $\text{Vect}(\ca M)$ as a practical/formal characterization of $\fr{diff}(\ca M)$.} Since $G$ acts on the left of $\ca M$, there is an anti-homomorphism between the Lie algebra of $G$ and the vector field algebra of $\ca M$, that is, $[V_\xi, V_{\xi'}] = V_{[\xi', \xi]}$. This is perhaps aesthetically unpleasing, but completely transparent. Confusion  may arise, however, when we start writing $\xi$ as a shorthand for $V_\xi$. In this case, one solution is to always specify which bracket is being used by writing $[\xi, \xi']_\fr{diff} := [\xi, \xi']$ and $[\xi, \xi']_{\ca M} := [V_\xi, V_{\xi'}]$, so we would have $[\xi, \xi']_\fr{diff} = - [\xi, \xi']_{\ca M}$. In this manner, we can write \eqref{coadsigma0} more precisely as
\be\label{coadsigma}
\coad_\zeta \ac\sigma(\xi) = - \ac\sigma(\ad_\zeta \xi) = - \ac\sigma([\zeta, \xi]_\fr{diff}) = \ac\sigma([\zeta, \xi]_{S^1})
\ee
Thus, if we denote $\coad_\zeta \ac\sigma = \sigma' d\t^2$, we have
\be
\int\!d\t\, \sigma' f = \int\!d\t\, \sigma \left( z \frac{\partial f}{\partial \t} - f \frac{\partial z}{\partial \t} \right) = \int\!d\t \left( - z\frac{\partial \sigma}{\partial \t} - 2\sigma\frac{\partial z}{\partial \t} \right) f
\ee
Since this is valid for any $\xi \in \adiff$ (i.e., for any $f \in C^\infty(S^1, \bb R)$), we conclude that
\be
\coad_\zeta \ac\sigma = - \left( z\frac{\partial \sigma}{\partial \t} + 2\sigma\frac{\partial z}{\partial \t} \right) d\t^2
\ee
Therefore we see that the collapsed directions under the projection $J$ (i.e., the kernel of $J_*$) are given, according to \eqref{kerJ*}, by
\be
\hat\eta = \left( z \partial_\t ,  (z \partial_\t \sigma + 2\sigma \partial_\t z)d\t^2 \right)
\ee
where $z$ is any linear combination of $1, \, \sin\theta, \, \cos\theta$. These directions precisely match the null directions \eqref{etanull} of $\hat\omega$ on $\hat{\ca S}$, confirming that $J$ is the desired quotient map to the reduced phase space.

\subsubsection{Proof of $(ii)$: The natural symplectic form is the physical one}
\label{subsubsec:canon=sympl}

Now that we have shown that the reduced phase space $\wt{\ca P}$ has topology $T^*[\diff/\psl]$, we must figure out what is the correct symplectic structure on it. The symplectic form $\wt\omega$ on $\wt{\ca P}$ must be such that $J^*\wt w = \hat\omega$, where $\hat\omega$ is the symplectic form on $\hat{\ca S}$. The natural guess is the canonical symplectic form associated with the cotangent bundle structure of $T^*[\diff/\psl]$. Let us review how it is constructed and then show that this is indeed the correct choice.

Let $T^*\ca M$ be the cotangent bundle of a manifold $\ca M$ and let $\pi: T^*\ca M \rightarrow \ca M$ be the projection map (which maps a dual vector to the point it is based at). There is a natural 1-form on $T^*\ca M$, called {\sl canonical potential 1-form} $\theta$, defined as
\be
\theta(\eta) := \sigma(\pi_*\eta)
\ee
where $\eta \in T_\sigma(T^*\ca M)$. In words, ``given a vector $\eta$ at $\sigma \in T^*\ca M$, we project it down to $\ca M$ and act with $\sigma$ on it''. Taking the exterior derivative of $\theta$ yields a 2-form,
\be
\omega = d\theta
\ee
which is called the {\sl canonical symplectic 2-form} on $T^*\ca M$. Note that it is indeed a symplectic form since it is closed and non-degenerate. 

So let $\theta$ be canonical potential 1-form on $T^*[\diff/\psl]$ and let us evaluate its pull-back to $\hat{\ca S}$, $\hat\theta = J^*\theta$. If $\hat\eta$ is a vector at $(\psi, \ac\sigma) \in \hat{\ca S}$, then
\be
\hat\theta(\hat\eta) = \theta(J_*\hat\eta) = J(\psi, \ac\sigma)(\pi_*J_*\hat\eta) = J(\psi, \ac\sigma)(q_* p_{1*} \hat\eta)
\ee
where we have used the definition of $\theta$ and the relation $\pi \circ J = q \circ p_1$. But note that $p_{1*} \hat\eta$ is obviously a vector at $\psi \in \diff$ which projects to $q_* p_{1*} \hat\eta$ under $q_*$, so by the definition of $J$ we have
\be
\hat\theta(\hat\eta) = \ac\sigma (l_{\psi^{-1}*} p_{1*} \hat\eta)
\ee
If $\hat\eta$ is expressed as $(\xi, \hat\alpha) \in \adiff \oplus \whddiff$, we have that $p_{1*} \hat\eta = l_{\psi*}\xi$, so
\be\label{hatthetaformula}
\hat\theta(\hat\eta) = \ac\sigma (\xi)
\ee
Since the exterior derivative commutes with the pull-back, we have $\hat\omega := J^*(\omega) = J^*(d\theta) = d(J^*\theta) = d\hat\theta$. The exterior derivative of a 1-form is generally given by
\be
d\hat\theta(\hat\eta, \hat\eta') = \hat\eta[\hat\theta(\hat\eta')] - \hat\eta'[\hat\theta(\hat\eta)] - \hat\theta([\hat\eta, \hat\eta'])
\ee
where $\hat\eta[\hat\theta(\hat\eta')]$ denotes the directional derivative of the scalar field $\hat\theta(\hat\eta')$ along the vector $\hat\eta$, and $[\hat\eta, \hat\eta']$ denotes the Lie brackets of the vector fields $\hat\eta$ and $\hat\eta'$. Note that since $d\hat\theta$ is a tensor it depends only on the values of $\hat\eta$ and $\hat\eta'$ at the point where it is evaluated, so the right-hand side of this formula is independent of how $\hat\eta$ and $\hat\eta'$ are extended to vector fields in a neighborhood of the evaluation point. A particularly convenient choice is to take $\hat\eta = (\xi, \hat\alpha)$ and $\hat\eta' = (\xi', \hat\alpha')$ with $\xi$, $\xi'$, $\hat\alpha$ and $\hat\alpha$ being constants in a neighborhood of $(\psi, \ac\sigma)$. In other words, $\hat\eta$ is extended into a vector field in such a way that $p_{1*}\hat\eta$ is a left-invariant field on $\diff$ and $p_{2*}\hat\eta$ is a fixed point in $\whddiff$. With this choice we note that $\hat\theta(\hat\eta') = \ac\sigma (\xi')$ is a function on $\diff \times \whddiff$ depending only on $\ac\sigma$ (not on $\psi$). Thus,
\begin{align}
\hat\eta[\hat\theta(\hat\eta')] &= (\xi, \hat\alpha)[\ac\sigma (\xi')] \nonumber\\ 
&= (\xi, 0)[\ac\sigma (\xi')] + (0, \hat\alpha)[\ac\sigma (\xi')] \nonumber\\ 
&= \hat\alpha(\xi')
\end{align}
where the first term on the second line vanishes because we can take a horizontal curve $t \mapsto (\psi_t, \ac\sigma)$ as tangent to $(\xi, 0)$, so that $\frac{d}{dt}\ac\sigma (\xi') = 0$; for the second term on the second line we can take the vertical curve $t \mapsto (\psi, \ac\sigma + t \hat\alpha)$ as tangent to $(0, \hat\alpha)$, so that $\frac{d}{dt}(\ac\sigma + t \hat\alpha) (\xi') = \hat\alpha (\xi')$. Analogously, $\hat\eta'[\hat\theta(\hat\eta)] = \hat\alpha'(\xi)$. Lastly, consider the term $\hat\theta([\hat\eta, \hat\eta'])$ which is given by $\ac\sigma(l_{\psi^{-1}*} p_{1*} [\hat\eta, \hat\eta'])$. Given our choice, we have $p_{1*}\hat\eta = l_{\psi*}\xi$ and $p_{1*}\hat\eta' = l_{\psi*}\xi'$, so that both fields are projected nicely\footnote{Note that, in general, vector fields cannot be pushed through a (non-injective) map. Let $f : \ca M \rightarrow \ca N$ be a smooth map and let $V$ be a vector field on $\ca M$. For each $x \in \ca M$ we can define the push-forward $f_*V_x$ to $y = f(x) \in \ca N$. But if $f$ is not injective, there may exist $x' \ne x$ such that $f(x') = y$, and in general the push-forward $f_* V_{x'}$ to $y$ will not coincide with $f_*V_x$. If a vector field $V$ on $\ca M$ satisfies $f_*V_x = f_*V_{x'}$ for all $x,\, x' \in f^{-1}(y)$, for all $y \in \ca N$, then we say that $V$ is projected {\sl nicely} by $f$ to $\ca N$. The Lie brackets behave nicely under nice projections, in the sense that $f_*[V, V'] = [f_*V, f_*V']$.} to $\diff$, which implies that $p_{1*} [\hat\eta, \hat\eta'] =  [p_{1*}\hat\eta, p_{1*}\hat\eta'] = [l_{\psi*}\xi, l_{\psi*}\xi'] = l_{\psi*}[\xi,\xi']_{\fr{diff}}$, where in the last step we used the definition of the Lie algebra product as being homomorphic to the Lie bracket algebra of left-invariant fields. Thus,
\be
\hat\theta([\hat\eta, \hat\eta']) = \ac\sigma([\xi,\xi']_{\fr{diff}}) = - \ac\sigma([\xi,\xi']_{S^1})
\ee
So we have
\be
\hat\omega(\hat\eta, \hat\eta') = \hat\alpha(\xi') - \hat\alpha'(\xi) + \ac\sigma([\xi,\xi']_{S^1})
\ee
which matches exactly with the symplectic form on $\hat{\ca S}$ given in \eqref{hatomega}, confirming that the correct symplectic form $\wt\omega$ on $\wt{\ca P}$ is precisely the canonical symplectic form
\be
\wt\omega = \omega = d\theta
\ee
where $\theta$ is the canonical potential 1-form on $T^*[\diff/\psl]$. This concludes the reduction process from $\ca P = \text{Riem}(\Sigma, \gamma) \times \text{Sym}(\Sigma, (2,0))$ to $\wt{\ca P} = T^*[\diff/\psl]$.

\subsection{Diagrammatic summary}
\label{subsec:sumdiagram}

\begin{figure}[h!]
\hspace{-0.5cm}\begin{tikzpicture}[commutative diagrams/every diagram]
\node (P) at (0,0) {$\ca P = \text{Riem}(\Sigma, \gamma) \times \text{Sym}(\Sigma, (2,0))$};
\node (bP) at (-3,-2) {$\overline{\ca P} := \text{Diff}^+(D) \times_\gamma C^\infty(D, \bb R^+) \times \text{Sym}(\Sigma, (2,0))$};
\node (bS) at (0,-4) {$\overline{\ca S} := \text{Diff}^+(D) \times \text{Sym}(D, (2,0); TT[\bar h])$};
\node (hS) at (0,-6) {$\wh{\ca S} := \diff \times \whddiff$};
\node (tP) at (0,-8) {$\wt{\ca P} = T^*[\diff/\psl]$};
\node (S) at (3.5,-2) {$\ca S$};
\path[commutative diagrams/.cd, every arrow, every label]
(P) edge[bend right=10] node[swap] {enlargement} (bP)
(bP) edge[bend right=10] node[swap] {constraints} (bS)
(bS) edge node {bulk-diffeos quotient} (hS)
(hS) edge node {$\psl$ quotient} (tP)
(P) edge[bend left=10]  node {constraints} (S)
(S) edge[bend left=10]  node {enlargement} (bS);
\end{tikzpicture}
\caption{A diagrammatic summary of the phase space reduction process for the diamond. From $\ca P$ to $\bar{\ca P}$ the phase space is enlarged by introducing ``conformal coordinates'' $(\Psi, \Omega, \bar\sigma^{ab})$, and the constraints define a submanifold $\bar{\ca S}$. The order of these two steps can be interchanged, first going to $\ca S$ by imposing the constraints and then enlarging to $\bar{\ca S}$ by introducing conformal coordinates $(\Psi, \bar\sigma^{ab})$. From $\bar{\ca S}$ to $\hat{\ca S}$ the bulk diffeomorphisms are removed, and to $\wt{\ca P}$ the remaining $\psl$ action is quotiented out.}
\label{red_diag}
\end{figure}
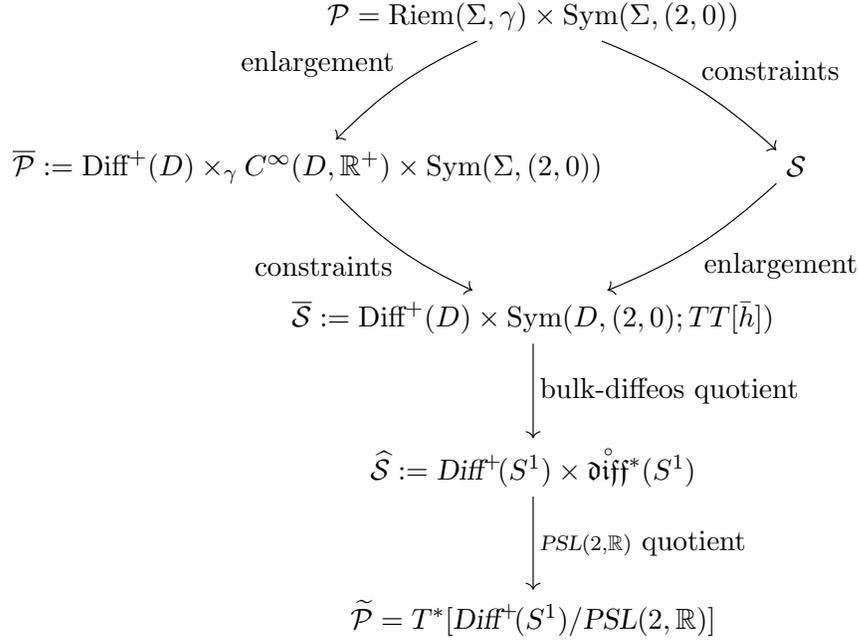

In Fig. \ref{red_diag} we illustrate the reduction process via conformal coordinates in a diagrammatic way. 

\newpage
\section*{Acknowledgments} 

I am especially grateful to Ted Jacobson for motivating this project, for collaboration on parts of it, and for useful discussions and valuable insights.
I also thank Stefano Antonini, 
Luis Apolo, Abhay Ashtekar, Batoul Banihashemi,
Steve Carlip, Gong Cheng, Laurent Freidel,
Marc Henneaux, Jim Isenberg, 
Alex Maloney, 
Blagoje Oblak,
Pranav Pulakkat, Gabor Sarosi, Antony Speranza, Raman Sundrum, 
Yixu Wang and Edward Witten for helpful discussions.

This research was supported in part by the National Science Foundation under Grants PHY-1708139 and 
PHY-2012139 at the University of Maryland; and
in part by the National Science Foundation under Grants No. NSF PHY-1748958 and PHY-2309135 at the Kavli Institute for Theoretical Physics. I am grateful for the hospitality of Perimeter Institute for Theoretical Physics where part of this work was carried out. (Research at Perimeter Institute is supported by the Government of Canada through the Department of Innovation, Science and Economic Development and by the Province of Ontario through the Ministry of Research, Innovation and Science.)

\newpage
\appendix

\section{Glossary, symbols and conventions}
\label{app:GSandC}

This appendix features a quick reference guide to recurrent terms and symbols found in this paper. It is organized roughly in the order of appearance in the text, but also in such a way that later entries only refers to terms and symbols already introduced in earlier entries. Each entry is indicated by an \underline{\emph{underlined italic name}} followed by the explanation; in cases where relevant symbols are introduced in the explanation, those symbols are displayed on the left margin for easier reference.
Despite our efforts to keep the notation uniform throughout, there may be instances where a symbol is used with a different meaning for a specific section. For example, $\Lambda$ is mostly used to denote the cosmological constant, but sometimes it can be used to denote a Weyl scaling factor -- we hope the context will prevent confusion in those instances. Some general conventions are also referenced in this section.
\\

\begin{itemize}[
	label=,
     labelsep=1.7em,
     itemsep=7pt,
     leftmargin=2em,
     itemindent=-2em
]

\item[] \underline{\emph{Causal diamonds}}~~ The domain of dependence of an acausal spacelike disc is referred to as a causal diamond.

\item[$T^a_{\,\,b}$, $T^\mu_{\,\,\nu}$, $T^i_{\,\,j}$] \underline{\emph{Tensors}}~~ The abstract index notation is used for tensors, where subscript Latin letters indicate covariant tensor slots and upperscript Latin letters indicate contravariant slots (typically we reserve for this purpose letters from the first half of the alphabet like $a$, $b$, $c$ etc). That is, if $V$ denotes the space of vectors at a point $p$ on a manifold, and $V^*$ denotes the dual vector space (i.e., space of 1-forms, or linear maps from $V$ to $\bb R$) at $p$, then a tensor $T$ at $p$ of type $\tensortype{n}{m}$ is a linear map from $V^m \times (V^*)^n$ to $\bb R$ and it is denoted by $T_{a_1 a_2 \cdots a_m}^{b_1 b_2 \cdots b_n}$; a 1-form $\alpha$ is a linear map from $V$ to $\bb R$ and therefore has one covariant index and is denoted as $\alpha_a$; a vector $\xi$ can be seen as a tensor with one contravariant index, denoted by $\xi^a$, given the natural duality between vectors and double dual vectors ($V \sim V^{**}$) which identifies $\xi$ with a linear map from $V^*$ to $\bb R$ (this map associates any 1-form $\alpha$ to the real number $\alpha(\xi)$). In this notation, the tensor $T$ applied to $m$ vectors $\xi_1,\, \xi_2,\, \cdots \xi_m$ and $n$ dual vectors $\alpha^1,\, \alpha^2,\, \cdots \alpha^n$ is expressed by contracting the respective indices, $T(\xi_1,\xi_2, \cdots \xi_m; \alpha^1, \alpha^2,\cdots \alpha^n) = T_{a_1 a_2 \cdots a_m}^{b_1 b_2 \cdots b_n} (\xi_1)^{a_1} (\xi_2)^{a_2} \cdots (\xi_m)^{a_m} (\alpha^1)_{b_1} (\alpha^2)_{b_2} \cdots (\alpha^n)_{b_n}$. A basis of vectors is typically denoted as $e_\mu$, where the subscript label $\mu$ runs from $1$ to the dimension $d$ of the manifold; the associated dual basis is denoted by $e^\mu$, with a superscript label, and defined so that $e^\nu(e_\mu) = \delta^\nu_\mu$, where $\delta^\nu_\mu$ is the Kronecker delta; tensors can be decomposed in that basis with components $T_{\mu_1 \mu_2 \cdots \mu_m}^{\nu_1 \nu_2 \cdots \nu_n} := T(e_{\mu_1},e_{\mu_2}, \cdots e_{\mu_m}; e^{\nu_1}, e^{\nu_2},\cdots e^{\nu_n})$; in any basis, contraction between a contravariant and a covariant index is achieved by summing over the repeated indices, for example $T_{\mu\nu}^\nu := \sum_{\nu = 1}^d T_{\mu\nu}^\nu$. Typically we reserve Greek letters ($\mu$, $\nu$, etc) to denote tensors decomposed in a spacetime basis, and Latin letters from the second half of the alphabet (like $i$, $j$, etc) for tensors decomposed on a spatial slice. 

\item[$f_*$, $f^*$] \underline{\emph{Push-forward and pull-back}}~~ Given a smooth map between two manifolds, $f : \ca M \rightarrow \ca N$, the push-forward operator mapping vectors (or contravariant tensors) on $\ca M$ to $\ca N$ is denoted by $f_*$ and the pull-back operator mapping 1-forms (or covariant tensors) from $\ca N$ to $\ca M$ is denoted by $f^*$. If $f$ is a diffeomorphism (smooth invertible map) then $f^* = f^{-1}_*$ and any tensor can be pushed forward or pulled back.

\item[$\Sigma$, $D$] \underline{\emph{Cauchy slice}}~~ A Cauchy slice will typically be denoted by $\Sigma$. In this paper it is assumed to have the topology of a two-dimensional disc, $D$. (Because of this, sometimes $D$ is also used to denote the Cauchy slice.)

\item[$\partial\Sigma$, $\partial$] \underline{\emph{Corner}}~~ The boundary of any Cauchy slice, $\partial\Sigma$, is referred to as the corner of the diamond. It has the topology of a circle, $S^1$. (Sometimes we shall abbreviate and simply denote it by $\partial$ --- not to be confused with partial differentiation.)

\item[$\Lambda$] \underline{\emph{Cosmological Constant}}~~ The cosmological constant will be denoted by $\Lambda$. In this paper we assume $\Lambda \le 0$. (In some instances the symbol $\Lambda$ may be used for other purposes, such as denoting certain maps or Weyl factors.)

\item[$h_{ab}$, $g_{ab}$] \underline{\emph{(Physical) Metric}}~~ Metrics in this paper typically refer to the spatial disc $\Sigma$ and are denoted by $h_{ab}$. Spacetime metrics are normally denoted by $g_{ab}$. 

\item[$\bar h_{ab}$] \underline{\emph{(Reference) Metric}}~~ The metric on the reference round unit disc, $dr^2 + r^2 d\theta^2$, is denoted by $\bar h_{ab}$. (The bar notation will often be used to indicate objects associated with the reference disc, like $\bar\sigma^{ab}$ below.)

\item[$\gamma_{ab}$] \underline{\emph{Boundary metric}}~~ We consider ``Dirichlet condition'' for the metric induced on the corner, $h_{ab}|_{\partial \Sigma} = \gamma_{ab}$.

\item[$\text{Riem}(\Sigma,\gamma)$] \underline{\emph{Space of metrics}}~~ The space of all (sufficiently regular) Riemannian metrics on a manifold $\Sigma$ is denoted by $\text{Riem}(\Sigma)$. Its subspace consisting of metrics satisfying the boundary condition $h|_\partial = \gamma$ is denoted by $\text{Riem}(\Sigma,\gamma)$.

\item[$\ell$, $\lads$, $\ell_P$] \underline{\emph{Length scales}}~~ There are three main length scales: the corner length $\ell$ (determined by $\gamma_{ab}$), the Anti-de Sitter radius $\lads = 1/\sqrt{-\Lambda}$ (if $\Lambda < 0$), and the Planck length $\ell_P = \hbar G$. (The Planck constant is introduced here even though it is only relevant for the quantum part, as we shall consider units where $c = \hbar = 1$.)

\item[$\nabla$, $\bar\nabla$, $\boldsymbol\nabla$] \underline{\emph{Covariant derivatives}}~~ The covariant derivative on a Cauchy slice associated with a spatial metric $h$ is denoted by $\nabla$; and the one associated with the reference metric $\bar h$ by $\bar\nabla$. The covariant derivative on spacetime associated with a metric $g$ is denoted by $\boldsymbol\nabla$. 

\item[$\pounds$, $\delta$, $\ii$] \underline{\emph{Lie, exterior and interior derivatives}}~~ The Lie derivative along a vector field $X$ is denoted by $\pounds_X$. The exterior derivative of a form is denoted by $\delta$; and the interior derivative (a.k.a., interior product), with respect to a vector $X$, by $\ii_X$.

\item[$(\Psi, \Omega)$, $\text{Con}(\Sigma)$] \underline{\emph{Conformal transformation}}~~ A transformation on metrics labeled by a pair $(\Psi, \Omega)$, where $\Psi$ is a diffeomorphism and $\Omega$ is a positive scalar, that acts by multiplying the metric by the scalar (a.k.a., Weyl factor) and pushing-forward by the diffeomorphism, $(\Psi, \Omega)h_{ab} := \Psi_* \Omega h_{ab}$ is called a conformal transformation. Since $\Omega > 0$, we often write it as $e^\lambda$ or $e^\phi$, where $\lambda,\, \phi \in \bb R$. The space of conformal transformations acting on metrics on a manifold $\Sigma$ is denoted by $\text{Con}(\Sigma)$.
(In the literature the term ``conformal transformation'' is sometimes used in a restricted sense equivalent to our definition  of a ``conformal isometry'' --- see below.)

\item[$\text{Con}(\Sigma^*)$] \underline{\emph{Boundary-trivial conformal transformation (BTCT)}}~~ A conformal transformation whose multiplicative scalar is equal to $1$ at the boundary of the manifold and diffeomorphism acts as the identity at the boundary is said to be a boundary-trivial conformal transformation or BTCT. The space of BTCTs acting on metrics on a manifold $\Sigma$ is denoted by $\text{Con}(\Sigma^*)$.

\item \underline{\emph{Conformal equivalence}}~~ Two metrics related by a boundary-trivial conformal transformation will be said to be conformally equivalent.

\item[$\text{ConGeo}(\Sigma)$] \underline{\emph{Conformal geometries}}~~ The space of conformally-equivalent (see above) metrics on a manifold $\Sigma$, denoted by $\text{ConGeo}(\Sigma)$, is called the space of conformal geometries on $\Sigma$.

\item[$\text{ConIso}(h)$] \underline{\emph{Conformal isometry}}~~ A conformal transformation that leaves a given metric $h_{ab}$ invariant, $h_{ab} = \Psi_* \Omega h_{ab}$, is said to be a conformal isometry. The space of conformal isometries of a metric $h$ is denoted by $\text{ConIso}(h)$.

\item[$K^{ab}$, $\sigma^{ab}$, $\tau$] \underline{\emph{(Physical) Extrinsic curvature}}~~ The extrinsic curvature of the Cauchy slice (as embedded in the spacetime) is normally denoted by $K^{ab}$, its trace part by $-\tau := K := K^{ab}h_{ab}$, and its traceless part by $\sigma^{ab} := K^{ab} - \frac{1}{2}K h^{ab}$.

\item[$\bar\sigma^{ab}$] \underline{\emph{(Transformed) Extrinsic curvature}}~~ The ``conformal pull-back'' of the (traceless part of the) extrinsic curvatures to the reference disc are typically denoted by $\bar\sigma^{ab}$. If $(\Psi, \Omega)$ maps the metric on the reference disc to the physical metric, the $\bar\sigma^{ab} = \Omega^2 \Psi^* \sigma^{ab}$. It is usually understood that the momentum constraint is satisfied, so $\bar\sigma^{ab}$ are traceless and divergenceless (with respect to $\bar h$) symmetric tensors.

\item[$\chi$, $\bar\chi$] \underline{\emph{``Chi parameter''}}~~ A parameter featuring in the Lichnerowicz equation, that appears often in the text, is defined as $\chi = -2\Lambda + \tau^2/2$. In Sec. \ref{sec:approxH} a dimensionless version of the parameter is introduced, $\bar\chi = (\ell/2\pi)^2 \chi$.

\item[$\psi$, $\diff$] \underline{\emph{Boundary diffeomorphisms and its group}}~~ The restriction to the boundary circle of a diffeomorphism on the disc is normally denoted by the lowercase version of the symbol, $\psi = \Psi|_\partial$. Similarly, an extension to the disc of a boundary diffeomorphism is denoted by the uppercase version of the symbol. The group of (orientation-preserving) diffeomorphisms on the circle is denoted as $\diff$. The identity element is denoted by $I$.

\item[$\adiff$, $\ddiff$\!\!] \underline{\emph{Algebra of boundary diffeomorphisms}}~~ The Lie algebra of $\diff$ is denoted as $\adiff$ and consists of vector fields $\xi$ on $S^1$. The dual Lie algebra of $\diff$ is denoted by $\ddiff$ and consists of quadratic forms $\alpha$ on $S^1$. The pairing is $\alpha(\xi) := \int \alpha(\theta) d\theta^2 (\xi(\theta) \partial_\theta) = \int d\theta \alpha(\theta) \xi(\theta)$.

\item[$\psl$, $\apsl$\!\!\!\!\!\!] \underline{\emph{Projective special linear group and algebra}}~~ The subgroup of $\diff$ corresponding to the restriction of the (diffeomorphism part of the) group of conformal isometries of the unit round disc to the boundary defines $\psl$. Its Lie algebra, denoted $\apsl$, consists of the span of vectors $\partial_\theta$, $\sin\theta\,\partial_\theta$, $\cos\theta\,\partial_\theta$

\item[$\ac\sigma$, $\whddiff$] \underline{\emph{``Sigma-circle''}}~~ The subspace of $\ddiff$ that annihilates $\apsl$ is denoted by $\whddiff$ and its elements are typically denoted by $\ac\sigma$. There is a natural correspondence between $\ac\sigma$'s and $\bar\sigma^{ab}$ that satisfies the constrains of general relativity. The notation is supposed to bring to mind that ``sigma-circle'' describes a (traceless) extrinsic curvature ``sigma-bar'' as a quantity living on the boundary ``circle'', $S^1$.

\item[$\ca P$, $\wh{\ca S}$, $\wt{\ca P}$] \underline{\emph{Phase space}}~~ The pre-phase space (described by spatial metrics and extrinsic curvatures) for the causal diamonds is denoted by $\ca P$, and the fully reduced phase space by $\wt{\ca P} = T^*[\diff/\psl]$. In some occasions it is convenient to work with the partially reduced phase space $\wh{\ca S} = \diff \times \whddiff$.

\item[$J$, $q$] \underline{\emph{$\psl$ projections}}~~ The quotient by $\psl$ from $\wh {\ca S}$ to the reduced phase space $\wt{\ca P}$ is denoted by $J$. The quotient by $\psl$ from $\diff$ to $\diff/\psl$ is denoted by $q$.

\item[$\Omega$, $\wt\omega$] \underline{\emph{Symplectic form}}~~ The pre-symplectic form on $\ca P$ is denoted by $\Omega$; the symplectic form on the reduced phase space $\wt{\ca P}$ is denoted by $\wt\omega$ (or simply $\omega$).

\item[$\text{[}a\sim b\text{]}$] \underline{\emph{Classes of equivalence}}~~ The space of classes of equivalence of all objects $a$ and $b$, belonging to some space $S$, identified under the relation ``$\sim$'' are typically denoted as $[a\sim b; a,b\in S]$. Sometimes the space $S$ is clear from the context and omitted in the notation, $[a \sim b]$. Often the equivalence relation comes from a group $G$ acting on $S$; then $[a \sim ga; a\in S, g \in G]$ is also called the space of $G$-orbits on $S$.

\item[$\log$] \underline{\emph{Logarithm}}~~ The natural logarithm (base $e$) is denoted by $\log$.

\item[$l_g$, $r_g$] \underline{\emph{Group translations}}~~ The left group translation $l_g : G \rightarrow G$, by a group element $g \in G$, is defined as $l_g(g') := gg'$. The right group translation $r_g: G \rightarrow G$, by $g$, is defined as $r_g(g') := g'g$.

\item[$\text{Ad}$, $\ad$] \underline{\emph{Adjoint maps}}~~ The adjoint action of a group element $g \in G$ on the group $G$ is denoted by $\text{Ad}_g : G \rightarrow G$. The adjoint action of a group element $g \in G$ on its Lie algebra $\fr g$ is denoted by $\ad_g : \fr g \rightarrow \fr g$. The adjoint action of an algebra element $\xi \in \fr g$ on the Lie algebra $\fr g$ is denoted by $\ad_\xi : \fr g\rightarrow \fr g$. 

\item[$\coad$] \underline{\emph{Coadjoint maps}}~~ The coadjoint action of a group element $g \in G$ on its dual Lie algebra $\fr g^*$ is denoted by $\coad_g : \fr g \rightarrow \fr g$. The coadjoint action of an algebra element $\xi \in \fr g$ on the dual Lie algebra $\fr g^*$ is denoted by $\coad_\xi : \fr g^*\rightarrow \fr g^*$. 

\item[$\Xi$] \underline{\emph{Maurer-Cartan form}}~~ The Maurer-Cartan form is denoted by $\Xi$. It is a Lie algebra-valued 1-form on the Lie Group defined by $\Xi(X) = l_{g^{-1}*}X$, where $X \in T_gG$.

\end{itemize}

\section{A brief review of reduced phase space}
\label{sec:reduced}

One of the fundamental assumptions of physics is that the laws of nature should be deterministic. That is, the knowledge of the state of a (closed) system at a given time should allow us to completely know its state at any future (or past) time. More precisely, let the state of the system be denoted by $\psi$; if the system is at the state $\psi_0$ at the initial time $t=0$, then the equations of motion should uniquely determine $\psi(t) = F(\psi_0; t)$ as a function $F$ of the time $t$ and the initial state. Certain theories, however, are described by equations of motion that are not deterministic, so that to each initial condition there corresponds a class of solutions $[\psi]$ to the equations of motion that satisfies the initial conditions. The principle of determinism then implies one of two things: $(i)$ the theory is incomplete, so that further equations of motion or constraining conditions are necessary to make the time evolution unique; $(ii)$ there is a redundancy in the description, so that only the equivalence classes of solutions, compatible with the initial conditions, are really physical. Thus, assuming that the theory is complete leave us with option $(ii)$, meaning that the physical space of states is only a quotient of the prototypical space of states, under the projection $\psi \mapsto [\psi]$. Such theories are called {\sl gauge theories}, and states $\psi$ and $\psi'$ within the same equivalence class, $[\psi] = [\psi']$, are said to be ``gauge-equivalent'' or ``related by a gauge transformation''. 

A notorious example of a gauge theory is electromagnetism. The equations of motion are $d\hodge F = 4\pi \hodge J$, where $F := dA$ is the electromagnetic strength and $A$ is the electromagnetic potential. Since the equations of motion depend only on $F$, it is insensitive to the change $A \mapsto A + d\a$, where $\a$ is any real function on the spacetime. Therefore, for any solution $A$ of the equations of motion, and any function $\a$ that vanishes in a neighborhood of the spatial slice at $t=0$, we have that $A' = A + d\a$ is also a solution of the equations of motion, satisfying the same initial conditions. In this way, the theory is incomplete unless we admit that the change $A \mapsto A + d\a$ is not physically observable, i.e., the physical states are described by the equivalence classes $[A] = [A + d\a]$. 

The consequence of a gauge ambiguities for the Hamiltonian description is the appearance of constraints in the phase space. To see this, let us consider a system described by a finite set of configuration variables $q^i$, $i = 1, \ldots n$, associated with an action principle $S = \int\! dt\, L(q, \dot q)$. The equation of motion are
\be
\frac{d}{dt}\frac{\partial L}{\partial \dot q^i} - \frac{\partial L}{\partial q^i} = 0
\ee
From the chain rule it follows
\be
\frac{\partial^2 L}{\partial \dot q^j \partial \dot q^i} \ddot q^j = -\frac{\partial^2 L}{\partial  q^j \partial \dot q^i} \dot q^j + \frac{\partial L}{\partial q^i}
\ee
Note that a necessary and sufficient condition for this set of equations to have a unique solution, given initial values for $q$ and $\dot q$, is that we can solve algebraically for the second time derivatives in terms of the lower time derivatives, i.e., that we can put it in the form $\ddot q^i = E^i(q, \dot q)$. This is equivalent to say that 
\be\label{detL}
\text{det} \left(\frac{\partial^2 L}{\partial \dot q^j \partial \dot q^i}\right) \ne 0
\ee
where the argument of the determinant is understood as a $n \times n$ matrix with indices $i$ and $j$.
If we define the momenta $p$ conjugated to $q$ by
\be
p_i := \frac{\partial L}{\partial \dot q^i}
\ee
then the non-degeneracy condition translates into
\be\label{Jacobqp}
\text{det} \left(\frac{\partial p_i}{\partial \dot q^j }\right) \ne 0
\ee
This determinant is equal to the Jacobian of the Legendre transformation $(q, \dot q) \mapsto (q, p)$, mapping the space of ``positions and velocities'' to the phase space, and the non-degeneracy condition thus implies that this map is (locally) invertible. In a gauge system the equations of motion do not have unique solution, which requires \eqref{detL} to be zero and therefore implies that the transformation $(q, \dot q) \mapsto (q, p)$ is not (locally) invertible. In other words, the image of the Legendre transformation is a surface $\ca S$ (of dimension less than $2n$) in the phase space $\ca P$. Hence there are constraints $C_\a = 0$ on the phase space specifying the location of this surface. 

Let us assume that the constraint surface $\ca S$ is a manifold smoothly embedded in the phase space, and call $\rho : \ca S \rightarrow \ca P$ the embedding map. The symplectic form on $\ca P$, $\Omega = \delta p_i \wedge \delta q^i$, can be pulled back to the constraint surface, defining a ``symplectic form'' $\omega := \rho^* \Omega$ on $\ca S$. The reason for the quotations is that although $\omega$ is closed (since $\delta\omega = \delta(\rho^* \Omega) = \rho^* \delta\Omega = 0$) it may fail to be non-degenerate, i.e., there may exist certain {\sl null} directions $\xi$ along $\ca S$ such that $\ii_\xi \omega = 0$. If it happens that $\omega$ is degenerate, so it is called a pre-symplectic form, then the null directions are precisely the gauge directions. This is because given any Hamiltonian $H : \ca S \rightarrow \bb R$, the vector flow $X$ generated by $H$ is defined to be a solution of $\delta H = - \ii_X \omega$, but it is clear that for any solution $X$ then $X' = X + \xi$ is also a solution. Therefore $\xi$ are the directions corresponding to the ambiguities in the time evolution, and thus moving along $\xi$ must not correspond to a change in the physical states. An interesting fact about the gauge directions is that they are surface-integrable, in the sense that the vector field commutator of two null fields $\xi$ and $\xi'$ is another null field. This follows from the identities $\ii_{[X,Y]} = [\pounds_X, \ii_Y]$ and $\pounds_X = \delta\ii_X + \ii_X \delta$, which applied to $\omega$ gives
\be
\ii_{[\xi, \xi']} \omega = \pounds_{\xi} \ii_{\xi'} \omega - \ii_{\xi'} \left( \delta\ii_{\xi} \omega + \ii_{\xi} \delta\omega \right) = 0
\ee
where it was used that $\ii_{\xi} \omega = \ii_{\xi'} \omega = 0$ and $\delta\omega = 0$. The set of all null vector fields thus form an (infinite-dimensional) algebra, and the set of all gauge transformations (flows along null directions) form a group. The ``integrated'' surfaces whose tangent vectors at every point are null are called {\sl gauge orbits}. A simple linear algebra argument reveals that the dimension of such surfaces must be at most equal to the number of constraints (in fact, it matches the number of {\sl first class} constraints\footnote{A constraint is {\sl first-class} if its symplectic flow (under the original pre-symplectic form $\Omega$) is tangent to the constraint surface, and it is {\sl second-class} otherwise. A first-class constraint always leads to a degenerate $\omega$: if $C$ is a first-class constraint then its flow $X$, defined from $\delta C = - \ii_X \Omega$, will be tangent to $\ca S$; so if $T$ is any vector tangent to $\ca S$ we have $0 = \delta C(T) = -\Omega(X, T) = - \omega(X, T)$, implying that $X$ is a null direction for $\omega$.\label{firstsecondclass}}).

The {\sl reduced phase space}, $\widetilde{\ca P}$, is the space of physically distinct states, defined by the quotient of $\ca S$ under the gauge transformations, or in other words, the equivalence classes of gauge orbits. There is a natural symplectic form $\widetilde\omega$ (closed and non-degenerate) on the reduced phase space, having the property that its pull-back under the quotient map $J : \ca S \rightarrow \widetilde{\ca P}$ is equal to $\omega$, i.e., $\omega = J^* \widetilde\omega$. This can be seen by a constructive approach. Given $p \in \widetilde{\ca P}$, let $\bar p \in \ca S$ be a point in the pre-image of $p$ under $J$, i.e., $J(\bar p) = p$. Given two vectors $X, Y \in T_p \widetilde{\ca P}$, consider any two vectors $\bar X, \bar Y \in T_{\bar p}\ca S$ that project to $X$ and $Y$ under $J_*$, i.e., $J_*\bar X = X$ and $J_*\bar Y = Y$. Define,
\be\label{wideomegadef}
\left. \widetilde\omega(X, Y) \right|_p := \left. \omega(\bar X, \bar Y) \right|_{\bar p}
\ee
Now we must show that this definition is consistent, in the sense that it must not depend on the particular point $\bar p$ chosen in the pre-image of $J$, nor on the particular vectors chosen in the pre-image of $J_*$. Note that if $\phi$ is a gauge transformation on $\ca S$ implementing a flow along null vector fields, then by definition of the projection map we have $J \circ \phi = J$. This implies that the kernel of $J_*$ at $\bar p$ consists of null vectors $\xi$ at $\bar p$, i.e., $J_*\xi = 0$ if and only if $\ii_\xi \omega = 0$. Therefore $\bar X$ is defined up to the addition of a null vector, $\xi$. As replacing $\bar X$ by $\bar X + \xi$ in the right-hand side of \eqref{wideomegadef} does not affect the result, this definition is insensitive to the choices of vectors in the pre-image of $J_*$. Now we consider the ambiguity in the choice of $\bar p$. First note that gauge transformations $\phi$ are symmetries of $\omega$, $\phi^*\omega = \omega$, which follows from $\pounds_\xi \omega = \delta\ii_\xi \omega + \ii_\xi \delta\omega = 0$. Second note that for any other point $\bar p'$ in the pre-image of $p$ under $J$, there must exist a gauge transformation $\phi$ such that $\bar p' = \phi(\bar p)$. It follows from $J_* \phi_* = J_*$ that $\bar X' := \phi_* \bar X$ and $\bar Y' := \phi_* \bar Y$ are vectors at $\bar p'$ in the pre-image of $X$ and $Y$ under $J_*$. Hence,
\be
\left. \omega(\bar X', \bar Y') \right|_{\bar p'} = \left. \omega(\phi_*\bar X, \phi_*\bar Y) \right|_{\phi(\bar p)} = \left. \phi^*\omega(\bar X, \bar Y) \right|_{\bar p} = \left. \omega(\bar X, \bar Y) \right|_{\bar p}
\ee
showing that \eqref{wideomegadef} is independent of the choice of $\bar p \in J^{-1}(p)$. Therefore we have shown that there exists a antisymmetric 2-form $\widetilde\omega$ on $\widetilde{\ca P}$ satisfying $\omega = J^*\widetilde\omega$. To finish the story, we must show that $\widetilde\omega$ is closed and non-degenerate. Note that $0 = \delta\omega = J^*\delta\widetilde\omega$. But since $J$ is a projection map, so $J_*$ has maximum rank, then $J^*\delta\widetilde\omega = 0$ implies that $\delta\widetilde\omega = 0$, establishing closedness. Now let $\eta$ be a vector on $\widetilde{\ca P}$ such that $\ii_{\eta}\widetilde\omega = 0$, and let $\bar\eta$ be a vector on $\ca S$ that projects to $\eta$ under $J$, i.e., $J_*\bar\eta = \eta$. We have,
\be
\ii_{\bar\eta}\omega(\cdot) = \omega(\bar\eta, \cdot) = J^*\widetilde\omega(\bar\eta, \cdot) = \widetilde\omega(J_*\bar\eta, J_*(\cdot)) = \ii_\eta \widetilde\omega(J_*(\cdot)) = 0
\ee
implying that $\bar\eta$ is a null vector. But null vectors are in the kernel of $J_*$ and hence $\eta = 0$, establishing non-degeneracy.

To wrap up, let us return to the discussion at the beginning of this section about deterministic evolution. If the equations of motions are deterministic, then there is a one-to-one correspondence between solutions to the equations of motion and initial data. Since the phase space is the space of ``initial data'' (as it is isomorphic to the space of ``initial positions and velocities''), we can look at the phase space as the space of ``solutions to the equations of motion''. This perspective is taken as the basis of the {\sl covariant} construction of the phase space for field theories, which avoids having to pick an arbitrary time direction and spatial slice and simply consider the field configurations that satisfy the equations of motion in spacetime. In this construction, one starts with the space $\ca C$ of all field configurations on spacetime and define the phase space $\ca P$ as the submanifold consisting of configurations satisfying the equations of motion. The (pre)symplectic structure follows from the action, and it may be degenerate on $\ca P$. Accordingly, this degeneracy corresponds to gauge ambiguities and the reduced phase space is defined by quotienting over all the gauge transformations. In this way, we can think of the reduced phase space as the space of all {\sl physical} solutions to the equations of motion (where gauge-equivalent solutions are regarded as the same physical solution).

\section{The uniformization map}
\label{app:unimap}

In this appendix we shall explain how to construct an explicit conformal map that transforms a generic Riemannian disc, $(h_{ab}, D)$, into the reference Euclidean unit disc, $(\bar h_{ab}, D)$. In other words, we shall establish that the map considered in Sec. \ref{sec:Rcc},
\be\label{confmap2}
(\Psi, \Omega) \mapsto h_{ab} = \Psi_* \Omega \bar h_{ab}
\ee
is a surjection from $\text{Diff}^+(D) \times C^\infty(D, \bb R^+)$ onto $\text{Riem}(D)$. The construction also automatically implies that when the domain of the map is restricted to $\text{Diff}^+(D) \times_\gamma C^\infty(D, \bb R^+)$ the surjection is onto $\text{Riem}(D, \gamma)$. This is a particular form of the uniformization theorem, which says that every simply-connected Riemannian 2-manifold is conformally equivalent to the open unit disc, or the complex plane, or the Riemann sphere. In fact, the particular case we will consider is known as the {\sl Riemann mapping theorem}. The explicit construction of such a map permits us to write down explicitly the projection map from the ADM phase space, described by metrics $h_{ab}$ and (traceless) extrinsic curvatures $\sigma^{ab}$, into the reduced phase space $\wt{\ca P}$. More precisely, note that the first step of the reduction process discussed in Sec. \ref{sec:Rcc}, summarized in Fig. \ref{red_diag}, consists of an {\sl enlargement} of the phase space where $(h_{ab}, \sigma^{ab})$ is replaced by $(\Psi, \Omega, \bar\sigma^{ab})$, and the purpose of this appendix is to find a $(\Psi, \Omega)$ from a given $h_{ab}$ (and after having done that, one would simply find $\bar\sigma^{ab}$ by computing $\Omega^2 \Psi^* \sigma^{ab}$). There are many $(\Psi, \Omega)$ that can uniformize any given $h_{ab}$, a consequence of certain ambiguities in the construction (in fact, as we know, these ambiguities should correspond to a $\psl$ gauge). 

Given a manifold (with boundary) $D$ with the topology of a closed disc, let $h_{ab}$ be a generic Riemannian metric on $D$ and let $\bar h_{ab}$ be the metric that makes $D$ a Euclidean unit disc, i.e., in polar coodinates $\{ r, \theta \} \in [0,1] \times [0, 2\pi)$ we have $\bar h = dr^2 + r^2d\theta^2$. The construction of the map consist of two main steps:
\begin{enumerate}[label=(\roman*)]
\item Use a Weyl transformation to ``flatten'' $(h, D)$, so that it can be isometrically embedded as a region $R$ of the complex plane (with its natural, Euclidean metric);
\item Construct an analytical map that deforms $R$ into the unit complex disc $\bb D = \{z \in \bb C, |z| \le 1 \}$, which is isometric to $(\bar h, D)$. 
\end{enumerate}

\subsection{Step $(i)$: Flattening and embedding}

For the first step we must find $\Gamma : D \rightarrow \bb R^+$ such that
\be
\wh h_{ab} = \Gamma h_{ab}
\ee
is flat. In two dimensions, flatness follows from requiring that the Ricci scalar vanishes,
\be
0 = \wh R = \frac{1}{\Gamma} \left( R - \nabla^2 \log\Gamma \right)
\ee
where $R$ and $\nabla^2 = h^{ab} \nabla_a \nabla_b$ are respectively the Ricci scalar and Laplacian associated with $h$, and $\wh R$ is associated with $\wh h$. Therefore, $\Gamma$ is solution of
\be
\nabla^2 \log\Gamma = R 
\ee
Note that the boundary conditions are not specified, so there are many solutions for this equations. In fact, note that this is the familiar Poisson equation for a ``potential'' $\log \Gamma$ and a ``charge density'' $-R$, so we know that for any choice of (Dirichlet) boundary values the equation has a unique (real) solution for $\log \Gamma$. For concreteness, we may (arbitrarily) choose $\Gamma|_{\partial D} = 1$. Now that we have a ``flattened'' disc, $(\wh h, D)$, we construct an embedding isometry into the flat plane. To do so, choose an arbitrary point $p_0 \in D$ and an arbitrary orthonormal basis $\{e_1, e_2\}$, with respect to $\wh h$, at $p_0$. Since $\wh h$ is flat, this basis can be unambiguously extended to an orthonormal frame $\{e_1, e_2\}$ on the whole $D$ by demanding that $e_i$ is covariantly constant with respect to $\wh h$,
\be\label{frameconst}
\wh\nabla_a (e_i)^b = 0
\ee
and it matches with the basis chosen at $p_0$. Let $\{e^1, e^2 \}$ denote the dual frame, satisfying $e^i(e_j) = \delta^i_{\, j}$ everywhere on $D$. We can define {\sl Cartesian coordinates} on $D$ by integrating these dual vectors along arbitrary curves starting from $p_0$, that is, we assign coordinates $(x^1, x^2)$ to $p \in D$ via
\be\label{CartCoord}
x^i(p) := \int_{p_0}^p e^i
\ee
where the integral is along any curve in $D$ joining $p_0$ and $p$. Note that since $\{e_i\}$ was defined though \eqref{frameconst}, the dual frame also consist of covariantly constant 1-forms, $\wh\nabla_a (e^i)_b = 0$, which implies that $de^i = 0$ and therefore the choice of the curve joining $p_0$ to $p$ does not affect the result of the integral. It follows that $e^i = dx^i$, and also that $\wh h = \delta_{ij} e^i e^j = (dx^1)^2 + (dx^2)^2$, justifying the term ``Cartesian'' for these coordinates. In fact, this map $p \mapsto (x^1, x^2)$ is the isometric embedding into the Euclidean plane. Naturally, we can also express this map as an embedding in $\bb C$ by 
\be
\phi : D \rightarrow \bb C \,,\quad \phi(p) = x^1(p) + i x^2(p)
\ee
whose image shall be denoted by $R \subset \bb C$. This embedding is an isometry with respect to the Euclidean metric on $\bb C$,
\be
w = \frac{dz d\bar z + d\bar z dz}{2}
\ee
that is, $w = \phi_* \wh h$. Thus we have constructed a conformal tranformation from $(h, D)$ to $(w, R)$ where $w = \phi_* \Gamma h$. 

\subsection{Step $(ii)$: Deforming and identifying}

\begin{figure}
\centering
\includegraphics[scale = 0.65]{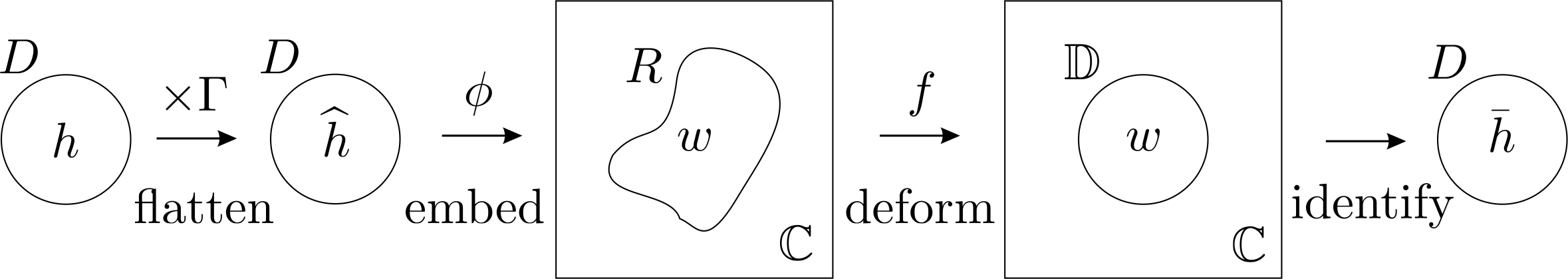}
\caption{Illustration of the uniformization algorithm. The space $(h, D)$ is flattened with a Weyl transformation, then isometrically embedded in the complex plane, then analytically deformed into the unit disc, and naturally identified with $(\bar h, D)$.}
\label{fig:unimap}
\end{figure}

Now we proceed to the second step, in which we analytically deform $R$ into  the unit complex disc $\bb D = \{z \in \bb C, |z| \le 1 \}$. That is, we want to construct an analytical map $f : R \rightarrow \bb C$ such that $\text{Im}(f) = \bb D$. The reason for requiring $f$ to be analytic is so that it generates a conformal transformation on $w$, which can be seen as follows
\be\label{conisoanalytic}
f^*w = f^* \left( \frac{dz d\bar z + d\bar z dz}{2} \right) = \frac{d(z \circ f) d(\bar z \circ f) + d(\bar z \circ f) d(z \circ f)}{2} = |f'|^2 w
\ee
where in the last step we have used that $d(z \circ f) = (\partial f/\partial z) dz + (\partial f/\partial \bar z) d\bar z = f' dz$ and $d(\bar z \circ f) = \overline{d(z \circ f)} = \overline{f'} d\bar z$. Thus $w = f_* |f'|^2 w$, which corresponds to a conformal transformation from $(w, R)$ to $(w, \bb D)$. Since $(w, \bb D)$ is naturally isometric to $(\bar h, D)$, we would have constructed a conformal transformation from $(h, D)$ to $(\bar h, D)$ given by $\bar h = f_* |f'|^2 w = f_* |f'|^2 \phi_* \Gamma h = (f \circ \phi)_* (|f'|^2 \circ \phi)\Gamma h$. (See Fig. \ref{fig:unimap} for an illustration of the algorithm.)
By inverting this transformation we would have the desired conformal map from $\bar h$ to $h$,
\be
h = (f \circ \phi)^{-1}_* \left[ \frac{1}{|f'|^2 (\Gamma \circ \phi^{-1})} \circ f^{-1} \right] \bar h
\ee
so, in the notation of \eqref{confmap2}, we identify $\Psi = (f \circ \phi)^{-1}$ and $\Omega$ as the function inside the square brackets. Thus it only remains to explain how to construct the map $f$. There is some freedom in the construction of this map, so for concreteness let us impose that $f(0) = 0$. With this choice, the point $p_0 \in D$ will end up at the origin of the unit disc. Moreover, we can try the following ansatz,
\be
f(z) = z e^{g(z)}
\ee
where $g(z)$ is some analytic function. Both $u(z) := \text{Re}[g(z)]$ and $v(z) := \text{Im}[g(z)]$ are therefore harmonic (real) functions on $\bb C$. Now we impose that the boundary of $R$ is mapped to the boundary of $\bb D$,
\be
|f(z)| = 1 \,\,\text{when}\,\, z \in \partial R \quad\Longrightarrow\quad u(z) = - \log|z| \,\,\text{when}\,\, z \in \partial R
\ee
This fixes the value of $u$ on $\partial R$, and since $u$ is harmonic (with respect to $w$), there is a unique solution for $u$ in the domain $R$. Also, $v(z)$ can be solved as
\be
v(z) = \int_0^z dv = \int_0^z \left( \frac{\partial v}{\partial x} dx + \frac{\partial v}{\partial y} dy \right) = \int_0^z \left( - \frac{\partial u}{\partial y} dx + \frac{\partial u}{\partial x} dy \right) 
\ee
where $v(x, y) := v(x + iy)$ and we made the arbitrary choice $v(0) = 0$. With $g$ fully determined in the domain $R$, $f$ is also determined and the construction is therefore complete.

\section{Embedding the diamond in $\ads$}
\label{app:embed}

Since the diamond should be a region of $\ads$ it is interesting to have an explicit algorithm for, given any point in the phase space, constructing the corresponding diamond embedded in $\ads$. For this purpose, we make use of the fact that $\ads$ is a maximally-symmetric spacetime and, in particular, possesses a time-translation and a spatial-rotation symmetry.

\subsection{General strategy}

Consider the usual global coordinates on $\ads$, $\{t, r, \phi\}$, in which the metric takes the form
\be
g = - \left(1 +\frac{r^2}{\lads^2} \right) dt^2 + \left(1 +\frac{r^2}{\lads^2} \right)^{-1} dr^2 + r^2 d\phi^2
\ee
In this coordinate system, we can identify the time-translation symmetry as generated by $\xi^t = \partial/\partial t$ and the spatial-rotation as generated by $\xi^\phi = \partial/\partial\phi$. Observe that if we know these two Killing fields, in a generic connected patch of $\ads$, we can locally construct these coordinates as follows. Note that
\be
dt_a = \frac{(\xi^t)_a}{(\xi^t)_b (\xi^t)^b}
\ee
so, up to an arbitrary assignment of time $t_0$ to a point $x_0$, the time coordinate $t$ of any other point is given by
\be
t(x) = t_0 + \int_{x_0}^x dt = t_0 + \int_0^1 ds   \frac{(\xi^t)_a \dot\gamma^a}{(\xi^t)_b (\xi^t)^b}
\ee
where $\gamma(s)$ is any curve from $x_0$ at $s=0$ to $x$ at $s=1$, and $\dot\gamma$ is the vector tangent to it. 
Similarly,
\be
d\phi_a = \frac{(\xi^\phi)_a}{(\xi^\phi)_b (\xi^\phi)^b}
\ee
so we can assign angular coordinate $\phi$ to points as
\be
\phi(x) = \phi_0 + \int_{x_0}^x d\phi = \phi_0 + \int_0^1 ds   \frac{(\xi^\phi)_a \dot\gamma^a}{(\xi^\phi)_b (\xi^\phi)^b}
\ee
where $\phi_0$ is an arbitrary angle coordinate for $x_0$, assumed not to be at the spatial origin (where the angle is undefined). The $r$ coordinate of the point $x$ can be extracted from the norm of either $\xi^t$ or $\xi^\phi$ as
\be
r(x) = \lads \sqrt{- (\xi^t)_a (\xi^t)^a - 1 } = \sqrt{(\xi^\phi)_b (\xi^\phi)^b}
\ee
Of course, the condition that $x_0$ is not at the spatial origin is simply that $r(x_0) \ne 0$. 

Another useful fact is that a Killing field is always uniquely determined from its value and first (anti-symmetric) derivative at any single point. This is so because a Killing field has no symmetric part for its first derivative, which implies that its second derivative is locally determined from its value as $\boldsymbol\nabla_a \boldsymbol\nabla_b \xi_c = \ca R_{cbad}\xi^d$,\footnote{We use the boldface symbol $\boldsymbol\nabla$ to denote the three-dimensional covariant derivative associated with the $\ads$ metric $g$.} and thus we can write the first-order system of equations
\begin{align}
v^a\boldsymbol\nabla_a\xi_b &= v^a \chi_{ab} \nonumber\\
v^a \boldsymbol\nabla_a \chi_{bc} &= v^a {\ca R_{cba}}^d \xi_d \label{Killing1order}
\end{align}
where $\chi_{ab} := \nabla_a \xi_b = \nabla_{[a} \xi_{b]}$ and $v^a$ is any vector field. If one takes $v = \dot\gamma$ as the tangent vector field along a curve $\gamma$, it is possible to integrate the equations from the initial values of $\xi$ and $\chi$. Using \eqref{caRAdS3} we can rewrite the second equation as
\be\label{Killing1orderchi}
v^a \boldsymbol\nabla_a \chi_{bc} = \frac{2}{\lads^2} v_{[b}\xi_{c]}
\ee
which is slightly simpler.

The final observation is that these equations can be restricted to any surface embedded in $\ads$, and can be solved as long as we know the induced metric $h_{ab}$ and the extrinsic curvature $K^{ab}$ on the surface. In particular, we can use the ADM data $(h_{ab}, \sigma^{ab}, \tau)$ on the CMCs to solve for the Killing fields and construct the desired coordinate system on it (which automatically provide the embedding of the CMCs into $\ads$). Consider a generic spacelike surface $\Sigma$ embedded in $\ads$, with normal vector field $n^a$. The induced metric can be expressed as $h_{ab} = n_a n_b + g_{ab}$,\footnote{More precisely, ${h_a}^b$ is the orthogonal projector to $\Sigma$, but $h_{ab}$ coincides with the induced metric when rectricted to $\Sigma$.} and the extrinsic curvature is given by $K_{ab} = {h_a}^c \boldsymbol\nabla_c n_b$. Now we decompose the objects $\xi$ and $\chi$ in their orthogonal (``hat'') and tangent (``bar'') components, as
\be
\xi_a = g_{ab} \xi^b = (-n_a n_b + h_{ab}) \xi^b = \widehat\xi n_a + \overline\xi_a
\ee
where
\begin{align}
\widehat\xi &:= - n_a \xi^a \nonumber\\
\overline\xi_a &:= h_{ab}\xi^b
\end{align}
and
\be
\chi_{ab} = {g_a}^{a'}{g_a}^{b'} \chi_{a'b'} = (-n_a n^{a'} + {h_a}^{a'})(-n_b n^{b'} + {h_b}^{b'})\chi_{a'b'} = 2 n_{[a} \widehat\chi_{b]} + \overline\chi_{ab}
\ee
where 
\begin{align}
\widehat\chi_a &:= {h_a}^{a'}\chi_{a'b'} n^{b'}  \nonumber\\
\overline\chi_{ab} &:={h_a}^{a'}{h_b}^{b'}\chi_{a'b'}
\end{align}
If $v$ is tangent to $\Sigma$, the variation of $\xi$ becomes
\begin{align}
v^a\boldsymbol\nabla_a\xi_b &= v^a\boldsymbol\nabla_a(\widehat\xi n_b + \overline\xi_b) \nonumber\\
&= v^a\boldsymbol\nabla_a\widehat\xi n_b + \widehat\xi v^a (\boldsymbol\nabla_a n_b) + v^a\boldsymbol\nabla_a\overline\xi_b \nonumber\\
&= v^a\nabla_a\widehat\xi n_b + \widehat\xi v^a K_{ab} + v^a\boldsymbol\nabla_a\overline\xi_b 
\end{align}
where in the last line it was used that $v^a\boldsymbol\nabla_a\widehat\xi = v^a \nabla_a\widehat\xi$, since the derivative of a scalar is independent of the connection, and also that $v^a K_{ab} = v^a {h_a}^c \boldsymbol\nabla_c n_b = v^a \boldsymbol\nabla_a n_b$, since $v$ is tangent to $\Sigma$. For the last term we use the relation between the 3d derivative $\boldsymbol\nabla$ associated with $g$ and the 2d derivative $\nabla$ on $\Sigma$ associated with $h$,
\be
\nabla_a \overline\xi_b = {h_a}^{a'}{h_b}^{b'} \boldsymbol\nabla_{a'}\overline\xi_{b'}
\ee
which gives
\be\label{nablavec3d2d}
v^a \nabla_a \overline\xi_b = v^a (n_b n^{b'} + {g_b}^{b'}) \boldsymbol\nabla_{a}\overline\xi_{b'} = v^a \boldsymbol\nabla_a \overline\xi_b - n_b v^a (\boldsymbol\nabla_{a} n^{b'}) \overline\xi_{b'} = v^a \boldsymbol\nabla_a \overline\xi_b - n_b v^a {K_a}^c \overline\xi_c
\ee
where we used that $v^a \boldsymbol\nabla_a (n^b \overline\xi_b) = 0$. Thus the first equation in \eqref{Killing1order} becomes
\be
v^a\nabla_a\widehat\xi n_b + \widehat\xi v^a K_{ab} + v^a \nabla_a \overline\xi_b + n_b v^a {K_a}^c \overline\xi_c = v^a \chi_{ab} = -n_b v^a \widehat\chi_a + v^a \overline\chi_{ab}
\ee
which naturally decomposes into orthogonal and tangent parts,
\begin{align} 
v^a\nabla_a\widehat\xi &= -v^a \widehat\chi_a - v^a {K_a}^c \overline\xi_c \nonumber\\
v^a \nabla_a \overline\xi_b &= -\widehat\xi v^a K_{ab} + v^a \overline\chi_{ab} \label{Killing1xisurface}
\end{align}
We now look at the variation of $\chi$, which is given by
\begin{align}
v^a \boldsymbol\nabla_a \chi_{bc} &= v^a \boldsymbol\nabla_a (2 n_{[b} \widehat\chi_{c]} + \overline\chi_{bc}) \nonumber\\
&= 2 v^a \left(\boldsymbol\nabla_a n_{[b}\right)  \widehat\chi_{c]} + 2 n_{[b} v^a \boldsymbol\nabla_a \widehat\chi_{c]} + v^a \boldsymbol\nabla_a\overline\chi_{bc} \nonumber\\
&= 2 v^a K_{a[b} \widehat\chi_{c]} + 2 n_{[b} v^a \left(\nabla_a \widehat\chi_{c]} + n_{c]} {K_a}^d \widehat\chi_d\right) + v^a \boldsymbol\nabla_a\overline\chi_{bc}
\end{align}
where the middle term in the last line was manipulated analogously to \eqref{nablavec3d2d}. For the last term we have
\begin{align}
v^a \nabla_a\overline\chi_{bc} &= {h_b}^{b'}{h_c}^{c'}v^a \boldsymbol\nabla_a\overline\chi_{b'c'} \nonumber\\
&= (n_b n^{b'} + {g_b}^{b'})(n_c n^{c'} + {g_c}^{c'})v^a \boldsymbol\nabla_a\overline\chi_{b'c'} \nonumber\\
&= v^a \boldsymbol\nabla_a\overline\chi_{bc} - 2 n_{[b} n^d v^a \boldsymbol\nabla_a\overline\chi_{c]d} \nonumber\\
&= v^a \boldsymbol\nabla_a\overline\chi_{bc} + 2 n_{[b}  v^a (\boldsymbol\nabla_a n^d) \overline\chi_{c]d} \nonumber\\
&= v^a \boldsymbol\nabla_a\overline\chi_{bc} + 2 n_{[b}  v^a {K_a}^d \overline\chi_{c]d}
\end{align}
where in the forth line we used that $v^a \boldsymbol\nabla_a (n^d\overline\chi_{cd}) = 0$. So equation \eqref{Killing1orderchi} becomes
\be
2 v^a K_{a[b} \widehat\chi_{c]} + 2 n_{[b} v^a \nabla_a \widehat\chi_{c]} +v^a \nabla_a\overline\chi_{bc} - 2 n_{[b}  v^a {K_a}^d \overline\chi_{c]d}
 = -\frac{2}{\lads^2}  \left( \widehat\xi n_{[b} + \overline\xi_{[b} \right)v_{c]}
\ee
which again can be decomposed into orthogonal and tangent parts,
\begin{align}
v^a \nabla_a \widehat\chi_b &= v^a {K_a}^c \overline\chi_{bc} - \frac{1}{\lads^2} \widehat\xi v_b   \nonumber\\
v^a \nabla_a\overline\chi_{bc} &= -2 v^a K_{a[b} \widehat\chi_{c]} - \frac{2}{\lads^2}\overline\xi_{[b} v_{c]} \label{Killing1chisurface}
\end{align}
The system of equations \eqref{Killing1xisurface} and \eqref{Killing1chisurface} allows us to completely solve for $\xi$ on the surface $\Sigma$ given initial values for $\xi$ and $\chi$ at a point of $\Sigma$.

This leads to the following algorithm for embedding the diamond in $\ads$,
\begin{enumerate}
\item Given a point $p \in \wt{\ca P}$ in the reduced phase space, consider any point $(\psi, \ac\sigma) \in J^{-1}(p) \subset \wh{\ca S}$ in the pre-image of $p$ under $J$. 

\item For a given time $\tau$, solve the associated Lichnerowicz equation \eqref{lambdaLich} for $\lambda$. Given any extension of $\psi$ to a diffeomorphism $\Psi$ of the disk, define $(h_{ab}, \sigma^{ab}) \in \ca S$ as in \eqref{hconftrans} and \eqref{sigmaconftrans}. On the CMC at time $\tau$, with $K = -\tau$, we have $K^{ab} = \sigma^{ab} - \frac{1}{2}\tau h^{ab}$. 

\item Choose an arbitrary point $x_0$ of the disk to be the origin of the $\ads$ coordinate system. We can always choose a boost such that $\partial/\partial t$ is orthogonal to the corresponding CMC at $x_0$. That is, $(\xi^t)^a = n^a$ and $(\chi^t)_{ab} = 0$ at $x_0$. 

\item Solve the system of equations \eqref{Killing1xisurface} and \eqref{Killing1chisurface} on the disk along a curve from $x_0$ to a point $x_1 \in \partial D$ at the boundary using the initial data
\[
\wh\xi^t = 1 \,,\quad (\overline\xi^t)_a = 0 \,,\quad (\wh\chi^t)_a = 0 \,,\quad (\overline\chi^t)_{ab} = 0
\]

\item Continue solving the equations along the whole boundary circumference, starting from $x_1$.

\item With $(\xi^t)^a$ determined along the boundary, compute the time coordinate of the points $x$ of the boundary via
\[
t(x) = t(x_1) - \int_{\theta(x_1)}^{\theta(x)} d\theta \frac{(\overline\xi^t)_a (\partial_\theta)^a}{(\wh\xi^t)^2 - (\overline\xi^t)_b(\overline\xi^t)^b}
\]
where $\theta$ is an arbitrary angular coordinate on the disk. Note that $t(x_1)$ can also be computed, given $t(x_0) = 0$, but it is unimportant. 

\item At the coordinate origin, the angular Killing field vanishes and its first derivative is given by $(\chi^\phi)_{ab} = (e_1)_a (e_2)_b - (e_2)_a (e_1)_b$, where $e_1$ and $e_2$ are two orthogonal unit vectors orthogonal to $\partial/\partial t$. Note that, when restricted to the surface, it matches the volume element associated with the induced metric. 

\item Solve the system of equations \eqref{Killing1xisurface} and \eqref{Killing1chisurface} on the disk along a curve from $x_0$ to $x_1$ using the initial data
\[
\wh\xi^\phi = 0 \,,\quad (\overline\xi^\phi)_a = 0 \,,\quad (\wh\chi^\phi)_a = 0 \,,\quad (\overline\chi^\phi)_{ab} = \vartheta_{ab}
\]

\item Continue solving the equations along the whole boundary circumference, starting from $x_1$.

\item With $(\xi^\phi)^a$ determined along the boundary, compute the angular coordinate of the points $x$ of the boundary via
\[
\phi(x) = \phi(x_1) + \int_{\theta(x_1)}^{\theta(x)} d\theta \frac{(\overline\xi^\phi)_a (\partial_\theta)^a}{(\overline\xi^\phi)_b(\overline\xi^\phi)^b - (\wh\xi^\phi)^2}
\]
Note that $\phi(x_1)$ can chosen arbitrarily, but it is unimportant.

\item Define the radial coordinate for points of the boundary as
\[
r(x) = \lads \sqrt{(\wh\xi^t)^2 - (\overline\xi^t)_a(\overline\xi^t)^a - 1 } = \sqrt{(\overline\xi^\phi)_a(\overline\xi^\phi)^a - (\wh\xi^\phi)^2}
\]

\item With the boundary of one CMC successfully embedded in $\ads$, the diamond is determined as the domain of dependence of the interior of the boundary (where the ``interior'' is any spacelike disk attached to the boundary).

\end{enumerate}

\subsection{Numerical implementation and pictures}

We have implemented the algorithm proposed above in {\sl Mathematica} (Ver. 12). The document is named {\sl Embedding\_AdS3}. The program takes the input state $(\psi, \wh\sigma) \in \wh{\ca S}$ and plots the boundary loop (corner) of the diamond as embedded in $\ads$, in usual coordinates $\{t,x = r\cos\phi, y = r\sin\phi \}$.

\begin{figure}[h!]
\centering
\includegraphics[scale = 0.4]{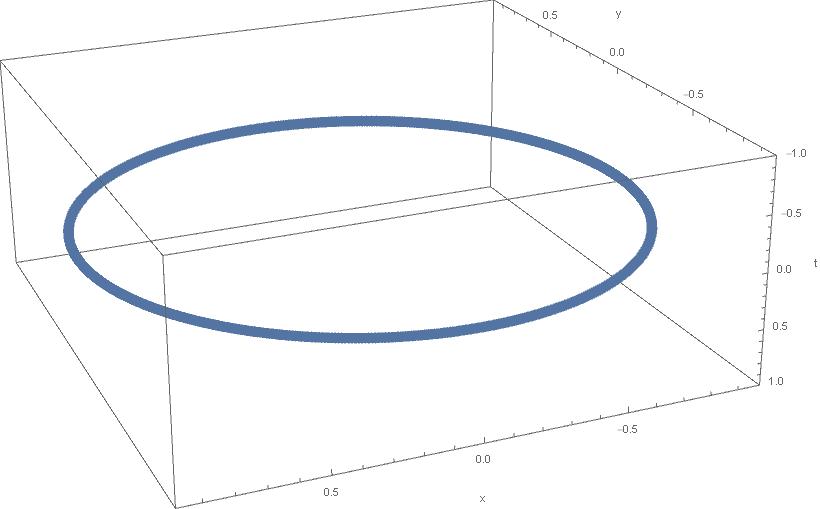}
\caption{Corner of the ``symmetric'' diamond: $\psi(\theta) = \theta,\,\wh\sigma(\theta) = 0$}
\label{fig:symnb}
\end{figure}

\begin{figure}[h!]
\centering
\includegraphics[scale = 0.4]{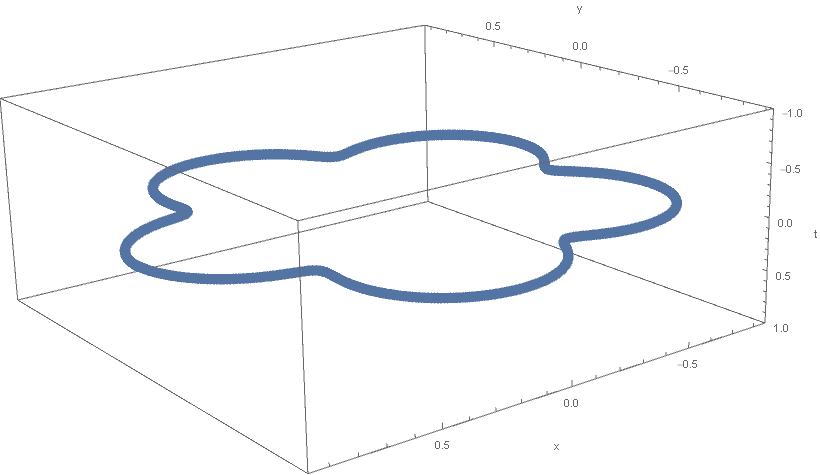}
\caption{Corner of a ``zero-momentum'' diamond: $\psi(\theta) = \theta + 0.15 \sin(5\theta),\,\wh\sigma(\theta) = 0$}
\label{fig:psinb}
\end{figure}

\begin{figure}[h!]
\centering
\includegraphics[scale = 0.4]{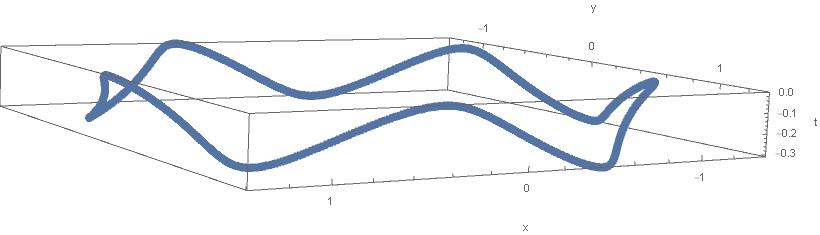}
\caption{Corner of a ``pure-momentum'' diamond: $\psi(\theta) = \theta ,\,\wh\sigma(\theta) = 10 \cos(5\theta)$}
\label{fig:sigmanb}
\end{figure}

We have plotted the corner of the diamond for some special states $(\psi, \wh\sigma)$. In Fig. \ref{fig:symnb} we have the case $\psi(\theta) = \theta,\,\wh\sigma(\theta) = 0$, which corresponds to the ``symmetric'' diamond; note that the corner of the diamond is embedded as a planar circle. In Fig. \ref{fig:psinb} we have the case $\psi(\theta) = \theta + 0.15 \sin(5\theta),\,\wh\sigma(\theta) = 0$, which corresponds to a ``zero-momentum'' diamond; note that the corner is still planar\footnote{The fact that ``zero-momentum'' diamonds are all planar can be seen from the time-reversal symmetry of the system.} but now oscillates spatially (5 times).  In Fig. \ref{fig:sigmanb} we have the case $\psi(\theta) = \theta ,\,\wh\sigma(\theta) = 10 \cos(5\theta)$, which corresponds to a ``pure-momentum'' diamond; note that the corner oscillates in a lightlike direction (5 times). Finally, in Fig. \ref{fig:P0nb} we have the case $\psi(\theta) = \theta + 0.25 (\cos(2\theta) - 1) ,\,\wh\sigma(\theta) = \sin(2\theta)$, which corresponds to a ``spinning'' diamond; although it may not be easy to see, the corner has a twist ($P_0 = 1.94$).

\begin{figure}[h!]
\centering
\includegraphics[scale = 0.4]{diamond_P0_nb.jpg}
\caption{Corner of a ``spinning'' diamond ($P_0 = 1.94$): $\psi(\theta) = \theta + 0.25 (\cos(2\theta) - 1) ,\,\wh\sigma(\theta) = \sin(2\theta)$}
\label{fig:P0nb}
\end{figure}

For artistic purposes, we have also developed a {\sl Mathematica} program that produces mathematically accurate pictures of causal diamonds (in Minkowski space) given a (spatial, acausal) boundary loop. The future horizon is obtained by shooting geodesic light rays from the boundary loop, going inward and to the future; a light ray is terminated if it meets (or come sufficiently close, given a suitable threshold parameter) to the prolongation of any other light rays coming from the loop. The past horizon is produced in a similar fashion, using light rays going inwards and to the past
\begin{figure}[h!]
\centering
\includegraphics[scale = 0.4]{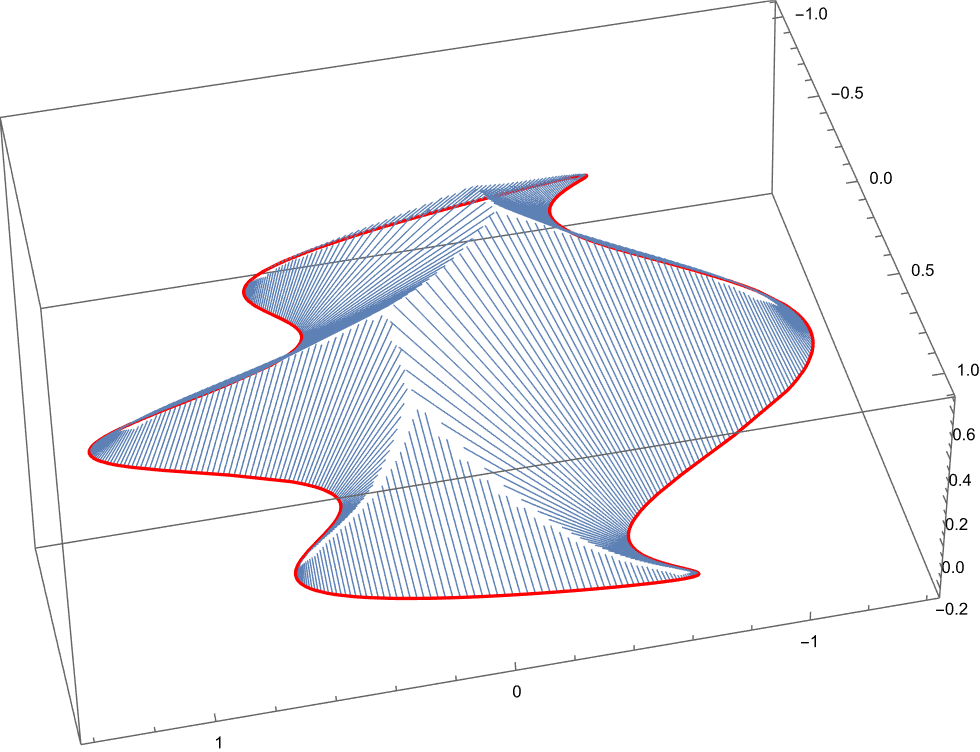}
\caption{The future horizon of a causal diamond with a given boundary loop. The light rays are shown in blue and the loop in red.}
\label{fig:blue1}
\end{figure}
\begin{figure}[h!]
\centering
\includegraphics[scale = 0.4]{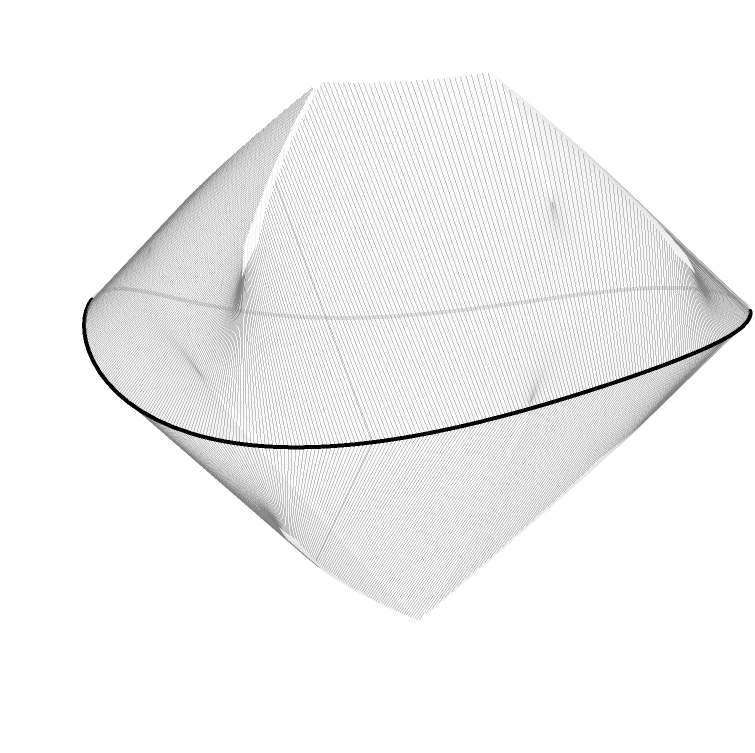}
\caption{Another diamond with a finer spacing between the light rays.}
\label{fig:grey3}
\end{figure}
\begin{figure}[h!]
\centering
\includegraphics[scale = 0.4]{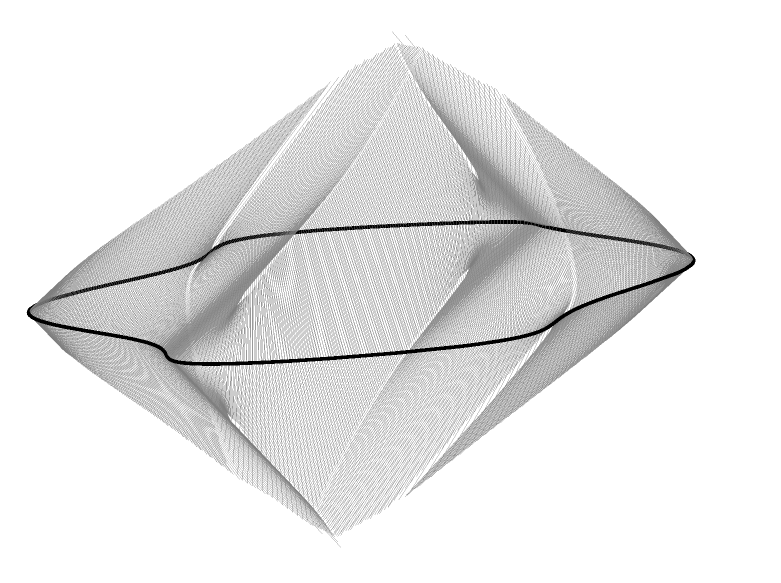}
\caption{A diamond with a more complicated boundary loop.}
\label{fig:grey1}
\end{figure}

\section{The reduced Hamiltonian}
\label{sec:redH}

The Hamiltonian description of a dynamical system involves the specification of a phase space (its topology and symplectic structure) and a Hamiltonian function which generates time evolution. Having already completed the specification of the reduced phase space $\wt{\ca P}$ for the diamond, we now proceed to computing the Hamiltonian $H$ which generates time-evolution according to our gauge-fixed choice of time, i.e., evolution along the CMC foliation.

\subsection{General review}
\label{subsec:genreview}

\begin{figure}
\centering
\includegraphics[scale = 0.8]{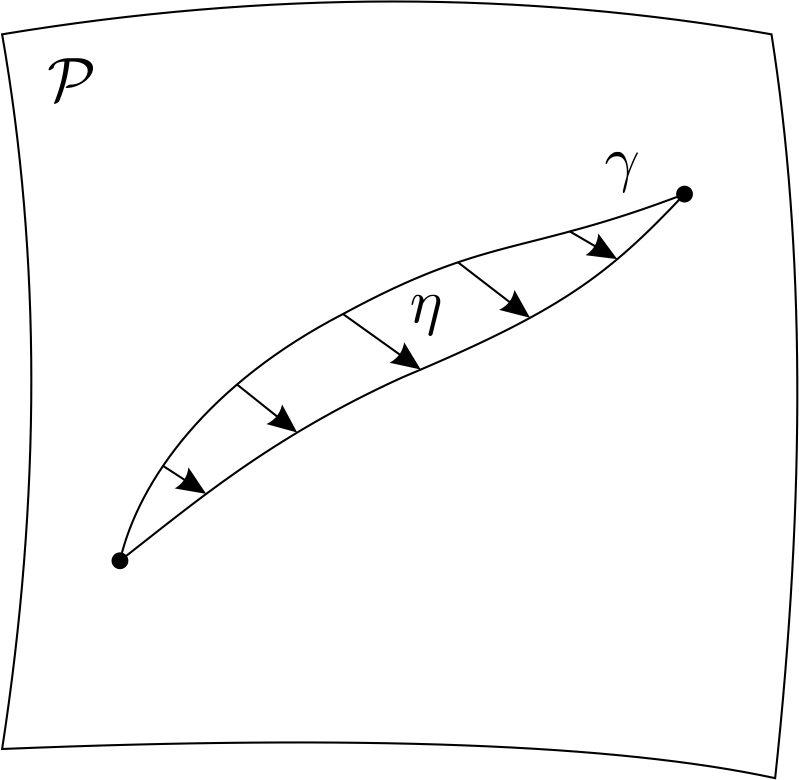}
\caption{A curve $\gamma$ on the phase space $\ca P$ is deformed by the vector field $\eta$, while its endpoints remain fixed.}
\label{fig:Hred}
\end{figure}

First, let us recall how every symplectic motion on a phase space corresponds to an action principle. Let $\ca P$ be a phase space with symplectic 2-form $\omega$. Locally, let $\theta$ be the symplectic potential 1-form for $\omega$, i.e., $\omega = d\theta$. Given any function $H : \ca P \rightarrow \bb R$, define the action principle
\be\label{actionH}
S[\gamma] = \int_\gamma \left( \theta - H dt \right)
\ee
where $\gamma : [0,1] \rightarrow \ca P$ is a curve on the phase space.\footnote{Rigorously, in order to think of $dt$ as a legitimate 1-form, the integral should be defined on the extended phase space, $\ca P \times \bb R$, where the ``time'' is adjoined to the usual phase space. In this way, the integral is evaluated along the lifted curve $\Gamma(t) := (\gamma(t), t)$. However we shall not be so pedantic about this.} The domain of $S$ is given by all smooth curves $\gamma$ with fixed endpoints. Note that this is just a covariant way to write the usual phase space action principle: from Darboux's theorem there are always local coordinates $\{p_i, q^i\}$ such that $\omega = dp_i \wedge dq^i$ and $\theta = p_i dq^i$; then if $\gamma(t) = (p_i(t), q^i(t))$, so the tangent vector is $\dot\gamma = \dot p_i \frac{\partial}{\partial p_i} + \dot q^i \frac{\partial}{\partial q^i}$, we have $S[\gamma] = \int_0^1 (p_i \dot q^i - H) dt = \int_0^1 L dt$, where $L$ is the Lagrangian associated with $H$. Now we consider an infinitesimal deformation of $\gamma$ by a vector field $\eta$ and evaluate the corresponding variation of the action. [See Fig. \ref{fig:Hred}.]
More precisely, if $\Phi_\eta(s)$ is the diffeomorphism associated with flowing along $\eta$ by a parameter-length $s$, define
\be
\delta S := \left. \frac{d}{ds} S[\Phi_\eta(s) \circ \gamma] \right|_{s=0}
\ee
From the definition of $S$ we have
\be
\frac{d}{ds} S[\Phi_\eta(s) \circ \gamma] = \frac{d}{ds} \int_{\Phi_\eta(s) \circ \gamma} \left( \theta - H dt \right) = \frac{d}{ds} \int_\gamma \Phi_\eta(s)^* \left( \theta - H dt \right)
\ee
and since the derivative is evaluated at $s=0$ this gives a Lie derivative,
\be
\delta S = \int_\gamma \pounds_\eta \left( \theta - H dt \right)
\ee
We have
\begin{align}
\pounds_\eta\theta &= \ii_\eta d\theta + d\ii_\eta \theta \nonumber\\
\pounds_\eta(Hdt) &= (\pounds_\eta H) dt = (\ii_\eta dH) dt 
\end{align}
which leads to
\be\label{deltaSEoM+BT}
\delta S = \int_\gamma \left( \ii_\eta \omega - (\ii_\eta dH) dt \right) + \ii_\eta \theta \big|_{\partial\gamma}
\ee
where the boundary term vanishes because $\eta = 0$ at the endpoints of $\gamma$ (recall that we are only allowing deformations that keep the endpoints fixed). This integral can be expressed as
\be
\delta S = \int_0^1\! dt \left( \ii_{\dot\gamma} \ii_\eta \omega - \ii_\eta dH \right)
\ee
where $\dot\gamma$ is the vector field tangent to $\gamma$. Thus we have
\be
\delta S = - \int_0^1\! dt\, \ii_\eta \left( \ii_{\dot\gamma} \omega + dH \right)
\ee
The variational principle says that $\delta S$ must vanish for all first-order deformations of the curve $\gamma$, i.e., for all $\eta$, so we conclude that
\be
dH = - \ii_{\dot\gamma} \omega
\ee
That is, the time evolution associated with the action $S$ (i.e., the curve $\gamma(t)$ satisfying the equations of motion) is precisely the symplectic motion generated by the Hamiltonian $H$.

\subsection{The CMC Hamiltonian}
\label{subsec:CMCHam}

We have learned that if one starts with an action $S$ which can be written in the form \eqref{actionH} for some phase space function $H$, then $H$ is the Hamiltonian which generates $t$-evolution. This tells us how to obtain the Hamiltonian on the reduced phase space. We begin with the Einstein-Hilbert action \eqref{GRaction}, written in the ADM form,
\be
S = \int dt \left( \int_\Sigma d^2\!x\, \pi^{ab} \dot h_{ab} - H \right)
\ee
where $H$ is the Hamiltonian given in \eqref{GRhamiltonian}. The ADM decomposition must be associated with our gauge-fixed choice of foliation, by CMC slices, so $t$ must be some monotonic function of the CMC time $\tau = - K$. However we will leave this relationship unspecified for now, simply keeping in mind that $\tau$ is some function of $t$, i.e., $\tau(t)$. As we must consider only paths restricted to the constraint surface, where $H=0$, we have
\be
S = \int dt \int_\Sigma d^2\!x\, \pi^{ab} \dot h_{ab}
\ee
The goal is to write this action as an integral over curves on the reduced phase space. To do so, let us first ``pull-back'' this action to the enlarged space $\bar{\ca S}$. Given a curve $\gamma(t) = (h_{ab}(t), \sigma^{ab}(t))$ on the contraint surface $\ca S$, there are curves $\bar\gamma(t) = (\Psi(t), \bar\sigma^{ab}(t))$ on $\bar{\ca S}$ which map into $\gamma$ under the transformation \eqref{barStoS}, i.e., $\gamma(t) = T_t( \bar\gamma(t))$. Note that we have a subscript on $T_t$ to emphasize that this map is time-dependent since $T$ depends on $\tau$ through $\lambda$, solution of \eqref{lambdaLich}. That there are many $\bar\gamma$ associated to each $\gamma$ is just a consequence of the fact that $\bar{\ca S}$ has more gauge ambiguities than $\ca S$. From \eqref{pidef2} and \eqref{varthetarel} we have
\be
d^2\!x\, \pi^{ab} = \Psi_* \left[ \vartheta_{\bar h} \left( e^{-\lambda} \bar\sigma^{ab} + \frac{1}{2} \tau \bar h^{ab} \right) \right]
\ee
The ``velocity'' $\dot h_{ab}$ can be written as $\delta h_{ab}(\dot\gamma)$, where $\dot\gamma$ is the vector tangent to $\gamma$. If $\dot{\bar\gamma} = (\xi, \alpha^{ab})$ is the vector tangent to $\bar\gamma$, then from a computation similar to what led to \eqref{deltah} we obtain
\be
\delta h_{ab}(\dot\gamma) = \Psi_* \left[ - \pounds_\xi (e^{\lambda} \bar h_{ab}) + \dot\lambda e^\lambda \bar h_{ab} \right]
\ee
where $\dot\lambda := \frac{d}{dt}\lambda_t$, with $\lambda_t$ being the solution of \eqref{lambdaLich} for the curve $(\Psi(t), \bar\sigma^{ab}(t); \tau(t))$. Putting these together we get
\be
S = \int dt \int  \vartheta_{\bar h} \left( e^{-\lambda} \bar\sigma^{ab} + \frac{1}{2} \tau \bar h^{ab} \right) \left( - \pounds_\xi (e^{\lambda} \bar h_{ab}) + \dot\lambda e^\lambda \bar h_{ab} \right)
\ee
where the $\Psi_*$ goes away because $\Psi$ is an automorphism of $D$ and the integral is thus invariant. Distributing the product we have
\be\label{S3terms}
S = \int dt \int  \vartheta_{\bar h} \left( - \bar\sigma^{ab} \pounds_\xi \bar h_{ab} - \frac{1}{2} \tau \bar h^{ab} \pounds_\xi (e^{\lambda} \bar h_{ab}) + \tau \frac{d}{dt} e^\lambda \right)
\ee
Let us analyze each of these three terms. The first term can be written as a total spatial derivative, since $\bar\sigma^{ab}$ is transverse with respect to $\bar h_{ab}$, and so we can apply Gauss' theorem,
\be
- \int  \vartheta_{\bar h} \bar\sigma^{ab} \pounds_\xi \bar h_{ab} = - 2 \int  \vartheta_{\bar h} \bar\sigma^{ab} \nabla_a \xi_b =  - 2 \int  \vartheta_{\bar h}  \nabla_a \left(\bar\sigma^{ab}  \xi_b \right) = - 2 \int\! d\theta\, n_a \bar\sigma^{ab} \xi_b
\ee
From the association $\ac\sigma^{ab} \leftrightarrow \ac\sigma$, defined in \eqref{hatsigmadef}, we identify
\be\label{sigmahatsigmabarrelation}
- \int  \vartheta_{\bar h} \bar\sigma^{ab} \pounds_\xi \bar h_{ab} = \ac\sigma(\hat\xi)
\ee
where $\hat\xi$ is the restriction of $\xi$ to $\partial D$. This reveals that this term depends only on the curve $\hat\gamma(t) := (\psi(t), \ac\sigma(t))$  projected from $\bar{\ca S}$ to $\hat{\ca S}$, where $\psi(t) = \Psi(t)|_{\partial D}$ and $\bar\sigma^{ab}(t) \leftrightarrow \ac\sigma(t)$. In particular, note that the tangent vector to $\hat\gamma$ is given by $\dot{\hat\gamma} = (\hat\xi, \hat\alpha)$, where $\alpha^{ab} \leftrightarrow \hat\alpha$. In fact, this term really looks like a $p\dot q$ term on $\hat{\ca S}$.  Our goal, however, is to have a $p\dot q$ term on $\wt{\ca P}$. From formula \eqref{hatthetaformula}, relating $\hat\theta = J^*\theta$ and $\ac\sigma$, we can express this term as
\be
\ac\sigma (\hat\xi) = J^*\theta(\dot{\bar\gamma}) = \theta(J_*\dot{\bar\gamma})
\ee
This further reveals that this term depends only on the curve $\wt\gamma := J \circ \bar\gamma$  projected from $\hat{\ca S}$ to $\wt{\ca P}$, since $J_*\dot{\bar\gamma} = \dot{\wt\gamma}$ is precisely the vector tangent to $\wt\gamma$. Thus we have
\be
- \int  \vartheta_{\bar h} \bar\sigma^{ab} \pounds_\xi \bar h_{ab} = \theta(\dot{\wt\gamma})
\ee
which is the desired $p\dot q$ term on the reduced phase space. The second term on \eqref{S3terms} can be worked out as follows,
\be
- \frac{1}{2} \tau \bar h^{ab} \pounds_\xi (e^{\lambda} \bar h_{ab}) = - \frac{1}{2} \tau \bar h^{ab} \left( \xi^c \nabla_c (e^\lambda)  \bar h_{ab} + e^{\lambda} 2 \nabla_a \xi_b \right) = -\tau \nabla_a \left( e^\lambda \xi^a \right)
\ee
so again we have a total spatial derivative and we can apply Gauss' theorem,
\be
- \frac{1}{2} \tau \int  \vartheta_{\bar h} \bar h^{ab} \pounds_\xi (e^{\lambda} \bar h_{ab}) =  -\tau \int  \vartheta_{\bar h} \nabla_a \left( e^\lambda \xi^a \right) = -\tau \int\!d\theta\, n_a e^\lambda \xi^a = 0
\ee
and we conclude that it vanishes since $\xi$ is tangent to the boundary. Finally, the third term on \eqref{S3terms} can be integrated by parts (in time) to yield
\be
\int\!dt\, \tau \frac{d}{dt} \int \! \vartheta_{\bar h} e^\lambda = - \int\! dt\,  \frac{d\tau}{dt} \int \! \vartheta_{\bar h} e^\lambda + \left[ \tau \int  \vartheta_{\bar h} e^\lambda \right]_0^1
\ee
Therefore, up to a time-boundary term (which does not affect the equations of motion), we have cast the action in the form
\be
S[\gamma] = \int_{\wt\gamma} \left( \theta - \int \! \vartheta_{\bar h} e^\lambda \, d\tau \right)
\ee
where $\wt\gamma$ is the projection of $\gamma$ to the reduced phase space. We conclude that the reduced Hamiltonian conjugated to the time variable $\tau$ must be given by
\be\label{Hred}
\wt H = \int \! \vartheta_{\bar h} e^\lambda
\ee
Although it should follow just from consistency that this $\wt H$ is a well-defined function on $\wt{\ca P}$, we will show it explicitly. Before presenting a technical proof, let us give a physical interpretation for this Hamiltonian, which should by itself clarify why $\wt H$ is well-defined on the reduced phase space. From \eqref{varthetarel} and \eqref{hconftrans} we have 
\be
\wt H = \int \! \vartheta_{e^\lambda\bar h} = \int \! \vartheta_{\Psi^{-1}_* h} 
\ee
since $\vartheta$ is the covariant volume element, it transforms nicely under a diffeomorphism, i.e., $\vartheta_{\Psi^{-1}_* h} = \Psi^{-1}_* \vartheta_{h}$. Using the invariance of the integral under automorphisms, we get
\be
\wt H = \int \! \vartheta_{h} 
\ee
That is, $\wt H$ is equal to the {\sl area of the CMC slice} with mean curvature $K = -\tau$. Since this is a gauge-invariant property of the spacetime, it should be a well-defined function on the reduced phase space. This Hamiltonian associated with evolution along CMC slices is known as the {\sl York Hamiltonian}~\cite{york1972role}.

\subsection{The Hamiltonian is well-defined}
\label{subsec:Hwelldef}

Now, just for completeness, we will prove explicitly that $\wt H$ in \eqref{Hred} is a well-defined function on $\wt{\ca P}$. First we note that $\wt H$ is a well-defined function on $\hat{\ca S}$, $\wt H(\psi, \ac\sigma)$, since $\lambda$ is the unique solution of \eqref{lambdaLich} where $\sigma^{ab}$ is determined by $\ac\sigma$ via the mode expansions in \eqref{sigmaab} and \eqref{hatsigma} and the boundary value of $\lambda$ is determined by $\psi = \Psi|_{\partial D}$ via \eqref{lambdabdycond}. We therefore just need to show that $\wt H$ projects nicely through $J$ to $\wt{\ca P}$, i.e., that $\wt H(\psi, \ac\sigma) = \wt H(\psi', \ac\sigma')$ whenever $J(\psi, \ac\sigma) = J(\psi', \ac\sigma')$. 

Let us go back, temporarily, to the notation used in the theorem proven in point $(i)$ after \eqref{redphase}. We wish to have an explicit characterization of the pre-image under $J$ of points in $T^*(G/H)$; in particular, we want to know how the gauge group $H$ acts on $G \times \ac{\fr g}^*$, for the orbits of $H$ are precisely the pre-images under $J$ of points in $T^*(G/H)$. Consider a transformation on $G \times \ac{\fr g}^*$, $(g, \sigma) \mapsto (g', \sigma')$, such that
\be
J(g, \sigma) = J(g', \sigma')
\ee
This requires that $[g] = [g']$, and so there must exist $h \in H$ such that $g' = gh$. Also $J(g, \sigma)(V) = J(gh, \sigma')(V)$ for all vectors $V$ at $[g] \in G/H$. This implies 
\be
\sigma(\Xi(X)) = \sigma'(\Xi(X'))
\ee
where $X$ and $X'$ are vectors at $g \in G$ and $gh \in G$, respectively, such that $q_*X = q_*X' = V$. Note that, given $X$ at $g$, we can always choose $X' = r_{h*}X$ as the vector at $gh$. This follows from the fact that $q(g) = q(gh) = q\circ r_h(g)$, so $q_*X' = q_*(r_{h*}X) = q_*X = V$. Thus we have
\be
\sigma(l_{g^{-1}*}X) = \sigma'(l_{(gh)^{-1}*} r_{h*} X) = \sigma'(l_{h^{-1}*} l_{g^{-1}*} r_{h*} X) = \sigma'(\text{Ad}_{h^{-1}*} l_{g^{-1}*}X) 
\ee
As we can choose any $V$ at $[g]$, in particular we can choose $V = q_* l_{g*} \xi$, for any $\xi \in \fr g$, so we conclude that $\sigma' = \text{Ad}_{h}^*\sigma$ or, using the definition of the coadjoint action introduced in \eqref{coaddef},
\be
\sigma' = \text{coad}_{h^{-1}}\sigma
\ee
Thus we can define the (right) $H$-action $\Gamma_h$ on $G \times \ac{\fr g}^*$ by
\be
\Gamma_h(g, \sigma) := (gh, \text{coad}_{h^{-1}}\sigma)
\ee
and this is precisely the gauge group of $G \times \ac{\fr g}^*$, that is, $J(g, \sigma) = J(g', \sigma')$ if and only if $(g', \sigma') = \Gamma_h (g, \sigma)$ for some $h \in H$. Note that this result can be straightforwardly used to derive the kernel of $J_*$, as characterized in \eqref{kerJ*general}. 

Particularizing to $\diff \times \whddiff$, we consider the transformation
\be\label{pslgaugeaction}
(\psi, \ac\sigma) \mapsto (\psi', \ac\sigma') = (\psi \varphi, \text{coad}_{\varphi^{-1}}\ac\sigma)
\ee
where $\varphi \in \psl$. We are interested in computing how $\lambda$ changes under this transformation. Recall that $\lambda$ is the solution of 
\be
\bar\nabla^2 \lambda + e^{-\lambda} \bar\sigma^{ab} \bar\sigma_{ab} - e^\lambda \chi = 0
\ee
with boundary conditions
\be
\left. e^\lambda \bar h \right|_{\partial D} = \psi^{-1}_* \gamma
\ee
The transformed $\lambda$, $\lambda'$, will be the solution of 
\be
\bar\nabla^2 \lambda' + e^{-\lambda'} {\bar\sigma}^{\prime ab} {\bar\sigma'}_{ab} - e^{\lambda'} \chi = 0
\ee
with boundary conditions
\be
\left. e^{\lambda'} \bar h \right|_{\partial D} = {\psi}^{'-1}_* \gamma
\ee
Let us first study how the boundary conditions for $\lambda$ and $\lambda'$ are related. Recall that $\psl$ action on $S^1$ is simply the boundary action of the diffeomorphism part of the conformal isometries of $D$, that is, given $\varphi \in \psl \subset \diff$, there exists a unique $\Phi \in \text{Diff}^+(D)$ satisfying $\Phi|_{\partial D} = \varphi$ and a unique $\Omega \in C^\infty(D, \bb R^+)$ such that
\be
\bar h = \Phi_* \Omega \bar h
\ee
We have
\be
\left. e^{\lambda'} \bar h \right|_{\partial D} = {\varphi}^{-1}_*  {\psi}^{-1}_* \gamma = {\varphi}^{-1}_*\left. e^\lambda \bar h \right|_{\partial D}
\ee
so
\be
\left. \bar h \right|_{\partial D} = e^{-\lambda} \varphi_* \left. e^{\lambda'} \bar h \right|_{\partial D} = \varphi_* \left. e^{-\lambda \circ \varphi} e^{\lambda'} \bar h \right|_{\partial D}
\ee
so we can identify
\be
\left. e^{-\lambda \circ \varphi + \lambda'} \right|_{\partial D} = \Omega \Big|_{\partial D} 
\ee
which gives
\be
\lambda'\Big|_{\partial D}  = \left( \lambda \circ \varphi + \log\Omega \right) \Big|_{\partial D} 
\ee
This suggests a convenient change of variables from $\lambda'$ to $\wt\lambda$ defined as
\be
\wt\lambda  := \Phi_* (\lambda' - \log\Omega )
\ee
since this implies that $\wt\lambda$ is subjected to the same boundary conditions as $\lambda$, that is,
\be
\wt\lambda \Big|_{\partial D}  = \lambda \Big|_{\partial D} 
\ee
It is natural to wonder if $\wt\lambda$ also satisfies the same equation as $\lambda$, which would imply that $\wt\lambda = \lambda$. To investigate this possibility, consider the action of the Lichnerowicz operator (for $\ac\sigma$) on $\wt\lambda$,
\be\label{Lichoperator}
L\wt\lambda := \bar\nabla^2 \wt\lambda + e^{-\wt\lambda} \bar\sigma^{ab} \bar\sigma_{ab} - e^{\wt\lambda} \chi
\ee
The first term can be manipulated as
\be
\nabla^2_{(\bar h)} \wt\lambda = \nabla^2_{(\bar h)}\Phi_* (\lambda' - \log\Omega ) = \Phi_* \left[ \nabla^2_{(\Phi^{-1}_*\bar h)} (\lambda' - \log\Omega ) \right] = \Phi_* \left[ \nabla^2_{(\Omega\bar h)} (\lambda' - \log\Omega ) \right]
\ee
where $\nabla^2_{(h)}$ denotes Laplacian covariantly associated with $h$, and it transforms in a simple manner under the Weyl scaling
\be
\nabla^2_{(\bar h)} \wt\lambda = \Phi_* \left[ \Omega^{-1} \nabla^2_{(\bar h)} (\lambda' - \log\Omega ) \right] = \Phi_* \left[ \Omega^{-1} \nabla^2_{(\bar h)}\lambda'  \right]
\ee
in which we used that $\nabla^2_{(\bar h)} \log\Omega = 0$.\footnote{As explained before, this follows from the fact that $\Phi^{-1}_*\bar h$ and $\Omega\bar h$ are both flat and the Ricci curvature transforms like $R_{(\Omega\bar h)} = \Omega^{-1} \left(R_{(\bar h)} - \nabla^2_{(\bar h)}\log\Omega \right)$.} The second term gives,
\be
e^{-\wt\lambda} \bar\sigma^{ab} \bar\sigma_{ab} = e^{-\Phi_* (\lambda' - \log\Omega )} \bar\sigma^{ab} \bar\sigma_{ab} = \Phi_*\left[  e^{-\lambda'} \Omega \, \Phi^{-1}_*\! \left(\bar\sigma^{ab} \bar\sigma_{ab} \right) \right]
\ee
and the $\bar\sigma^{ab} \bar\sigma_{ab}$ term can be written as 
\be
\Phi^{-1}_*\! \left(\bar\sigma^{ab} \bar\sigma_{ab} \right) = \Phi^{-1}_*\! \left(\bar h_{ac} \bar h_{bd} \bar\sigma^{ab} \bar\sigma^{cd} \right) = \Omega \bar h_{ac} \Omega \bar h_{bd} (\Phi^{-1}_*\bar\sigma^{ab}) (\Phi^{-1}_* \bar\sigma^{cd})
\ee
and so
\be
e^{-\wt\lambda} \bar\sigma^{ab} \bar\sigma_{ab} =  \Phi_*\left[ \Omega^3 e^{-\lambda'}  \bar h_{ac} \bar h_{bd} (\Phi^{-1}_*\bar\sigma^{ab}) (\Phi^{-1}_* \bar\sigma^{cd}) \right]
\ee
The third term is easy to compute,
\be
 e^{\wt\lambda} \chi =  e^{\Phi_* (\lambda' - \log\Omega )} \chi = \Phi_* \left[ \Omega^{-1} e^{\lambda'} \chi    \right]
\ee
Now let us see how the transformation $\ac\sigma \mapsto \ac\sigma'$ is expressed as $\bar\sigma^{ab} \mapsto {\bar\sigma}^{\prime ab}$. Given a vector field $\hat\xi$ on $S^1$, representing an element of the algebra of $\diff$, we have
\be
\ac\sigma'(\hat\xi) = \text{Ad}^*_{\varphi}\ac\sigma(\hat\xi) = \ac\sigma(\text{Ad}_{\varphi *}\hat\xi) = \ac\sigma(\varphi_*\hat\xi)
\ee
where we used that $\text{ad}_{\varphi}\hat\xi = \varphi_*\hat\xi$. This formula for the adjoint action on a group of diffeomorphisms is intuitive because $\text{ad}_{\varphi}$ is a linear map on the space of vectors fields $\hat\xi$ and it carries a representation of the group, so it is natural to guess that it should act as the push-forward $\varphi_*$.\footnote{Here is the proof. Let $\ca M$ be a manifold and $\text{Diff}(\ca M)$ be its group of diffeomorphisms. Let the action of $\text{Diff}(\ca M)$ on $\ca M$ be denoted by $\Gamma_p(\psi) := \psi(p)$, where $p \in \ca M$, so the vector field induced by $\xi \in \fr{diff}(\ca M)$ is given by $X^\xi_p := \Gamma_{p*}\xi$.  Given two diffeomorphisms $\phi$ and $\psi$, note that $\psi \circ \phi = \psi\phi\psi^{-1}\psi = \text{Ad}_\psi \circ \psi$, which can also be expressed as $\Gamma_{\psi(p)} \circ \text{Ad}_\psi = \psi \circ \Gamma_p$. Taking the derivative of this expression and acting on $\xi$ gives $X^{\text{Ad}_\psi \xi}_{\psi(p)} = \psi_* (X^\xi_p)$, where $\text{ad}_\psi := \text{Ad}_{\psi*}$. Under the identification $\xi \sim X^\xi$ this reads $\text{ad}_\psi \xi = \psi_* \xi$.} From the identification \eqref{hatsigmadef} we obtain
\be
\int\! \vartheta_{\bar h}\, \bar\sigma^{\prime ab} \pounds_\xi \bar h_{ab} = \int\! \vartheta_{\bar h}\, \bar\sigma^{ab} \pounds_{\Phi_*\xi} \bar h_{ab}
\ee
where $\xi$ is any extension of $\hat\xi$ to $D$. In principle this equality is also valid for an arbitrary extension of $\varphi$ to $D$, but we chose to extend it as $\Phi$ for convenience. Pulling out the diffeomorphism on the right side of the equality gives
\be
\int\! \vartheta_{\bar h}\, \bar\sigma^{ab} \pounds_{\Phi_*\xi} \bar h_{ab} = \int\! \Phi_* \left[ \vartheta_{\Phi^{-1}_*\bar h}\, (\Phi^{-1}_*\bar\sigma^{ab}) \pounds_{\xi} (\Phi^{-1}_* \bar h_{ab}) \right] = \int\! \vartheta_{\Omega\bar h}\, (\Phi^{-1}_*\bar\sigma^{ab}) \pounds_{\xi} (\Omega \bar h_{ab})
\ee
where we used the invariance of the integral under a diffeomorphism. Note that
\be
(\Phi^{-1}_*\bar\sigma^{ab}) \pounds_{\xi} (\Omega \bar h_{ab}) = (\Phi^{-1}_*\bar\sigma^{ab}) \left[ \pounds_{\xi}\Omega \bar h_{ab} + \Omega \pounds_\xi \bar h_{ab} \right] = \Omega (\Phi^{-1}_*\bar\sigma^{ab}) \pounds_\xi \bar h_{ab}
\ee
where the first term inside the brackets vanishes because of the transverseness of $\bar\sigma^{ab}$, i.e., $(\Phi^{-1}_*\bar\sigma^{ab})\bar h_{ab} = \Omega^{-1} \Phi^{-1}_* [ \bar\sigma^{ab}  \bar h_{ab} ] = 0$. Thus, 
\be
\int\! \vartheta_{\bar h}\, \bar\sigma^{\prime ab} \pounds_\xi \bar h_{ab} = \int\! \vartheta_{\bar h}\, (\Omega^2 \Phi^{-1}_*\bar\sigma^{ab}) \pounds_{\xi} \bar h_{ab}
\ee
so we can identify
\be
\bar\sigma^{\prime ab} = \Omega^2 \Phi^{-1}_*\bar\sigma^{ab}
\ee
This identification is legitimate because $\Omega^2 \Phi^{-1}_*\bar\sigma^{ab}$ is traceless and transverse with respect to $\Phi_*\Omega\bar h = \bar h$ and it is associated with  $\ac\sigma'$ via \eqref{hatsigmadef}. With this expression, we can rewrite the second term in \eqref{Lichoperator} as
\be
e^{-\wt\lambda} \bar\sigma^{ab} \bar\sigma_{ab} = e^{-\Phi_* (\lambda' - \log\Omega )} \bar\sigma^{ab} \bar\sigma_{ab} = \Phi_*\left[ \Omega^{-1} e^{-\lambda'} {\bar\sigma}^{\prime ab} {\bar\sigma'}_{ab}  \right]
\ee
Putting the three terms together we obtain
\be
L\wt\lambda = \Phi_*\Omega^{-1} \left[  \bar\nabla^2 \lambda' + e^{-\lambda'} {\bar\sigma}^{\prime ab} {\bar\sigma'}_{ab} - e^{\lambda'} \chi \right] = 0
\ee
so we conclude that $\wt\lambda$ is in fact equal to $\lambda$, which implies that
\be
\lambda' = \lambda \circ \Phi + \log\Omega
\ee
The transformed Hamiltonian is then
\be
\wt H(\psi', \ac\sigma') = \int \! \vartheta_{\bar h} e^{\lambda'} = \int \! \vartheta_{\bar h} \Omega  e^{\Phi^{-1}_*\lambda} = \int \! \vartheta_{\Phi^{-1}_*\bar h} \Phi^{-1}_* e^{\lambda} = \int \! \Phi^{-1}_*\! \left[ \vartheta_{\bar h} e^{\lambda} \right] = \wt H(\psi, \ac\sigma)
\ee
finishing the proof that $\wt H$ is indeed a well-defined function on $\wt{\ca P}$.

\section{Approximations for the Hamiltonian}
\label{sec:approxH}

The Hamiltonian generating evolution along CMC slices is, as we have just seen, a $\tau$-dependent function on the reduced phase space given by the area of the CMC slice with mean curvature $K = -\tau$. Despite its simple geometric interpretation, this function is extremely complicated: in order to evaluate $\wt H$ at a point $p \in \wt{\ca P}$, we need to take any $(\psi, \ac\sigma) \in J^{-1}(p)$ and solve the differential equation \eqref{lambdaLich} for $\lambda$, in which $\ac\sigma$ enters through the term $\bar\sigma^{ab}\bar\sigma_{ab}$ and $\psi$ enters as a boundary condition, and then integrate $e^\lambda$ over the disc $D$. It is thus worthwhile to investigate whether there are certain regimes in which the Hamiltonian can be approximated by something simpler. There are three independent length scales in our problem,
\begin{itemize}
\item the boundary length, $\ell$;
\item the AdS length, $\ell_\text{AdS} := \frac{1}{\sqrt{-\Lambda}}$;
\item the Planck length, $\ell_P := \hbar G$;
\end{itemize}
and we recall that we are considering $c=1$. As usual in quantum field theory, we shall also consider $\hbar = 1$, so $\ell_P = G$. We wish to explore the different limits of these scales.

\subsection{Reintroducing the physical scales}

Let us explicitly reintroduce these scales in our formulas. As we wish to take $\bar h_{ab}$ as corresponding to the unit-radius (dimensionless) disc, $\bar h = dr^2 + r^2 d\theta^2$, let us redefine the conformal map in \eqref{hconftrans} as
\be\label{hconftrans2}
h_{ab} = \left( \frac{\ell}{2\pi} \right)^2 \Psi_* e^\lambda \bar h_{ab}
\ee
and the boundary condition \eqref{lambdabdycond} reads
\be
 \left. e^\lambda \bar h \right|_{\partial D} = \left( \frac{\ell}{2\pi} \right)^{-2} \psi^{-1}_* \gamma
\ee
In this way, the length scale is fully encoded in $h_{ab}$, while $\psi$ and $\lambda$ are dimensionless. In particular, if $\gamma = (\ell/2\pi)^2 d\theta^2$, then 
\be\label{lambdabdycond2}
\left. e^\lambda d\t^2 \right|_{\partial D} = \psi^{-1}_* d\t^2
\ee
There is no reason not to make that choice since $\gamma$ can always be parametrized by proper length, and we can define the unit disc by a uniform rescaling of the boundary.
Similarly, a reasonable redefinition of the map in \eqref{sigmaconftrans} would be
\be\label{sigmaconftrans2}
\sigma^{ab} =  \left( \frac{\ell}{2\pi} \right)^{-3} \Psi_* e^{-2\lambda} \bar\sigma^{ab}
\ee
so that $\bar\sigma^{ab}$ is dimensionless, since $\sigma^{ab}$ has dimensions of $\text{\sl length}^{-3}$.\footnote{A brief comment on how we are assigning dimensions to tensors. The length of a curve $\gamma$ is given by $\int dt \sqrt{h_{ab} \dot\gamma^a \dot\gamma^b}$, and this must have dimension $L$. The tangent vector $\dot\gamma$ can be defined for its action of functions as $\dot\gamma(f) = \frac{d}{dt} f(\gamma(t))$, so we see that it is natural to associated dimension $T^{-1}$ to $\dot\gamma$, where $T$ is the dimension of the parameter $t$. Therefore, from the formula for the length of $\gamma$, we have $L \sim T \sqrt{h_{ab} T^{-1} T^{-1}}$, so we conclude that the metric $h_{ab}$ must have dimension $L^2$. This is consistent with a typical expression like $h = dx^2 + dy^2$, where the coordinates $x$ and $y$ have dimension $L$. The extrinsic curvature is defined by a formula like $K_{ab} = \frac{1}{2} \pounds_n h_{ab}$ where $n$ is a unit vector (and so $n$ must have dimension $L^{-1}$). Therefore, $K_{ab} \sim L^{-1} L^2 \sim L$. Since $h^{ab} h_{bc} = \delta^a_c$, and $\delta^a_c$ is naturally dimensionless, we must have $h^{ab} \sim L^{-2}$. Note that this gives $K = K_{ab}h^{ab} \sim L L^{-2} \sim L^{-1}$, which is consistent with the interpretation that $K$ gives the rate (per unit of normal length) that the volume $\delta \vartheta$ of a piece of the surface changes, $K = \frac{1}{\delta \vartheta}\frac{d\delta \vartheta}{dl}$. Lastly, $K^{ab} = h^{ac} h^{bd} K_{cd} \sim L^{-2} L^{-2} L \sim L^{-3}$, so $\sigma^{ab}$ have dimension $L^{-3}$.} However, a more natural redefinition turns out to be
\be\label{sigmaconftrans3}
\sigma^{ab} =  16\pi\ell_P \left( \frac{\ell}{2\pi} \right)^{-4} \Psi_* e^{-2\lambda} \bar\sigma^{ab}
\ee
which also leads to a dimensionless $\bar\sigma^{ab}$. The reason for introducing this $\sim\! G$ factor is because the correct expression for the conjugate momentum, when the overall $(16\pi G)^{-1}$ factor in the action is not omitted, is
\be\label{pidef3}
\pi^{ab} = \frac{1}{16\pi G}\sqrt{\text{det}(h)} \left( \sigma^{ab} + \frac{1}{2} \tau h^{ab} \right)
\ee
so this $(16\pi G)^{-1}$ factor appears in the rightmost side of the expression \eqref{Omegarep} for the pre-symplectic form $\Omega$. In this way, the inclusion of the $16\pi\ell_P$ in \eqref{sigmaconftrans} exactly cancels this factor in the corresponding (pre-)symplectic form on $\bar{\ca S}$, $\bar\omega$, given in \eqref{baromega}. In fact, \eqref{baromega} is correct as it stands, as $\ell$ will not appear either. This is because $\ell$, as introduced above, can be completely absorbed in an effective redefinition of $\lambda$, 
\begin{align}
h_{ab} &=  \Psi_* e^{\lambda + 2\log(\ell/2\pi)} \bar h_{ab} \nonumber\\
\sigma^{ab} &=  16\pi\ell_P \Psi_* e^{-2(\lambda + 2\log(\ell/2\pi))} \bar\sigma^{ab} \label{conftrans2}
\end{align}
and $\lambda$ does not appear explicitly in \eqref{baromega}. This means that all formulas in the remaining of that section are correct as they stand. In particular, the association $\bar{\sigma}^{ab} \leftrightarrow \ac\sigma$, the projection map $J$, and the fact that the symplectic form on $\wt{\ca P}$ is equal to the canonical 2-form associated with its cotangent bundle structure all remain unchanged. There are only two formulas that need to be updated in view of \eqref{conftrans2}, the Lichnerowicz equation \eqref{lambdaLich} and the reduced Hamiltonian \eqref{Hred}. Since the Lichnerowicz equation is a direct consequence of the Hamiltonian constraint, \eqref{sigmaH2}, which is independent of the overall factor $(16\pi G)^{-1}$ in the action, we have
\be
\bar\nabla^2 \left(\lambda + 2\log(\ell/2\pi) \right)+ e^{-\left(\lambda + 2\log(\ell/2\pi) \right)} \big( 16\pi\ell_P \bar\sigma^{ab} \big) \big( 16\pi\ell_P \bar\sigma_{ab} \big) - e^{\left(\lambda + 2\log(\ell/2\pi) \right)} \chi = 0
\ee
which gives
\be\label{lambdaLich2}
\bar\nabla^2 \lambda + \left( \frac{32\pi^2\ell_P}{\ell} \right)^2 \bar\sigma^{ab} \bar\sigma_{ab} \, e^{-\lambda}  -  \left( \frac{\ell}{2\pi} \right)^2 \left( \frac{2}{\ell^2_\text{\sl AdS}} + \frac{\tau^2}{2} \right) e^\lambda = 0
\ee
where we wrote $\chi$ explicitly to display the AdS length. The reduced Hamiltonian gets a factor $(16\pi G)^{-1}$ from the action, and another from the effective redefinition of $\lambda$, so it becomes
\be\label{Hred2}
\wt H = (16\pi \ell_P)^{-1} \int \! \vartheta_{\bar h} e^{\left(\lambda + 2\log(\ell/2\pi) \right)} = \frac{\ell^2}{64\pi^3 \ell_P} \int \! \vartheta_{\bar h} e^\lambda
\ee
Note that it has dimension of $\text{\sl length}$ because it is conjugated to a time, $\tau$, which has dimension of $\text{\sl length}^{-1}$. While the overall factor in the Hamiltonian is not particularly interesting, the different length scales still enter implicitly through $\lambda$, which is a solution of \eqref{lambdaLich2}. 

Let us now study more carefully the behavior of the Lichnerowicz equation \eqref{lambdaLich2} in different regimes. For reference, let us cast it in the more compact form
\be\label{lambdaLich3}
\bar\nabla^2 \lambda + \kappa e^{-\lambda}  -  \bar\chi e^\lambda = 0
\ee
where
\be\label{Lichkappadef}
\kappa := \left( \frac{32\pi^2\ell_P}{\ell} \right)^2 \bar\sigma^{ab} \bar\sigma_{ab}
\ee
is some sort of kinetic energy term, since it is quadratic in $\bar\sigma^{ab}$ which is a ``momentum'' variable, and
\be\label{Lichbarchidef}
\bar\chi := \left( \frac{\ell}{2\pi} \right)^2 \left( \frac{2}{\ell^2_\text{\sl AdS}} + \frac{\tau^2}{2} \right)
\ee
is a dimensionless version of $\chi$, just a (time-dependent) constant. It is worth noting that this equation can be derived from a variational principle. The underlying functional is given by
\be\label{LichIfunctional}
I[\lambda] := \int_D \! \vartheta_{\bar h} \left( \frac{1}{2} |\bar\nabla \lambda |^2 + \kappa e^{-\lambda} + \bar\chi e^\lambda \right)
\ee
where $|\bar\nabla \lambda |^2 = \bar h^{ab} \bar\nabla_a \lambda \bar\nabla_b \lambda$. In the extremization, the variable $\lambda : D \rightarrow \bb R$ is varied with its boundary value fixed by \eqref{lambdabdycond2}. Computing the variation of $I$ up to second order in $\delta \lambda$ gives
\be
\delta I = - \int \! \vartheta_{\bar h} \left( \bar\nabla^2 \lambda + \kappa e^{-\lambda}  -  \bar\chi e^\lambda \right) \delta\lambda + \frac{1}{2} \int \! \vartheta_{\bar h} \left( |\bar\nabla \delta\lambda |^2 + \kappa e^{-\lambda} \delta\lambda^2 + \bar\chi e^\lambda \delta\lambda^2 \right)
\ee
so from the first-order term we recover \eqref{lambdaLich3} and from the second-order term we conclude that the extremum is a (global) minimum of $I$. There is a simple pictorial interpretation for this variational principle: we imagine $\lambda$ as specifying the configuration of a membrane in $D \times \bb R$. This membrane is a cross-section of this (solid) cylindrical space, fixedly attached to the boundary (as determined by $\psi$), which can be deformed up and down in the bulk. The first term in the functional is an {\sl elastic potential energy} of a ``Hookean'' membrane\footnote{The term ``Hookean'' is used here to describe an elastic membrane that sustains a  tension proportional to the local stretching factor, i.e., $\tau = k (\delta a'/\delta a)$ where $\delta a'$ is the area of a piece of membrane with relaxed area $\delta a$. If $\delta a'$ is further stretched by an infinitesimal normal displacement $\delta s$, its area will change as $\delta a' \rightarrow \delta a' + \int dl \delta s$, where $dl$ is the length element along the boundary of $\delta a'$. The work done to stretch this piece is $\delta U = \int\!dl\, \tau \delta s = \int\!dl\, k (\delta a'/\delta a) \delta s = (k/\delta a) \int d(\delta a') (\delta a') = \frac{k}{2} \delta a'^2/\delta a$. The total potential energy of the membrane is thus $U =  \int \delta a \frac{k}{2} \left( \frac{\delta a'}{\delta a}\right)^2$. Now consider the membrane in the Euclidean cylinder $D \times \bb R$, and say that $z = \lambda = 0$ is the relaxed configuration. Then for a generic configuration $z = \lambda(x)$, obtained by deforming the membrane vertically, $(x, 0) \rightarrow (x, \lambda(x))$, a relaxed piece of area $\delta a = d^2\!x$ will be stretched to $\delta a' = \sqrt{1 + |\nabla \lambda|^2} d^2\!x$, so the potential energy (for $k=1$) will be given by $U = \int_D d^2\!x\, \frac{1}{2} \left( 1 + |\nabla \lambda|^2 \right) = \frac{\pi}{2} + \int_D d^2\!x\, \frac{1}{2} |\nabla \lambda|^2$.} and, in the minimization process, it offers a resistance against stretching the membrane too much; the second term, proportional to the {\sl kinetic energy} $\kappa$, tries to pull the membrane upwards; and the third term, acting as a time-dependent {\sl external force}, tries to pull it downwards. This picture provides a clear way to see that the Hamiltonian tends to increase with the kinetic energy $\kappa$, just as in typical systems.

\subsection{Liouville equation}

The Lichnerowicz equation, \eqref{lambdaLich3}, is a quite complicated non-linear partial differential equation. It is remarkable that it can be solved exactly in the case of zero momentum. In 1853, Liouville decided to study the equation
\be
\frac{\partial^2 \log \Omega}{\partial u \partial v} - \frac{\bar\chi}{4} \Omega = 0
\ee
managing to find a general analytic solution for it \cite{liouville1853equation}. This is equation, which we shall call the {\sl Liouville equation}, is quite reminiscent of \eqref{lambdaLich3} when the kinetic term is absent,
\be\label{Lichzeromom}
\bar\nabla^2 \lambda  - \bar\chi e^\lambda = 0
\ee
In fact, they are related by a change of variables $\Omega = e^\lambda$ and
\begin{align}
u &= x + iy \nonumber\\
v &= x - iy
\end{align}
where $x$ and $y$ are Cartesian coordinates on $(D, \bar h)$. Since $u$ and $v$ are just the complex coordinates associated with $(x, y)$, we shall use the more standard notation of $z = x + iy$ and $\bar z = x - iy$. The general solution for this equation is given by
\be
\lambda = \log\left[ \frac{8}{\bar\chi} \frac{f'(z) g'(\bar z) }{\left( 1 - f(z) g(\bar z) \right)^2}  \right]
\ee
where $f$ and $g$ are arbitrary meromorphic and anti-meromorphic functions on $\bb D \sim D$, and $f'$ and $g'$ are their respective first derivatives. 
Since $\lambda$ must be smooth and real on $D$, we must impose some additional conditions for $f$ and $g$. To ensure reality, the simplest condition is to take $g(\bar z) = \overline{f(z)}$, so we would get
\be\label{Lichzeromomsol}
\lambda = \log\left[ \frac{8}{\bar\chi} \frac{|f'(z)|^2 }{\left( 1 - |f(z)|^2 \right)^2}  \right]
\ee
To ensure smoothness, we may require that $f'$ is nowhere vanishing on $\bb D$ and that the image of $f$ is either entirely in $\bb D$ or in $\bb C - \bb D$, so the denominator never vanishes.

An interesting solution is the one corresponding to the {\sl symmetric diamond}, i.e., the state $(\psi, \ac\sigma) = (\id, 0)$. This is the diamond whose maximal surface is an Euclidean disc. This case is particularly simple so that we do not even need to use the general solution described above, as the corresponding Lichnerowicz reduces to an ordinary differential equation for $\lambda_0(r, \theta) = \lambda_0(r)$, in polar coordinates,
\be
\frac{1}{r} \frac{d}{dr}\left( r \frac{\lambda_0}{dr} \right) - \bar\chi e^{\lambda_0} = 0
\ee
satisfying boundary conditions $\lambda(1) = 0$ and $\lambda'_0(0) = 0$ (where the derivative condition is so that $\lambda_0(r,\theta)$ is regular at the origin). The solution is given by
\be\label{lambda0}
\lambda_0(r, \theta) = \log\!\left( \frac{8C}{(\bar\chi C - r^2)^2} \right) \,,\quad C = \frac{\bar\chi + 4 + \sqrt{8\bar\chi + 16}}{\bar\chi^2}
\ee
Notice that this result can also be obtained from the general solution \eqref{Lichzeromomsol} for the function
\be
f(z) = \frac{z}{\sqrt{\bar\chi C}}
\ee
In fact, the choice $f(z) = \sqrt{\bar\chi C}/z$ also leads to the same result, revealing that different $f$s may correspond to the same solution. 

We can compute explicitly the Hamiltonian for the symmetric diamond. Inserting \eqref{lambda0} into \eqref{Hred2} we get
\be\label{wtH0exact}
\wt H_0 = \frac{\ell^2}{64\pi^3 \ell_P} 2\pi \int_0^1 \! rdr\, e^{\lambda_0(r)} = \frac{\ell^2}{32\pi^2 \ell_P \left( 1 + \sqrt{\sfrac{\bar\chi}{2} + 1} \right)}
\ee
Observe that, as expected, the Hamiltonian is maximal at $\tau = 0$, decreasing monotonically as $\tau^2$ increases.

\subsection{Large boundary length}

An interesting regime is that of a large boundary length compared to the $\ads$ length, i.e., $\ell \gg \lads$. Basically we are interested in the limit where $\bar\chi$ is large, presumably when compared to the scales of the kinetic term $\kappa e^\lambda$ and the boundary values $\lambda|_\partial = 2 \log \psi'$, but the precise conditions for the approximations to hold will be defined later. Note that this is also achieved for very late or earlier times, when $|\tau| \gg \ell^{-1}$. In this limit we expect that the contribution of the kinetic term should be subdominant when compared to the ``external force'' $\bar\chi e^\lambda$. Thus, it should in principle be possible to perturbatively expand the Hamiltonian around the zero-momentum solution, which we know exactly. Instead of tackling the full problem, let us consider a simpler version of it where we expand around the symmetric diamond state $(\psi, \ac\sigma) = (\id, 0)$. 

Let us assume that, in a certain neighborhood of the state $(\psi, \ac\sigma) = (\id, 0)$, the solution to the Lichnerowicz equation \eqref{lambdaLich3} can be written as
\be
\lambda = \lambda_0 + \delta\lambda
\ee
where $\lambda_0$ is the solution associated with the $(\id, 0)$, given in \eqref{lambda0}, and $\delta\lambda \ll 1$ is a small correction. The Lichnerowicz equation then reads
\be
\bar\nabla^2 (\lambda_0 + \delta\lambda) + \kappa e^{-\lambda_0} e^{-\delta\lambda}  -  \bar\chi e^{\lambda_0}e^{\delta\lambda} = 0
\ee
which can be approximated, to first order in $\delta\lambda$, as
\be\label{deltalambdaLich}
\bar\nabla^2 \delta\lambda -  \bar\chi e^{\lambda_0}\delta\lambda = - \kappa e^{-\lambda_0}
\ee
where we expanded $e^{\delta\lambda} \approx 1 + \delta\lambda$ and neglected the (higher order) term $\kappa\delta\lambda$. Since $\lambda_0|_\partial = 0$, the boundary condition for $\lambda$ must be entirely supplied by $\delta\lambda|_\partial = 2 \log \psi'$. Since $\delta\lambda \ll 1$ we must require
\be
2 \log \psi' \ll 1
\ee
that is, the diffeomorphism must be sufficiently close to the identity.
Notice that if $\kappa = 0$ and $\psi' = 1$ the perturbation would vanish and the solution would be $\lambda = \lambda_0$, as expected. It is convenient to decompose the perturbation into two parts depending on the ``source'' causing $\delta\lambda$ not to vanish. Namely, write $\delta\lambda = \delta\lambda_q + \delta\lambda_p$, where $\delta\lambda_q$ and $\delta\lambda_p$ are defined by the equations
\begin{align}
\bar\nabla^2 \delta\lambda_q -  \bar\chi e^{\lambda_0}\delta\lambda_q &=0 \,,\qquad\qquad\! \delta\lambda_q|_\partial = 2 \log \psi'  \label{deltalambdaq}\\
\bar\nabla^2 \delta\lambda_p -  \bar\chi e^{\lambda_0}\delta\lambda_p &= - \kappa e^{-\lambda_0} \,,\quad \delta\lambda_p|_\partial = 0 \label{deltalambdap}
\end{align}
In this way, $\delta\lambda_q$ satisfies a homogeneous equation ``sourced'' by a nontrivial boundary condition (so it is associated with ``position'' variables $q = \psi$) and $\delta\lambda_p$ satisfies an inhomogeneous equation, ``sourced'' by the kinetic term, with a trivial boundary condition (so it is associated with ``momentum'' variables $p = \ac\sigma$). Let us analyze the typical behaviors of these two equations below.

\begin{figure}
\centering
\includegraphics[scale = 0.5]{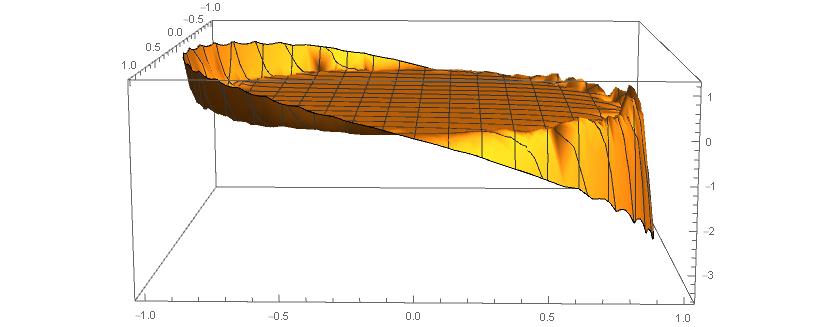}
\caption{Typical solution for $\delta\lambda_q$ in the large $\bar\chi$ limit.}
\label{fig:lambdaq}
\end{figure}

The position-sourced perturbation equation, \eqref{deltalambdaq}, can be cast as a minimization problem associated with the functional
\be
I_q[u] = \frac{1}{2} \int\!d^2x \left( |\bar\nabla u|^2 + \bar\chi e^{\lambda_0} u^2 \right)
\ee
The minimizer $u : D \rightarrow \bb R$ of $I_q$, satisfying the boundary condition $u|_\partial = 2 \log \psi'$,  is the solution $\delta\lambda_q$ of \eqref{deltalambdaq}. Using the ``elastic membrane'' interpretation discussed around \eqref{LichIfunctional}, we see that the term $|\bar\nabla u|^2$ tries to keep the membrane as flat as possible and term $u^2$ tried to keep it as close to $u=0$ as possible. The $u^2$ term comes accompanied by the factor $\bar\chi e^{\lambda_0}$. In the limit $\bar\chi \gg 1$ we can approximate $C$ in \eqref{lambda0} by $(\bar\chi + \sqrt{8\bar\chi})/\bar\chi^2$, so we have
\be
\bar\chi e^{\lambda_0} \approx \frac{8}{\left(1 - r^2 + \sqrt{8/\bar\chi} \right)^2}
\ee
Note that this factor is $\sim 10$ in most of the interior of the disk, growing to a huge value $\sim \chi$ near the boundary. This means that, if $\bar\chi$ is large enough, it will bring the membrane to zero as soon as it leaves the boundary, which would then just remain close to zero in the interior. See an example in Fig. \ref{fig:lambdaq}.

The momentum-sourced perturbation equation, \eqref{deltalambdap}, can also be cast as a minimization problem, associated with the functional
\be
I_p[u] = \frac{1}{2} \int\!d^2x \left( |\bar\nabla u|^2 + \bar\chi e^{\lambda_0} u^2 - 2\kappa e^{-\lambda_0}u \right) 
\ee
The minimizer, satisfying the boundary condition $u|_\partial = 0$,  is the solution $\delta\lambda_p$ of \eqref{deltalambdap}. In the membrane interpretation, the terms $|\bar\nabla u|^2$and $u^2$ act like before, and the term linear in $u$ tries to pull the membrane up. The coefficient of the linear term is ($-2$ times)
\be
\kappa e^{-\lambda_0} \approx  \frac{\kappa\bar\chi }{8} \left(1 - r^2 + \sqrt{8/\bar\chi} \right)^2
\ee
Note that this factor is of order $\sim \kappa\bar\chi/8$ in most of the interior of the disc, going to $\sim \kappa$ near the boundary. Therefore the kinetic term may actually have a significant effect near the center of the disc. To estimate its influence, we combine the $u^2$ and the $u$ terms by ``completing the square'',
\be
\bar\chi e^{\lambda_0} u^2 - 2\kappa e^{-\lambda_0}u = \bar\chi e^{\lambda_0} \left(u - \frac{\kappa e^{-\lambda_0}}{\bar\chi e^{\lambda_0}} \right)^2 - \frac{\kappa^2 }{\bar\chi}  e^{-3\lambda_0}
\ee
Hence we see that the effect of these terms combined in the functional is to pull the membrane toward the ``equilibrium'' configuration $u_\text{eq}$,
\be
u_\text{eq} := \frac{\kappa e^{-\lambda_0}}{\bar\chi e^{\lambda_0}} = 4\pi^2 \ell_P^2 \chi \bar\sigma^2 \left(1 - r^2 + \sqrt{8/\bar\chi} \right)^4
\ee
where we used the definition \eqref{Lichkappadef} of $\kappa$ in terms of  $\bar\sigma^2 := \bar\sigma^{ab} \bar\sigma_{ab}$ and the definition \eqref{Lichbarchidef} of $\bar\chi$ in terms of $\chi$. 
The term $|\bar\nabla u|^2$ will offer some resistance to deforming membrane from $u = 0$ (the flattest configuration) to $u_\text{eq}$, so we expect that $u$ will be (roughly) bounded by $u_\text{eq}$. Nonetheless, we still expect that the minimizer $u$ will be of the same magnitude as $u_\text{eq}$, so in order for our approximation to be legitimate (i.e., $u \ll 1$) we need to assume $u_\text{eq} \ll 1$, which implies
\be
\bar\sigma^2 \ll  \frac{(\ell/\ell_P)^2}{16\pi^4 \bar\chi}
\ee
If, for example, $\tau^2$ is not large so that $\chi \sim 2/\lads^2$, then the right-hand side becomes $\sim \lads^2/8\pi^2 \ell_P^2$, which can be satisfied for a large range of $\ac\sigma$ if $\lads \gg \ell_P$. Let us now consider the behavior of the solution near the boundary. There the coefficient $2\kappa e^{-\lambda_0}$ is suppressed by a factor of $\sim 8/\bar\chi$ while the coefficient $\bar\chi e^{\lambda_0}$ is enhanced and attains values of order $\sim\bar\chi$. Consequently, the equation tries to make the membrane very flat as it approaches the boundary. In particular, we expect the normal derivative of $\delta\lambda_p$ to be very small near the boundary. 

Let us now proceed to find an approximated expression for the Hamiltonian $\wt H$ in this regime. It is interesting that we can obtain a general result for $\wt H$ without the need to fully solve \eqref{deltalambdaLich}. Expanding $e^{\delta\lambda} \approx 1 + \delta\lambda$ in \eqref{Hred2} we get
\be
\wt H \approx \wt H_0 + \frac{\ell^2}{64\pi^3 \ell_P} \int \! d^2x\, e^{\lambda_0} \delta\lambda
\ee
where $\wt H_0$ is given exactly in \eqref{wtH0exact}, but can also be approximated in the large $\bar\chi$ limit as
\be
\wt H_0 \approx \frac{\ell^2}{16\pi^2 \ell_P \sqrt{2\bar\chi}}
\ee
Considering the decomposition of $\delta\lambda$ in $\delta\lambda_q$ and $\delta\lambda_p$, we can define two parts for the Hamiltonian perturbation,
\begin{align}
\delta \wt H_q &:= \frac{\ell^2}{64\pi^3 \ell_P} \int \! d^2x\, e^{\lambda_0} \delta\lambda_q \label{wtHq}\\
\delta \wt H_p &:= \frac{\ell^2}{64\pi^3 \ell_P} \int \! d^2x\, e^{\lambda_0} \delta\lambda_p \label{wtHp}
\end{align}
so that $\wt H \approx \wt H_0 + \delta \wt H_q  + \delta \wt H_p$. The slick observation here is that $\delta\lambda$ is integrated against a rotationally-symmetric function, $e^{\lambda_0}$, implying that only the zero Fourier mode of $\delta\lambda$ contributes to the Hamiltonian. Moreover, since $\delta\lambda$ satisfies a linear equation associated with the rotationally-symmetric operator $\bar\nabla^2 -  \bar\chi e^{\lambda_0}$, the Fourier modes are all independent and we can therefore solve specifically for the zero mode. Consider the zero modes of the ``sources'' given by
\begin{align}
\Upsilon_0 &:= \frac{1}{2\pi} \int\!d\theta\, 2 \log \psi'(\theta) \\
K_0(r) &:= \frac{1}{2\pi} \int\!d\theta\, \kappa(r,\theta) e^{-\lambda_0(r)}
\end{align}
and the zero modes of the perturbations
\begin{align}
\delta\lambda_q^{(0)} &:= \frac{1}{2\pi} \int\!d\theta\, \delta\lambda_q \\
\delta\lambda_p^{(0)} &:= \frac{1}{2\pi} \int\!d\theta\, \delta\lambda_p
\end{align}
Integrating \eqref{deltalambdaq} and \eqref{deltalambdap} in $\theta$, using the polar expression for the Laplacian, we get
\begin{align}
\frac{1}{r} \frac{\partial}{\partial r} \left( r \frac{\partial \delta\lambda_q^{(0)}}{\partial r} \right) -  \bar\chi e^{\lambda_0} \delta\lambda_q^{(0)} &= 0 \,,\qquad\quad\!\! \delta\lambda_q^{(0)}|_\partial = \Upsilon_0  \label{deltalambdaq0}\\
\frac{1}{r} \frac{\partial}{\partial r} \left( r \frac{\partial \delta\lambda_p^{(0)}}{\partial r} \right) -  \bar\chi e^{\lambda_0} \delta\lambda_p^{(0)} &= - K_0  \,,\quad \delta\lambda_p^{(0)}|_\partial = 0 \label{deltalambdap0}
\end{align}
Note that the term proportional to $\bar\chi$ is precisely what appears in the integrands of \eqref{wtHq} and \eqref{wtHp}. Therefore, if we integrate the equations above in $rdr$ we should get
\begin{align}
\delta \wt H_q &= \frac{\ell^2}{64\pi^3 \ell_P} 2\pi \int_0^1 \! rdr\, \frac{1}{\bar\chi} \left[ \frac{1}{r} \frac{\partial}{\partial r} \left( r \frac{\partial \delta\lambda_q^{(0)}}{\partial r} \right) \right] \\
\delta \wt H_p &= \frac{\ell^2}{64\pi^3 \ell_P} 2\pi \int_0^1 \! rdr\, \frac{1}{\bar\chi} \left[ \frac{1}{r} \frac{\partial}{\partial r} \left( r \frac{\partial \delta\lambda_p^{(0)}}{\partial r} \right) + K_0 \right]
\end{align}
which gives
\begin{align}
\delta \wt H_q &= \frac{\ell^2}{32\pi^2 \ell_P} \frac{1}{\bar\chi} \left.\frac{\partial \delta\lambda_q^{(0)}}{\partial r}\right|_{r = 1}  \\
\delta \wt H_p &= \frac{\ell^2}{32\pi^2 \ell_P} \left( \frac{1}{\bar\chi} \left.\frac{\partial \delta\lambda_p^{(0)}}{\partial r}\right|_{r = 1} + \frac{1}{16\pi}  \int_0^1 \! dr\, r(1-r^2)^2 \int\!d\theta\, \kappa \right)
\end{align}
where in the last integral, involving $\kappa$, we approximated $(1-r^2 + \sqrt{8/\bar\chi})^2 \approx (1-r^2)^2$.

In our study of the typical behavior of $\delta\lambda_q$ and $\delta\lambda_p$, we noted that $\delta\lambda_q$ grows steeply near the boundary in the limit of large $\bar\chi$. In fact we can show that, in this limit, 
\be\label{deltalambdaqder0}
\left.\frac{\partial \delta\lambda_q^{(0)}}{\partial r}\right|_{r = 1} = \Upsilon_0 \sqrt{\frac{\bar\chi}{2}}
\ee
The proof goes as follows. The function multiplying $\delta\lambda_q^{(0)}$ in \eqref{deltalambdaq0} ca be approximated by
\be
\bar\chi e^{\lambda_0} = \frac{8}{(1+\sqrt{8/\bar\chi} - r^2)^2} \approx \frac{8}{(1-\wt r^2)^2}
\ee
where
\be
\wt r := \frac{r}{1 + \sqrt{2/\bar\chi}}
\ee
In this new variable $\wt r$, equation \eqref{deltalambdaq0} becomes approximately
\be
\frac{1}{\wt r} \frac{d}{d \wt r} \left( \wt r \frac{d \delta\lambda_q^{(0)}}{d \wt r} \right)- \frac{8 \delta\lambda_q^{(0)}}{(1 - \wt r^2)^2} = 0
\ee
whose general solution reads
\be
\delta\lambda_q^{(0)} = \frac{c_1 (1 + \wt r^2) + c_2 (2 + \log \wt r - \wt r^2 \log \wt r)}{1 - \wt r^2}
\ee
for real coefficients $c_1$ and $c_2$. In order for $\delta\lambda_q^{(0)}$ to be regular at $r = 0$ ($\wt r = 0$) it is necessary that $c_2 = 0$. The boundary condition is that $\delta\lambda_q^{(0)} = \Upsilon_0$ at $r = 1$, or $\wt r = 1/(1 + \sqrt{2/\bar\chi})$, which implies
\be
c_1 = \Upsilon_0 \sqrt{\frac{2}{\bar\chi}}
\ee
The first derivative in $r$ gives
\be
\frac{d\delta\lambda_q^{(0)}}{dr} \approx \frac{d\delta\lambda_q^{(0)}}{d\wt r} = \Upsilon_0 \sqrt{\frac{2}{\bar\chi}} \frac{4 \wt r}{(1-\wt r^2)^2}
\ee
Note that at $r = 0$ we have $d\delta\lambda_q^{(0)}/dr = 0$, as expected (to avoid a kink at the origin). At $r=1$, we obtain \eqref{deltalambdaqder0}.

The above result yields,
\be
\delta \wt H_q = \frac{\ell \lads}{16\pi \ell_P \sqrt{2\chi}} \Upsilon_0
\ee
where we have written $\bar\chi$ in term of $\chi$. We should note however that, to first order in $\delta\lambda$, $\Upsilon_0$ actually vanishes. This follows from the fact that $\psi$ is a diffeomorphism, so
\be
2\pi = \int\!d\theta\, \psi'(\theta) = \int\!d\theta\, e^{\log \psi'} \approx \int\!d\theta\, (1 + \log \psi') = 2\pi + \pi \Upsilon_0
\ee
Hence, in our approximations, 
\be
\delta \wt H_q = 0
\ee
We have indicated that $\delta\lambda_p$ tends to flatness near the boundary, so the derivative term is subdominant in this case. 
Therefore we have
\be
\delta \wt H_p = 2\pi \ell_P \int_0^1 \! dr\, r(1-r^2)^2 \int\!d\theta\, \bar\sigma^2 
\ee
Using the Fourier expansion of $\bar\sigma^{\mu\nu}$, given in \eqref{sigmaab}, we obtain
\be
\bar\sigma^2 := \bar\sigma^{\mu\nu}\bar\sigma_{\mu\nu} = \sum_{n,m \ge 0} 4 r^{n+m} (a_n a_m + b_n b_m) \cos[(n-m)\theta]
\ee
whose zero mode is
\be
\frac{1}{2\pi} \int\!d\theta\, \bar\sigma^2 = \sum_{n \ge 0} 4 r^{2n} (a_n^2 + b_n^2)
\ee
Finally, using the result
\be
\int_0^1 \! dr\, r^{2n+1} (1-r^2)^2 = \frac{1}{(n+1)(n+2)(n+3)}
\ee
we obtain
\be
\delta \wt H_p = 16\pi^2 \ell_P \sum_{n \ge 0} \frac{a_n^2 + b_n^2}{(n+1)(n+2)(n+3)}
\ee
Hence, the Hamiltonian in this regime becomes approximately
\be
\wt H \approx 16\pi^2 \ell_P \sum_{n \ge 0} \frac{a_n^2 + b_n^2}{(n+1)(n+2)(n+3)} + \frac{\ell \lads}{8\pi \ell_P \sqrt{2\chi}}
\ee
That is, in this regime, the system behaves as a (countably infinite) collection of independent free ``Newtonian particles'', each with kinetic energy $p_n^2/2M_n$, where the ``effective mass'' is given by
\be
M_n \sim \frac{(n+1)(n+2)(n+3)}{32\pi^2 \ell_P}
\ee
However, we should only read off the ``effective masses'' after expressing the coefficients $a_n$ and $b_n$ in terms of the canonical momenta $P_n$, which is a topic for Part II. This detail aside, the masses above tend to grow with the mode number $n$, which is satisfying since this means that the modes associated with a more oscillatory diamond corner\footnote{As we are going to discuss later, the ``momentum variables'' are related to how the corner loop of the diamond wiggles in a lightlike direction, while the ``position variables'' are related to how it wiggles in a spacelike direction. This interpretation also provides a way to see that ``larger $p$'' tends to increase the Hamiltonian (since a more lightlike boundary can enclose a larger area for a given perimeter)  and a ``larger $q$'' tends to decrease the Hamiltonian (since more spatial wiggling tends to shrink the enclosed area).} are more massive, and thus more difficult to ``excite''. Finally, we remark that this approximate form of the Hamiltonian should be interpreted with care, as we must not forget that the exact Hamiltonian, although bounded from below, does not have a minimum (similar to an  exponential, $e^x > 0$). Therefore, the symmetric diamond is not the ``ground state'', as one could perhaps think by just looking at this approximated Hamiltonian, but there are states with lower ``energy''. 

It would be interesting to carry out the Hamiltonian expansion up to higher orders, where higher terms in $e^{\delta\lambda}$ would be kept. At second order we expect a non-trivial potential $V(\psi)$ to appear. Note that, at first order, one could be erroneously led to believe that one is expanding around a minimum of the Hamiltonian, since $p^2/2M$ is minimized at $p=0$. However, we know, non-perturbatively, that the symmetric state $(\psi, \ac\sigma) = (I, 0)$ is not a minimum of the Hamiltonian, but rather a saddle point. In fact, since the Hamiltonian corresponds to the area of the CMC slice, it is bounded from below (by zero) but it has no minimum (since there is no regular state with exactly zero area). This feature is reminiscent of Liouville field theory, whose potential has an exponential form, $V(\phi) = e^{2b\phi}$, and has shown many links to low-dimensional gravity~\cite{seiberg1990notes,balog1998coadjoint,krasnov2001three,nakayama2004liouville,faddeev2014zero,erbin2015notes,li2020liouville}.

\bibliographystyle{ieeetr}
\bibliography{CausalDiamonds}

\end{document}